\documentclass[]{aa}
\usepackage[toc,page]{appendix}
\usepackage[ruled,vlined,linesnumbered]{algorithm2e}
\makeatletter

\renewcommand{\theAlgoLine}{%
  \@arabic{\numexpr\value{algocf}+1\relax}.\arabic{AlgoLine}}
\makeatother

\usepackage{amssymb}
\usepackage{multirow}
\usepackage{natbib}
\usepackage{url}
\usepackage{caption}
\usepackage{subcaption}
\usepackage{graphicx}
\usepackage{soul}
\usepackage{xcolor}
\usepackage[procnames]{listings}

\usepackage{graphicx}
\usepackage{enumerate}   
\usepackage{enumitem}   
\usepackage{amssymb}
\usepackage{amsmath}
\usepackage{verbatim}
\usepackage{natbib}
\usepackage{rotate}
\usepackage{lscape}
\usepackage{aalongtable}
\usepackage{supertabular}
\usepackage{mathbbol}
\usepackage{bm}
\usepackage{mathtools}
\usepackage{enumerate}
\usepackage{booktabs}

\usepackage[varg]{txfonts}

\let\oldsqrt\sqrt
\def\sqrt{\mathpalette\DHLhksqrt}
\def\DHLhksqrt#1#2{%
\setbox0=\hbox{$#1\oldsqrt{#2\,}$}\dimen0=\ht0
\advance\dimen0-0.2\ht0
\setbox2=\hbox{\vrule height\ht0 depth -\dimen0}%
{\box0\lower0.4pt\box2}}

\def\aap{A\&A}

\def\u{{u}}
\def\v{{v}}
\def\w{{w}}
\def\l{{l}}
\def\m{{m}}
\def\n{{n}}

\def\RIME/{{\sc rime}}
\def\CLEAN/{{\sc clean}}
\def\Sec/{Sect.}
\def\MTMSCLEAN/{{\sc mtms-clean}}
\def\SSDCLEAN/{{\sc ssd-clean}}
\def\FACTOR/{{\sc factor}}
\def\KMS/{{\sc kms}}
\def\WSCLEAN/{{\sc ssd-clean}}
\def\DDFacet/{{\sc ddf}acet}
\def\SSD/{{\sc ssd}}
\def\BDA/{{\sc bda}}
\def\SSDGA/{\SSD/{\sc ga}}
\def\PMP/{{\sc hmp}}
\def\FW/{{\sc fw}}

\def\Montblanc/{{\sc montblanc}}

\newcommand{\OVec}[1]{\mathrm{Vec}\left\{#1\right\}}

\def\kron{\otimes}

\def\JonesMat{\bm{\mathrm{G}}}

\def\Unity{\bm{\mathrm{I}}}

\newcommand{\OMSOK}[1]{}

\renewcommand{\emph}[1]{"#1"}

\newcommand{\RA}[3]{$#1$h$#2$m$#3$s}
\newcommand{\DEC}[3]{$#1^{\circ}#2$\arcmin$#3$\arcsec}
\newcommand{\ANSW}[1]{#1}
\newcommand{\ANSWII}[1]{#1}

\def\PREFACTOR/{{\sc PreFactor}}
\def\LOTSS/{{\sc LoTSS}}
\def\LoLSS/{{\sc LoLSS}}
\def\LoTSS/{{\sc LoTSS}}
\def\LoTSSd/{{\sc LoTSS}-Deep Fields}
\def\SKSP/{{\sc sksp}}
\def\TGSS/{{\sc tgss}}
\def\PipeVI/{{\sc ddf-pipeline-v1}}
\def\PipeVII/{{\sc ddf-pipeline-v2}}
\def\PSF/{{\sc psf}}
\def\LOFAR/{LOFAR}
\def\LTA/{{\sc lta}}
\def\kMS/{k{\sc ms}}
\def\ddfpipe/{{\sc ddf-pipeline}}
\def\DI/{{\sc di}}
\def\DD/{{\sc dd}}
\def\DDE/{{\sc dde}}
\def\CRIME/{$\mathcal{C}$-{\sc rime}}
\def\IRIME/{$\mathcal{I}$-{\sc rime}}
\def\CIRIME/{$\mathcal{CI}$-{\sc rime}}
\def\HBA/{{HBA}}
\def\LBA/{{LBA}}
\def\HBADUALINNER/{{\sc hba\_dual\_inner}}
\def\Bootes/{Bo\"otes}
\def\LH/{Lockman Hole}
\def\ELAIS/{ELAIS-N1}
\def\SageCal/{{\sc sagecal}}
\def\SFR/{SFR}
\def\DRI/{DR1}
\def\AOFlagger/{{\sc AOFlagger}}
\def\OBIT/{{\sc obit}}

\def\sqdeg/{deg$^2$}
\def\uJypb/{$\mu$Jy.beam$^{-1}$}
\def\mJypb/{mJy.beam$^{-1}$}
\def\Mspy/{M$_{\odot}$yr$^{-1}$}
\def\LEdd/{L$_{\mathrm{Edd}}$}

\def\SkyX{\textbf{x}}
\def\SkyXnu{\SkyX_{\nu}}
\def\SkyXtnu{\SkyX_{t\nu}}
\def\SkyXsnu{\SkyX_{\vec{s}\nu}}
\def\SkyXnui{\SkyX_{\nu_i}}

\def\OpIm{\mathcal{I}}
\def\OpCal{\mathcal{K}}
\def\OpFit{\mathcal{F}}
\def\OpNorm{\mathcal{N}}
\def\OpBoot{\mathcal{B}}
\def\OpClust{\mathcal{C}}
\def\OpDyn{\mathcal{D}}

\def\JonesMatDI{\bm{\mathrm{G}}}
\def\JonesMat{\bm{\mathrm{J}}}
\def\MuellerMatDI{\bm{\mathcal{G}}}
\def\MuellerMat{\bm{\mathcal{J}}}
\def\MuellerMatBeam{\bm{\mathcal{B}}}
\def\JonesMatBeam{\bm{\mathrm{B}}}

\def\TextCaptionAlgI/{Overview of the algorithm implemented
in \PipeVI/ to produce the \LoTSS/-\DRI/ images.
The function $\OpIm$ represents the imaging step and takes as input
the visibility vector $\textbf{v}$
 together with the beam model $\JonesMatBeam_{\Omega_n}$ and
 \kMS/-estimated Jones matrices $\JonesMat_{\Omega_n}$ at locations
 $\Omega_n$. The function $\OpCal$ abstracts the \DD/ calibration
 step, and takes as arguments the visibilities $\textbf{v}$, the
 skymodel $\widehat{\SkyXnu}$, a solver mode (estimating for either $\mathrm{scalar}$ or
 $\mathrm{full}$ Jones matrices), a time-frequency solution interval
 (in $\mathrm{min}$ and $\mathrm{MHz}$), and a set of directions
 $\Omega_n$ in which to solve for.
 The extra functions $\OpClust$,
 $\OpNorm$, and $\OpBoot$ represent the clustering,
 normalisation (see text), and bootstrapping steps respectively.}

\def\TextCaptionAlgII/{Overview of the algorithm implemented
in \PipeVII/.
The function $\OpIm$ represents the imaging step and takes as input
the visibility vector $\textbf{v}$
 together with the beam model $\JonesMatBeam_{\Omega_n}$ and
 \kMS/-estimated Jones matrices $\JonesMat_{\Omega_n}$ at locations
 $\Omega_n$. The function $\OpCal$ abstracts the \DD/ calibration
 step, and takes as arguments the visibilities $\textbf{v}$, the
 skymodel $\widehat{\SkyXnu}$, a solver mode (estimating for either $\mathrm{scalar}$ or
 $\mathrm{full}$ Jones matrices), a time-frequency solution interval
 (in $\mathrm{min}$ and $\mathrm{MHz}$), and a set of directions
 $\Omega_n$ in which to solve for.
 The extra functions $\OpClust$, $\OpBoot$, and $\OpFit$ represent the clustering, bootstrapping and smoothing steps respectively.}

\def\ALGODRONEB/{
\setcounter{algocf}{-1}
\begin{algorithm}[t!]
\KwData{Visibilities $\textbf{v}$ calibrated from \DI/ effects
 using \PREFACTOR/.
}
 \KwResult{Deconvolved image $\widehat{\SkyXnu}$} 
\algrule
\tcc{{\bf On 60 LOFAR HBA subbands}}
\tcc{DI initial deconv and clustering}
$\widehat{\SkyXnu}\leftarrow\OpIm\left(\textbf{v}_6,\JonesMatBeam_{\Omega_r}\right)$\;
$\Omega_n\leftarrow \OpClust\left(\widehat{\SkyXnu}\right)$\;
\tcc{Phase only DD calibration}
$\widehat{\JonesMat}\leftarrow\varphi\circ\OpCal\left(\textbf{v}_6,\widehat{\SkyXnu},\JonesMatBeam_{\Omega_n}|\mathrm{scalar},1\mathrm{min},2\mathrm{MHz},\Omega_n\right)$\;
\tcc{Absolute flux density scale bootstrapping}
$\widehat{\textbf{v}}\leftarrow\OpBoot\left(\widehat{\textbf{v}_6}\right)$\;
$\widehat{\SkyXnu}\leftarrow\OpIm\left(\textbf{v}_6,\widehat{\JonesMat}\JonesMatBeam_{\Omega_n}\right)$\;
\tcc{DD calibration and imaging}
$\widehat{\JonesMat}\leftarrow\OpNorm\circ\OpCal\left(\textbf{v}_6,\widehat{\SkyXnu},\JonesMatBeam_{\Omega_n}|\mathrm{scalar},1\mathrm{min},2\mathrm{MHz},\Omega_n\right)$\;
$\widehat{\SkyXnu}\leftarrow\OpIm\left(\textbf{v}_6,\widehat{\JonesMat}\JonesMatBeam_{\Omega_n}\right)$\;
\algrule
\tcc{{\bf On 240 LOFAR HBA subbands}}
\tcc{Deep DD calibration and imaging}
$\widehat{\JonesMat}\leftarrow\OpNorm\circ\OpCal\left(\textbf{v}_{24},\widehat{\SkyXnu},\JonesMatBeam_{\Omega_n}|\mathrm{scalar},1\mathrm{min},2\mathrm{MHz},\Omega_n\right)$\;
$\widehat{\SkyXnu}\leftarrow\OpIm\left(\textbf{v}_{24},\widehat{\JonesMat}\JonesMatBeam_{\Omega_n}\right)$\;
 Facet-based astrometric correction \citep[see][for details]{Shimwell18}\;
\caption{\label{alg:DR1} \TextCaptionAlgI/ }
\end{algorithm}
}

\newcommand{\algrule}[1][.2pt]{\par\vskip.5\baselineskip\hrule height #1\par\vskip.5\baselineskip}

\def\ALGODRTWOB/{
\IncMargin{0.5em}
\begin{algorithm}[t!]
\SetAlgoLined
 \KwData{Visibilities $\textbf{v}$ calibrated from \DI/ effects
 using \PREFACTOR/.} 

\algrule
\tcc{{\bf On 60 LOFAR HBA subbands}}
\tcc{DI initial deconv and clustering}

$\widehat{\SkyXnu}\leftarrow\OpIm\left(\textbf{v}_6,\JonesMat_{\Omega_r}={\mathbf 1},\JonesMatBeam_{\Omega_r} \right)$\;\label{step:VII_I_DI0}

$\Omega_n\leftarrow\OpClust \left(\widehat{\SkyXnu}\right)$\;

\tcc{DI calibration and imaging}
$\widehat{\SkyXnu}\leftarrow\OpIm\left(\textbf{v}_6,\JonesMatBeam_{\Omega_n} \right)$\;\label{step:VII_I_DI1}
$\textbf{v}^c_6\leftarrow\OpCal\left(\textbf{v}_6,\widehat{\SkyXnu},\JonesMatBeam_{\Omega_n}|\mathrm{full},\delta t_0,\delta\nu_0,\Omega_0\right)$\;\label{step:VII_K_DI0}
$\widehat{\SkyXnu}\leftarrow\OpIm\left(\textbf{v}^c_6,\JonesMatBeam_{\Omega_n} \right)$\;\label{step:VII_I_DI2}

\tcc{Bootstrapping the flux density scale}
$\textbf{v}^c\leftarrow\OpBoot\left(\textbf{v}^c_6\right)$\;

\tcc{Phase only DD calibration and imaging}
$\widehat{\JonesMat}\leftarrow\varphi\circ\OpFit\circ\OpCal\left(\textbf{v}^c_6,\widehat{\SkyXnu},\JonesMatBeam_{\Omega_n}|\mathrm{scalar},1\mathrm{min},2\mathrm{MHz},\Omega_n\right)$\;\label{step:VII_K_DD0}
$\widehat{\SkyXnu}\leftarrow\OpIm\left(\textbf{v}^c_6,\widehat{\JonesMat}\JonesMatBeam_{\Omega_n}\right)$\;\label{step:VII_I_DD0}

\tcc{DD calibration and imaging}
$\widehat{\JonesMat}\leftarrow\OpFit\circ\OpCal\left(\textbf{v}^c_6,\JonesMatBeam_{\Omega_n},\widehat{\SkyXnu}|\mathrm{scalar},1\mathrm{min},2\mathrm{MHz},\Omega_n\right)$\;\label{step:VII_K_DD1}
$\widehat{\SkyXnu}\leftarrow\OpIm\left(\textbf{v}_6,\widehat{\JonesMat}\JonesMatBeam_{\Omega_n}\right)$\;

\tcc{DI calibration and imaging}
$\textbf{v}^c_6\leftarrow\OpCal\left(\textbf{v}_6,\widehat{\JonesMat}\JonesMatBeam_{\Omega_n},\widehat{\SkyXnu}|\mathrm{full},\delta
t_0,\delta\nu_0,\Omega_0\right)$\;
\label{step:VII_K_DI1}
$\widehat{\SkyXnu}\leftarrow\OpIm\left(\textbf{v}^c_6,\widehat{\JonesMat}\JonesMatBeam_{\Omega_n} \right)$\;\label{step:VII_I_DD1b}

\algrule
\tcc{{\bf On 240 LOFAR HBA subbands}}
\tcc{DD calibration}
$\widehat{\JonesMat}\leftarrow\OpFit\circ\OpCal\left(\textbf{v}_{24},\JonesMatBeam_{\Omega_n},\widehat{\SkyXnu}|\mathrm{scalar},1\mathrm{min},2\mathrm{MHz},\Omega_n\right)$\;
\label{step:VII_K_DD2}
\tcc{DI calibration}
$\textbf{v}_{24}^c\leftarrow\OpCal\left(\textbf{v}_{24},\widehat{\JonesMat}\JonesMatBeam_{\Omega_n},\widehat{\SkyXnu}|\mathrm{full},\delta
t_0,\delta\nu_0,\Omega_0\right)$\;
\label{step:VII_K_DI2}
\tcc{DD imaging}
$\widehat{\SkyXnu}\leftarrow\OpIm\left(\textbf{v}_{24}^c,\widehat{\JonesMat}\JonesMatBeam_{\Omega_n}\right)$\;
 
\tcc{DD calibration}
$\widehat{\JonesMat}\leftarrow\OpFit\circ\OpCal\left(\textbf{v}_{24}^c,\JonesMatBeam_{\Omega_n},\widehat{\SkyXnu}|\mathrm{scalar},1\mathrm{min},2\mathrm{MHz},\Omega_n\right)$\;\label{step:VII_K_DD3}
\tcc{Slow DD calibration}
$\widehat{\JonesMat_s}\leftarrow\OpCal\left(\textbf{v}_{24}^c,\widehat{\JonesMat}\JonesMatBeam_{\Omega_n},\widehat{\SkyXnu}|\mathrm{scalar},43\mathrm{min},2\mathrm{MHz},\Omega_n\right)$\;\label{step:VII_K_DD4}
\tcc{Final imaging steps}
 $\widehat{\SkyXnu}\leftarrow\OpIm\left(\textbf{v}_{24}^c,\widehat{\JonesMat_s}\widehat{\JonesMat}\JonesMatBeam_{\Omega_n}\right)$\;\label{step:VII_I_DD4}

 Facet-based astrometric correction \citep[see][for details]{Shimwell18}\;\label{step:VII_AstroCorr}
 \caption{\label{alg:DR2} \TextCaptionAlgII/ }
\end{algorithm}
}
 
\def\ALGODRTWOBM/{
\IncMargin{0.5em}
\begin{algorithm}[t!]
\SetAlgoLined
 \KwData{Outputs of Alg. \ref{alg:DR2b}: calibrated visibilities
 $\textbf{v}_{24}^c$, \DD/-Jones matrices $\widehat{\JonesMat_s}$, and
 $\widehat{\JonesMat}$}
 \KwResult{Deconvolved image $\widehat{\SkyXnu}$} 
\algrule
\tcc{{\bf Miscellaneous data products}}
\tcc{Low $20$\arcsec resolution image }
 $\widehat{\SkyXnu^{\mathrm{low}}}\leftarrow\OpIm\left(\textbf{v}_{24}^c,\widehat{\JonesMat_s}\widehat{\JonesMat}\JonesMatBeam_{\Omega_n}\right)$\;\label{step:VII_I_DD4_LR}
\tcc{Stokes I image in 3 frequency chunks over the whole hba bandwidth}
 $\widehat{\SkyXnui}\leftarrow\OpIm\left(\textbf{v}_{24}^c,\widehat{\JonesMat_s}\widehat{\JonesMat}\JonesMatBeam_{\Omega_n}\right)$\;\label{step:VII_I_DD4_HRi}
\tcc{Low $20$\arcsec resolution QU frequency cube}
 $\widehat{\SkyXnui^{\mathrm{low,IQUV}}}\leftarrow\OpIm\left(\textbf{v}_{24}^c,\widehat{\JonesMat_s}\widehat{\JonesMat}\JonesMatBeam_{\Omega_n}\right)$\;\label{step:VII_I_DD4_LR_QU}
\tcc{Very low $4.2$\arcmin resolution QU frequency cube}
 $\widehat{\SkyXnui^{\mathrm{vlow,QU}}}\leftarrow\OpIm\left(\textbf{v}_{24}^c,\widehat{\JonesMat_s}\widehat{\JonesMat}\JonesMatBeam_{\Omega_n}\right)$\;\label{step:VII_I_DD4_VLR_QU}
\tcc{Low $20$\arcsec resolution V continuum image}
 $\widehat{\SkyXnui^{\mathrm{low,V}}}\leftarrow\OpIm\left(\textbf{v}_{24}^c,\widehat{\JonesMat_s}\widehat{\JonesMat}\JonesMatBeam_{\Omega_n}\right)$\;\label{step:VII_I_DD4_LR_V}
\tcc{Make IQUV dynamic spectra}
 $\widehat{\SkyXtnu}\leftarrow\OpDyn\left(\textbf{v}_{24}^c,\textbf{x}_{s_t}, \widehat{\JonesMat_s}\widehat{\JonesMat}\JonesMatBeam_{\Omega_n}\right)$\;\label{step:VII_DynSpec}
 \caption{\label{alg:DR2b} La}
\end{algorithm}
}

\def\ALGODRTWOBDEEP/{
\IncMargin{0.5em}
\begin{algorithm}[t!]
\SetAlgoLined
 \KwData{Visibilities $\textbf{v}$ calibrated from \DI/ effects
 using \PREFACTOR/ of $n_p\times8$ hours observations (each with
 $240$ \LOFAR/-\HBA/ subbands),
 as
 well as the high resolution skymodel built in step \ref{step:VII_I_DD4}.}
 \KwResult{Deconvolved image $\widehat{\SkyXnu}$} 

\algrule
\tcc{{\bf On $n_p\times$240 LOFAR HBA subbands}}
\tcc{DD calibration}
$\widehat{\JonesMat}\leftarrow\OpFit\circ\OpCal\left(\textbf{v}_{n_p\times24},\JonesMatBeam_{\Omega_n},\widehat{\SkyXnu}|\mathrm{scalar},1\mathrm{min},2\mathrm{MHz},\Omega_n\right)$\;
\label{step:VII_K_DD2_DEEP}
\tcc{DI calibration}
$\textbf{v}_{n_p\times24}^c\leftarrow\OpCal\left(\textbf{v}_{n_p\times24},\widehat{\JonesMat}\JonesMatBeam_{\Omega_n},\widehat{\SkyXnu}|\mathrm{full},\delta
t_0,\delta\nu_0,\Omega_0\right)$\;
\label{step:VII_K_DI2_DEEP}
\tcc{DD imaging}
$\widehat{\SkyXnu}\leftarrow\OpIm\left(\textbf{v}_{n_p\times24}^c,\widehat{\JonesMat}\JonesMatBeam_{\Omega_n}\right)$\;
 
\tcc{DD calibration}
$\widehat{\JonesMat}\leftarrow\OpFit\circ\OpCal\left(\textbf{v}_{n_p\times24}^c,\JonesMatBeam_{\Omega_n},\widehat{\SkyXnu}|\mathrm{scalar},1\mathrm{min},2\mathrm{MHz},\Omega_n\right)$\;\label{step:VII_K_DD3_DEEP}
\tcc{Slow DD calibration}
$\widehat{\JonesMat_s}\leftarrow\OpCal\left(\textbf{v}_{n_p\times24}^c,\widehat{\JonesMat}\JonesMatBeam_{\Omega_n},\widehat{\SkyXnu}|\mathrm{scalar},43\mathrm{min},2\mathrm{MHz},\Omega_n\right)$\;\label{step:VII_K_DD4_DEEP}
\tcc{Final imaging steps}
 $\widehat{\SkyXnu}\leftarrow\OpIm\left(\textbf{v}_{n_p\times24}^c,\widehat{\JonesMat_s}\widehat{\JonesMat}\JonesMatBeam_{\Omega_n}\right)$\;\label{step:VII_I_DD4_DEEP}

\tcc{Absolute flux density scale correction \citep[see][for details]{LoTSSDeepII}}
$\widehat{\SkyXnu}\leftarrow \mathrm{f_c} \widehat{\SkyXnu}$\;
 Facet-based astrometric correction \citep[see][for details]{Shimwell18}\;\label{step:VII_AstroCorr_DEEP}

 \caption{\label{alg:DR2_DEEP} \TextCaptionAlgII/
 }
\end{algorithm}
}

\usepackage{verbatim}
\usepackage{alltt}
\usepackage{upgreek}

\authorrunning{C. Tasse}

\title{The LOFAR Two Meter Sky Survey: Deep Fields}

\subtitle{I - Direction-dependent calibration and imaging}

\author{
  Tasse, C.\inst{\ref{inst:GEPI},\ref{inst:RATTS},\ref{inst:USN}} \and
  Shimwell, T.\inst{\ref{inst:ASTRON},\ref{inst:LeidenStrw}} \and
  Hardcastle, M.J.\inst{\ref{inst:Herts}}  \and
O'Sullivan, S.P.\inst{\ref{inst:Dublin}}  \and
  van Weeren, R.\inst{\ref{inst:LeidenStrw}}  \and
Best, P.N.\inst{\ref{inst:SUPA}} \and
Bester, L.\inst{\ref{inst:SARAO},\ref{inst:RATTS}}\and
Hugo, B.\inst{\ref{inst:SARAO},\ref{inst:RATTS}}\and
Smirnov, O.\inst{\ref{inst:RATTS},\ref{inst:SARAO}}\and
Sabater, J.\inst{\ref{inst:SUPA}}\and
  Calistro-Rivera, G.\inst{\ref{inst:Garching}}\and
de Gasperin, F.\inst{\ref{inst:Hamburg}}\and
Morabito, L.K.\inst{\ref{inst:Durham}}\and
  R\"ottgering, H.\inst{\ref{inst:LeidenStrw}}  \and
Williams, W.L. \inst{\ref{inst:LeidenStrw}}  \and
Bonato, M.\inst{\ref{inst:INAF},\ref{inst:INAF-Padova},\ref{inst:ALMA-it}}\and
Bondi, M.\inst{\ref{inst:INAF}}\and
Botteon, A.\inst{\ref{inst:LeidenStrw},\ref{inst:INAF}}\and
Brüggen, M.\inst{\ref{inst:Hamburg}}\and
Brunetti, G.\inst{\ref{inst:INAF}}\and
Chy\.zy K.T.\inst{\ref{inst:Krakow}}\and
Garrett, M.A. \inst{\ref{inst:Jodrell},\ref{inst:LeidenStrw}}\and
  G\"urkan, G.\inst{\ref{inst:CSIRO}}\and
  Jarvis, M.J. \inst{\ref{inst:Oxford},\ref{inst:UWC}} \and
Kondapally, R.\inst{\ref{inst:SUPA}}\and
Mandal, S.\inst{\ref{inst:LeidenStrw}}  \and
Prandoni, I.\inst{\ref{inst:INAF}}\and
Repetti, A.\inst{\ref{inst:Heriot}}
Retana-Montenegro, E.\inst{\ref{inst:Durban}}\and
Schwarz, D.J.\inst{\ref{inst:Bielefeld}}\and
Shulevski, A.\inst{\ref{inst:Amsterdam}}\and
Wiaux, Y.\inst{\ref{inst:Heriot}}
  }

\institute{
  \label{inst:GEPI} GEPI, Observatoire de Paris, CNRS, Universit\'e Paris Diderot,
    5 place Jules Janssen, 92190 Meudon, France
  \and      
    \label{inst:RATTS} Centre for Radio Astronomy Techniques and Technologies, Department of
Physics and Electronics, Rhodes University, Grahamstown 6140, South
Africa
  \and      
    \label{inst:USN} USN, Observatoire de Paris, CNRS, PSL, UO, Nançay, France
  \and      
\label{inst:ASTRON}ASTRON, Netherlands Institute for Radio Astronomy, Oude
    Hoogeveensedijk 4, 7991 PD, Dwingeloo, The Netherlands
\and
  \label{inst:LeidenStrw} Leiden Observatory, Leiden University, PO
    Box 9513, NL-2300 RA Leiden, The Netherlands.
  \and
  \label{inst:Herts} Centre for Astrophysics Research, University of
  Hertfordshire, College Lane, Hatfield AL10 9AB, UK
    \and
    \label{inst:Dublin}
    School of Physical Sciences and Centre for Astrophysics \& Relativity,
Dublin City University, Glasnevin, D09 W6Y4, Ireland
  \and
  \label{inst:SUPA} SUPA, Institute for Astronomy, Royal Observatory,
  Blackford Hill, Edinburgh, EH9 3HJ, UK
    \and
  \label{inst:SARAO} South African Radio Astronomy Observatory, Observatory 7925, Cape
Town, South Africa
\and
    \label{inst:Garching} European Southern Observatory,
    Karl-Schwarzchild-Strasse 2, 85748, Garching bei M\"unchen,
    Germany
    \and
    \label{inst:Hamburg} Hamburger Sternwarte, University of
    Hamburg, Gojenbergsweg 112, 21029 Hamburg
  \and
  \label{inst:Durham} Centre for Extragalactic Astronomy, Department
    of Physics, Durham University, Durham, DH1 3LE, UK
  \and
  \label{inst:INAF} INAF - Istituto di Radioastronomia, Via P. Gobetti
    101, 40129, Bologna, Italy
  \and
    \label{inst:INAF-Padova} INAF-Osservatorio Astronomico di Padova,
    Vicolo dell'Osservatorio 5, I-35122, Padova, Italy.
   \and
\label{inst:ALMA-it} Italian ALMA Regional Centre, Via Gobetti 101,
    I-40129, Bologna, Italy
    \and
    \label{inst:Krakow} Astronomical Observatory, Jagiellonian University, ul. Orla 171, 30-244, Krak\'ow, Poland	
    \and
\label{inst:Jodrell} Jodrell Bank Centre for Astrophysics, University of Manchester, Alan
  Turing Building, Oxford Road, M13 9PL, UK
  \and
\label{inst:CSIRO} CSIRO Astronomy and Space Science, PO Box 1130, Bentley WA 6102, Australia
  \and
  \label{inst:Oxford} Astrophysics, Department of Physics, Keble Road, Oxford, OX1 3RH, UK
  \and
  \label{inst:UWC} Department of Physics \& Astronomy, University of the
                 Western Cape, Private Bag X17, Bellville, Cape Town, 7535, South
                 Africa
  \and
  \label{inst:Heriot} Institute of Sensors, Signals and Systems, Heriot-Watt University, Edinburgh EH14 4AS, United Kingdom
  \and
\label{inst:Durban} Astrophysics \& Cosmology Research Unit, School
    of Mathematics, Statistics \& Computer Science, University of
    KwaZulu-Natal, Durban, 3690, South Africa
    \and
    \label{inst:Bielefeld} Fakult\"at f\"ur Physik, Universit\"at
    Bielefeld, Postfach 100131, 33501 Bielefeld, Germany
    \and
    \label{inst:Amsterdam} Anton Pannekoek Institute for Astronomy,
    University of Amsterdam, Postbus 94249, 1090 GE Amsterdam, The
    Netherlands
}

\abstract{The Low Frequency Array (\LOFAR/) is an ideal instrument to
  conduct deep extragalactic surveys. It has a large field of view
  and is sensitive to large scale and compact emission. It is, however,
  very challenging to synthesize thermal noise limited maps at full
  resolution, mainly because of the complexity of the low-frequency
  sky and the direction dependent effects (phased array beams and
  ionosphere). In this first paper of a series we present a new
  calibration and imaging pipeline that aims at producing high
  fidelity, high dynamic range images with \LOFAR/ High Band Antenna data, while being
  computationally efficient and robust against the absorption of
  unmodeled radio emission. We apply this calibration and imaging
  strategy to synthesize deep images of the \Bootes/ and \LH/ fields
  at $\sim150$ MHz, totaling $\sim80$ and $\sim100$ hours of integration
  respectively and reaching unprecedented noise levels at these low frequencies of $\lesssim30$
  and $\lesssim23$ \uJypb/ in the inner $\sim3$ \sqdeg/. This approach
  is also being used to reduce the \LoTSS/-wide data for the
  second data release.}

\date{}

\begin{document}

\maketitle

 \ 
 \
\section{Introduction}

\begin{figure}[]
\begin{center}
\includegraphics[width=\columnwidth]{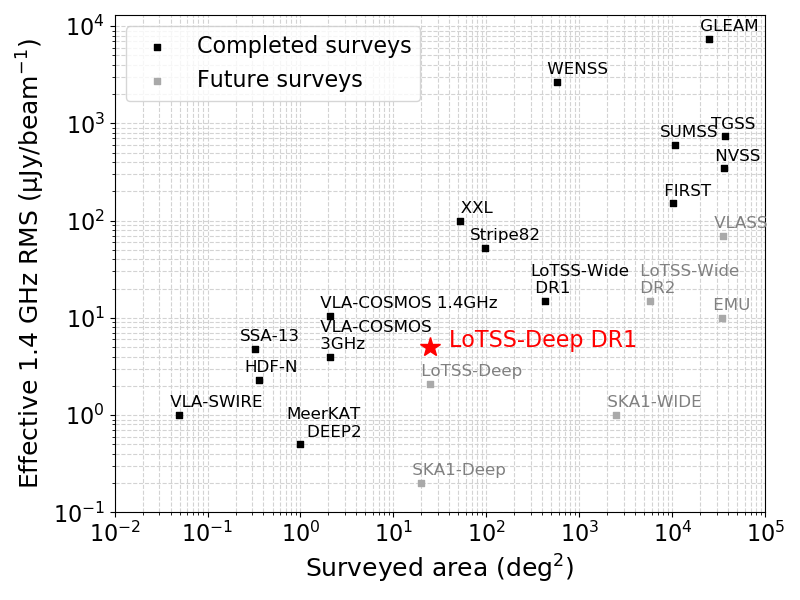}
\caption{\label{fig:SensitivityLoTSSDeep} This figure shows the effective noise in the
  \LoTSS/-Deep continuum maps as compared to other existing and future surveys. A spectral
  index of $\alpha=-0.7$ has been used to convert flux densities to
  the $1.4$ GHz reference frequency.}
\end{center}
\end{figure}

With its low observing frequency, wide fields of view, high
sensitivity, large fractional bandwidth and high spatial resolution,
the Low Frequency Array \citep[\LOFAR/, see][]{vanHaarlem13} is well suited to conduct deep extragalactic surveys. The
\LOFAR/ Surveys Key Science Project is building
tiered extragalactic surveys with 
\LOFAR/, of different depth and areas, and at frequencies ranging from $\sim30$ to $\sim200$
MHz. Specifically the \LOFAR/ \LBA/ Sky Survey (\LoLSS/) aims at
surveying the northern hemisphere using the \LOFAR/ \LBA/ antennas
while the \LOFAR/ Two Meter Sky Survey (\LoTSS/) uses the High Band
Antennas (\HBA/). Its widest
component (\LoTSS/-wide) has been described by
\citet{Shimwell17,Shimwell18}, and aims at reaching noise levels of
$\lesssim100$ \uJypb/ over the whole northern hemisphere. While the
bright sources identified in \LoTSS/-wide are largely radio
loud Active Galactic Nuclei (AGN), the population of faint
sources consists of star forming galaxies and radio quiet AGN \citep[see][and references therein]{Padovani16}. The \LoTSSd/ target noise levels of ultimately $\lesssim10$ \uJypb/, thereby entering a new fainter, higher redshift regime
where star forming galaxies and radio quiet AGN will outnumber the
population of radio loud AGN, and thereby probing the evolution of
those populations over cosmic time.
Fig. \ref{fig:SensitivityLoTSSDeep} is inspired by that of \citet{Smolcic17} and shows a sensitivity and surveyed area comparison between various
existing and future surveys. These include TGSS \citep{Intema17}, FIRST \citep{FIRST}, NVSS \citep{NVSS}, VLA-COSMOS \citep{VLACOSMOS,VLACOSMOS3}, VLASS \citep{VLASS}, EMU \citep{EMU}, VLA-SWIRE \citep{SWIRE}, SSA13 \citep{SSA13}, Stripe-82 \citep{Stripe82}, XXL \citep[][and references therein]{XXL}, DEEP2 \citep{DEEP2}, LOTSS-DR1 \citep{LOTSSDR1}, HDF-N \citep{Richards00}, WENSS \citep{WENSS}, GLEAM \citep{GLEAM}, and SKA \citep{Prandoni15}.

The depth of the \LoTSSd/ is unlikely to be routinely
surpassed at these low frequencies even into the era of the first
phase of the Square Kilometer Array \citep[SKA,][]{Dewdney_2009}
because, although the SKA will have the raw sensitivity to easily
reach such depths, the confusion noise of the SKA-low will likely
increase the image rms to values exceeding the target depth of the
LoTSS-deep images \citep[see e.g.][]{Zwart_2015,Prandoni_2015}. In order to construct the \LoTSSd/, we have selected the
\Bootes/, \LH/, and \ELAIS/ fields, together with the North Ecliptic
Pole (NEP). Each of them is covered by a wealth of
multiwavelength data, necessary to derive photometric redshifts and
low frequency radio luminosities, thereby providing an efficient way to estimate
Star Formation Rate (\SFR/ hereafter) in galaxies for
example. Together, these four multiwavelength fields allow us to probe
a total sky area of $\gtrsim30$ \sqdeg/, in order to probe all galaxy
environments at $z>1$.

It is, however, quite challenging to make thermal noise limited images
at low frequencies because of the presence of Direction Dependent Effects
(\DDE/), such as the ionospheric distortions, and
the complex primary beam shapes of phased arrays. We have shown \citep{Shimwell18} that
using a novel set of calibration and imaging algorithms developed by
\citet{Tasse14b}, \citet{Smirnov15} and \citet{Tasse18} we were able to estimate and
compensate for the \DDE/, and thus use LOFAR to produce thermal noise limited maps from 8-hour \LOFAR/ observations in a
systematic and automated way, while keeping the computational efficiency
high enough to be able to deal with the high \LOFAR/ data rates.

In
this first paper of a series we present an improved strategy to reach thermal noise
limited images after hundreds of hours of integration on the \Bootes/ and
\LH/ extragalactic fields, reaching $\sim30$ \uJypb/ noise levels,
while being more robust against absorbing faint unmodeled extended
emission and dynamic range issues. In Sec. \ref{sec:3GC} we introduce the \DD/ calibration and
imaging problem, together with the existing software that we use to
tackle it. \ANSWII{We describe
our \DDE/ calibration and imaging strategy (\PipeVII/) in Sec. \ref{sec:PipeVII} \citep[for completion in appendix \ref{sec:PipeVI} we describe \PipeVI/ that was presented in detail in][]{Shimwell18}}. In
Sec. \ref{sec:DeepObs} we use \PipeVII/ to synthesize deep images over
the \Bootes/ and \LH/ extragalactic fields and present these deep low
frequency images. The subsequent papers in this series will present the
deeper \ELAIS/ data products \citep[][in prep.]{LoTSSDeepII}, the multiwavelength
cross matching \citep[][in prep.]{LoTSSDeepIII} and the photometric redshifts and
galaxy characterisation \citep[][in prep.]{LoTSSDeepIV}.

\section{\LoTSS/ and the third generation calibration and imaging problem}
\label{sec:3GC}


Calibration and imaging techniques have greatly evolved since the
first radio interferometers have become operational. First generation
calibration is commonly refered as direction-independent (\DI/) calibration,
where calibration solutions are transferred to the target from an amplitude and/or phase calibrator field. Second
generation calibration is the innovation, beginning in the mid-1970s, of using the target field to calibrate itself
(self-calibration: \citealt{Pearson+Readhead84}). Third generation calibration and imaging consists
in estimating and compensating for direction-dependent effects (\DDE/).

As mentioned above, it is challenging to synthesize high
resolution thermal noise limited images with \LOFAR/ \citep{vanHaarlem13}. Specifically, \LOFAR/ (i)
operates at very low frequency ($\nu<250$ MHz), (ii) has very
large fields of view ({\sc fwhm} of $2$ -- $10$ degrees), and (iii) combines short ($\sim100$ m)
and long ($\sim2000$ km) baselines to provide sensitivity to both the
compact and extended emission. Because of the presence of the
ionosphere and the usage of phased array beams, the combination of (i) and (ii) make
the calibration problem direction-dependent by nature.
In Sec. \ref{sec:ME} we introduce the mathematical formalism used
throughout this paper, while in Sec. \ref{sec:DICalibration} and
\ref{sec:Algo3GC} we describe the algorithms and software used for
\DI/ and \DD/ calibration and imaging.

\subsection{Measurement equation formalism}
\label{sec:ME}

\def\XSky/{\textbf{x}_{\vec{s}\nu}}
\def\bl{\footnotesize\vec{b}}

The Radio Interferometry Measurement Equation \citep[\RIME/, see][]{Hamaker96} describes
how the underlying electric field coherence (the sky model), and the
the various direction-independent and
direction-dependent Jones
matrices (denoted $\JonesMatDI$ and $\JonesMat$ respectively), map to the measured
visibilities. In the following, we consider the electric field
in linear notation (along the $\textsc{x}$ and $\textsc{y}$ axes), at frequency $\nu$
in direction $\vec{s}=[\l,\m,\n=\sqrt{1-l^2-m^2}]^T$ (where $T$ is the matrix transpose) and write the $4\times1$
sky coherency matrix
as
$\XSky/=[\textsc{xx},\textsc{xy},\textsc{yx},\textsc{yy}]^T_{\vec{s}\nu}$. If
$\MuellerMatDI_{\bl}=\JonesMatDI^*_{qt\nu}\kron\JonesMatDI_{pt\nu}$ and $\MuellerMat^{\vec{s}}_{\bl}={\JonesMat^{\vec{s}*}_{qt\nu}}\kron\JonesMat^{\vec{s}}_{pt\nu}$ are the
direction-independent and direction-dependent $4\times4$ Mueller
matrices\footnote{As described by \citet{Hamaker96}, the Mueller
  matrices and the Jones matrices can be related to each other 
  using the Vec operator. In the context of the \RIME/, if
  $\JonesMat_{p}$ and $\JonesMat_{q}$ are $2\times2$ Jones matrices of
  antenna $p$ and $q$, and
  $\textbf{X}$ is the $2\times2$ source's coherency matrix then we
  have
  $\OVec{\JonesMat_{p}\textbf{X}\JonesMat_{q}}=\left(\JonesMat^*_{q}\kron\JonesMat_{p}\right)\OVec{\textbf{X}}$,
where $\kron$ is the Kronecker product.} on a baseline $\bl\leftrightarrow\{pqt\}\rightarrow[\u,\v,\w]^T$ between
antenna $p$ and $q$ at time $t$, then the
$4$-visibility $\textbf{v}_{\bl}$ is given by

\begin{alignat}{2}
\label{eq:ME}
\textbf{v}_{\bl}=&
\MuellerMatDI_{\bl}
\int_{\vec{s}}
\MuellerMat^{\vec{s}}_{\bl}
\MuellerMatBeam^{\vec{s}}_{\bl}\textbf{x}_{\vec{s}\nu}
k^{\vec{s}}_{\bl} \textrm{d}\vec{s}+\vec{n}_{\bl}
\\
\label{eq:kterm}
\text{with }k^{\vec{s}}_{\bl}=&\exp{\left(-2 i\pi \frac{\nu}{c}\left(\vec{b}_{pq,t}^T(\vec{s}-\vec{s}_0)\right)\right)}\\
\text{and }\vec{b}_{pq,t}=&
\begin{bmatrix}
\u_{pq,t} \\ 
\v_{pq,t} \\ 
\w_{pq,t} 
\end{bmatrix}=
\begin{bmatrix}
\u_{p,t} \\ 
\v_{p,t} \\ 
\w_{p,t} 
\end{bmatrix}-
\begin{bmatrix}
\u_{q,t} \\ 
\v_{q,t} \\ 
\w_{q,t} 
\end{bmatrix}\\
\text{and }
\vec{s}=&
\begin{bmatrix}
\l \\ 
\m \\ 
\n 
\end{bmatrix}
\text{ and }\vec{s}_0=
\begin{bmatrix}
0 \\ 
0 \\ 
1 
\end{bmatrix}
\end{alignat}

\noindent where $c$ is the speed of light in the vacuum, and
$\vec{n}_{\bl}$ is a
$4\times1$ random matrix following a normal distribution
$\mathcal{N}\left(0,\sigma_{\vec{b}}\right)$. Depending on the
context, in the rest of this paper we will either make use of the
antenna-based Jones matrices or the baseline-based Mueller matrices.

The elements of $\MuellerMatDI_{\bl}$ describe the direction-independent
effects such as the individual station electronics (the bandpass), or
their clock drifts and offsets. The $\MuellerMat^{\vec{s}}_{\bl}$ models
the \DDE/ including
the ionospheric distortion (phase shift, Faraday
rotation, scintillative decoherence) and phased array station beam
that depend on time, frequency, and antenna. Importantly, in order to estimate the intrisic flux densities we
use a description of the \LOFAR/ station primary beam that is built
from semi-analytic models\footnote{\url{https://github.com/lofar-astron/LOFARBeam}}, and
write it as $\MuellerMatBeam^{\vec{s}}_{\bl}$ in Eq. \ref{eq:ME}.

Solving for the third generation calibration and imaging problem consists of
estimating the terms on the right-hand side of Eq. \ref{eq:ME}, namely
the Mueller matrices $\MuellerMatDI_{\bl}$ and $\MuellerMat^{\vec{s}}_{\bl}$ and
the sky model $\SkyXnu$ from the set of visibilities
$\textbf{v}_{\bl}$. Due to the bilinear structure of the \RIME/,
instead of estimating all these parameters at once, inverting
Eq. \ref{eq:ME} is split into two steps. In the first step, the sky term $\SkyXnu$
is assumed to be constant, and the Jones matrices are estimated. The step
is referred as \emph{calibration} and as the \DD/-\CRIME/ system later in
this text (or simply \CRIME/ depending on the context). In the
second step the
Jones matrices are assumed to be constant, and the sky term $\SkyXnu$
is estimated. This step is commonly called \emph{imaging} and is referred as
solving the \DD/-\IRIME/ system later in the text. The \CRIME/ and \IRIME/ 
problems constitute two sub-steps in inverting the \RIME/
system. We will later describe the idea of alternating between
\DD/-\CRIME/ and \DD/-\IRIME/ as \DD/-self-calibration.

While
the vast majority of modern developments in the field of algorithmic research 
for radio interferometry aim at addressing either the
\DD/-\CRIME/ \citep[direction dependent calibration, see][]{Yatawatta08,Kazemi11,Tasse14b,Smirnov15} or \DD/-\IRIME/
\citep[direction-dependent imaging, see][]{Bhatnagar08,Tasse11,Tasse18}, in this article we aim at developing a
robust approach using existing \DD/-\CRIME/ and \DD/-\IRIME/ algorithms
to tackle the complete \RIME/ inversion problem.

\subsection{Direction-independent calibration}
\label{sec:DICalibration}

The standard \LOTSS/ \HBA/ observations consist of a $10$ minute scan on a
bright calibrator source (in general 3C\,196 or 3C\,295) before observing
the target field for 8 hours. On both calibrator and target fields, the visibilities of the 240 subbands (SB) are regularly distributed across
the $120$-$168$ MHz bandpass, with 64 channels per 195.3 kHz subband and 1 sec
integration time. The data are first flagged using \AOFlagger/\footnote{\url{https://sourceforge.net/p/aoflagger/wiki/Home}} \citep[][]{Offringa12} and
averaged to 16ch/sb and 1s.

\def\DTime{\Delta^t_p}
\def\DTEC{\Delta^T_p}
\def\AmpTerm{A_{p\nu}}

The interferometric data taken on the calibrator field are used to
estimate the direction independent Jones matrices $\JonesMatDI$ that are, to
first order, the same in the target and calibrator fields. These
include (i) the individual \LOFAR/ station electronics and (ii) the clock offsets and
drifts. This first pass of calibration is conducted using the \PREFACTOR/ software
package\footnote{\url{https://github.com/lofar-astron/prefactor}} \citep[][]{deGasperin19}.
Specifically, as the calibrator field essentially consists of a single bright source,
the measurement equation is direction independent and the
visibilities are modeled as

\def\MuellerCal{\bm{\mathcal{G}}^0}

\begin{alignat}{2}
\widehat{\textbf{v}^{\mathrm{cal}}_{\bl}}=& 
\widehat{\MuellerCal_{\vec{b}}}
\textbf{v}^{\mathrm{model}}_{\bl}
\end{alignat}

\noindent where $\textbf{v}^{\mathrm{model}}_{\bl}=\int_{\vec{s}}
\textbf{x}_{\vec{s}\nu}\ k^{\vec{s}}_{\bl}\ \textrm{d}\vec{s}$ is the
skymodel of the calibrator. We cannot just use $\MuellerCal_{\vec{b}}$
to calibrate the target field, since the ionosphere is different for the
calibrator and target fields. Instead, we want to extract (i) the bandpass and (ii) the clock
offsets from the calibrator field, these being valid for the target
field. The effective Mueller matrix of a given
baseline $\bm{\mathcal{G}_b}^0$ can be decomposed as the product of a
direction independent and direction dependent term

\begin{alignat}{2}
\widehat{\bm{\mathcal{G}}^0_{\vec{b}}}=&\widehat{\MuellerMatDI_{\vec{b}}}\widehat{\MuellerMat^0_{\vec{b}}}\\
\textrm{with }\widehat{\JonesMatDI_{p\nu}}=&\widehat{\AmpTerm}\exp{\left(i\nu\widehat{\DTime}\right)}\\
\textrm{and }\widehat{\JonesMat^0_{p\nu}}=&\exp{\left(iK\nu^{-1}\widehat{\DTEC}\right)}\label{eq:PhaseTEC}
\end{alignat}

\noindent where $K=8.44\times 10^{9}$ m$^3$s$^{-2}$, and $\AmpTerm$, $\DTime$ and $\DTEC$ are real-valued and
represent respectively the bandpass, the clock and ionospheric Total Electron
Content offset with respect to a reference antenna. The terms $\DTime$ and $\DTEC$
can be disentangled from the frequency dependent phases because the
former are linear with $\nu$ while the latter are non-linear.

Assuming the clocks and bandpass are the same
for the calibrator and for the target field, the corrected visibilities
$\textbf{v}^{\mathrm{c}}_{\bl}$ for the target field can be built from
the raw data $\textbf{v}_{\bl}$ as
$\textbf{v}^{\mathrm{c}}_{\bl}=\widehat{\MuellerMatDI_{\vec{b}}}^{-1}\textbf{v}_{\bl}$. In
order to calibrate for the remaining phase errors, the target field
is \DI/ calibrated against the TIFR GMRT Sky Survey (\TGSS/) catalogs \citep{Intema17} and visibilities are averaged to $2$ ch/sb and 8s.

\subsection{Direction-dependent calibration and imaging}
\label{sec:Algo3GC}

As discussed by \citet{Tasse14a} there are two families of calibration
algorithms. ``Physics-based'' solvers aim at estimating the
underlying Jones matrices whose product gives the effective $\JonesMatDI^{\vec{s}}_{pt\nu}$ and
$\JonesMat^{\vec{s}}_{pt\nu}$. Depending on the observing frequency
and instrumental setup, these can be the product of the ionospheric
Faraday rotation matrix, the scalar phase shift, and the individual
station primary beams. This approach has the great advantage of constraining
the free parameters to a low number, but it requires one to model
analytically the physics of the various effects over the $\{\vec{s}pt\nu\}$ space to be able to
disentangle them. The second family of algorithms directly estimate the effective
Jones matrices over piecewise constant domains in
$\{\vec{s}pt\nu\}$ space. These ``Jones-based'' solvers have the advantage of not requiring any
physical modeling of the DDE. However, this makes the number of
degrees of freedom increase dramatically, typically by a few orders
of magnitude. These additional degrees
of freedom can often make the inverse problem to be
ill-posed\footnote{Linear and non-linear problems can be ill-posed,
  meaning that non-unique solutions can be found.}. This means
in practice that the \DD/ solvers can overfit the data, leading to the
unmodeled sky flux being absorbed by the calibration solutions. This
happens differently at different scales, and has a greater effect on the extended
emission, which is measured only by the less numerous shorter baselines. 
Similar to linear problems, the situation depends on the sizes of the parameter space, and
also on the shape of the neighboring
domains in the $\{\vec{s}pt\nu\}$ spaces. Also, as explained by \citet{Shimwell18}, experience shows that we need to
split the sky model into a few tens of directions (``facets") to be able to
properly describe the spatial variation of the \DD/ Jones matrices.
This effect is amplified by the difficulty of properly modeling
the extended emission itself. Indeed, even in the absence of
calibration errors, the deconvolution problem consisting of inverting
Eq. \ref{eq:ME} by estimating $\textbf{x}_{\vec{s}\nu}$ for given
$\MuellerMatDI$, $\MuellerMat$ and $\textbf{v}$ is ill-posed. Furthermore the
situation is more severe when the Point Spread Function (\PSF/) is
less point-like (i.e. when the uv plane is not well sampled).
While the true measured visibilities are described by
Eq. \ref{eq:ME}, the (\emph{model}) visibilities
$\widehat{\textbf{v}_{\bl}}$ that are estimated\footnote{Throughout
  this paper the notation $\widehat{x}$ should be read as \emph{the estimate of
  $x$}.}
over the piecewise constant domains $p,\Omega_\varphi,\Delta t,\Delta \nu$ can be written as

\begin{alignat}{2}
\label{eq:ME_Facet0}
\widehat{\textbf{v}_{\bl}}=& 
\widehat{\MuellerMatDI_{\bl}}
\widehat{\textbf{v}^\Sigma_{\bl}}
\\
\label{eq:ME_Facet1}
\textrm{with }\widehat{\textbf{v}^\Sigma_{\bl}}=& 
\displaystyle\sum\limits_{\varphi}
\widehat{\textbf{v}^\varphi_{\bl}}
\\
\label{eq:ME_Facet2}
\textrm{and }\widehat{\textbf{v}^\varphi_{\bl}}=& 
\widehat{\MuellerMat^{\varphi}_{\bl}}
\MuellerMatBeam^{\varphi}_{\bl}
\int_{\vec{s}\in\Omega_\varphi}
\widehat{\SkyXsnu}
\ k^{\vec{s}}_{\bl}\ \textrm{d}\vec{s}
\end{alignat}


\noindent where $\Omega_\varphi$ is the set of directions $\vec{s}$ for
a facet $\varphi$, $\widehat{\textbf{x}_{\vec{s}\nu}}$ is the estimated
underlying sky, and $\widehat{\MuellerMatDI_{\bl}}$ and
$\widehat{\MuellerMat^\varphi_{\bl}}$ are the \DI/ and \DD/ Mueller
matrices for baseline $\vec{b}$, built from the corresponding
estimated Jones matrices in $p,\Delta t,\Delta \nu$.

Specifically, in order
to solve for the \DDE/, the size and shape of the domains
are critical. Intuitively, if the domains are too small, not enough
data points are used, and the solutions are subject to
ill-conditioning. On the other hand if the domains are too large the
true Jones matrices can vary within the domain and the piecewise
constant approximation cannot account for the physics that underlies
the measurement. Due to (i) the non-linear nature of Eq. \ref{eq:ME},
and (ii) the complexity of the background radio sky, optimising over the shape of these piecewise constant domains is a
difficult problem (and is non-differentiable to some extent).

\label{sec:Algo3GC}

The faceted Jones-based 
approach is to find sky $\widehat{\textbf{x}_{\vec{s}\nu}}$ as well as the \DI/
$\widehat{\JonesMatDI_{pt\nu}}$ and the \DD/ piecewise constant
$\widehat{\JonesMat^{\varphi}_{pt\nu}}$ for all $\{\vec{s}\varphi
pt\nu\}$ such that
$\widehat{\textbf{v}_{\bl}}\sim\textbf{v}_{\bl}$ (we remain
intentionally vague here, since the cost function that is minimised
depends on the specific \DD/ algorithm). In practice,
inverting Eq. \ref{eq:ME} (estimating the Jones matrices and sky
terms) is done by (i) solving for the Jones matrices assuming the sky
is known (the calibration step), and (ii) solving for the sky-term
assuming the Jones matrices are given (imaging step). Using this skymodel and repeating steps
(i) and (ii) is known as self-calibration, but in a third-generation approach we must
explicitly model the DD aspects.

Since the computing time evolves as $\sim n_d^3$, where $n_d$ is the
number of directions in the \DD/-solvers, the problem of
\DD/-calibration has in general been tackled using direction
alternating peeling-like techniques. Major breakthroughs have been
made in the field of \DD/-\CRIME/ solvers in the past decade by
\citet{Yatawatta08,Kazemi11} making this \DD/-calibration
computationally affordable. In addition, \citet{Tasse14b} and
\citet{Smirnov15} have described an alternative way to write the
Jacobian of the cost function by using the Wirtinger differentiation
method. The Jacobian and Hessian harbor a different structure and
shortcuts can be taken to invert the calibration problem. The net gain
over the classical method is not trivial, but can be as high as
$n_a^2$ \citep{Smirnov15} where $n_a$ is the number of antennas in the
interferometer.  This Jones-based approach is therefore fast, but is
still subject to the same flaws as any Jones-based solvers.

Only a very few \CIRIME/ algorithms using a full \DD/ self-calibration
loop have been described and implemented.  They include pointing
self-calibration \citep{Bhatnagar17}, or peeling-based techniques such
as {\sc MFImage} (implemented in the \OBIT/\footnote{\url{https://www.cv.nrao.edu/~bcotton/Obit.html}} package)
and \FACTOR/ \citep[][see also
  Sec. \ref{sec:ComparisionFactor}]{Weeren16}. Similar to peeling,
and developed for reducing \LOFAR/ data, \FACTOR/ is sequential along
the direction axis. Looping over the different facets it consists of
(i) subtracting all sources besides calibration sources in that one
facet, and (ii) \DI/-self-calibrating in that direction. In addition
to the ill conditioning issues discussed above on \DD/-\CRIME/ and
\DD/-\IRIME/ solvers, an expensive computational problem arises when
estimating the $\widehat{\JonesMat^{\varphi}_{pt\nu}}$.

The approach presented by \citet{Shimwell18} (also described in detail
Sec. \ref{sec:PipeVI} and referred to as \PipeVI/ in the following) is based on the
\kMS/ \DD/-\CRIME/ solver \citep{Tasse14b,Smirnov15} and \DDFacet/
\DD/-\IRIME/ imager \citep{Tasse18}, and is algebraically simultaneous
in directions. The direction dependent pipeline \ddfpipe/\footnote{\url{https://github.com/mhardcastle/ddf-pipeline}} is a high level wrapper
that mainly calls \DDFacet/\footnote{\url{https://github.com/saopicc/DDFacet}} and \kMS/\footnote{\url{https://github.com/saopicc/killMS}} for direction dependent
self-calibration. This type of algorithm has a number of
advantages. Specifically, the interaction terms between the different
directions are properly taken into account within the \DD/-\CRIME/
solver, {\it i.e.} the \DD/ affected sidelobes leaking from any facet
to any other facet are accounted for within the algebraic operations
of the algorithm. Another advantage compared to the \FACTOR/ approach
is that the data need only to be read rather than modified, making the
\ddfpipe/ more I/O efficient. 

\ANSW{As explained in \citet{Shimwell18}, the \PipeVI/ was however affected by a number of issues
  including (i) artifacts and dynamic range limit around the brightest
  radio sources, (ii) artificial and diffuse haloes around moderately
bright radio sources and (iii) unmodeled flux absorption.}

\section{Calibration and imaging robustness}
\label{sec:PipeVII}

\begin{figure*}[]
\begin{center}
\includegraphics[width=\columnwidth]{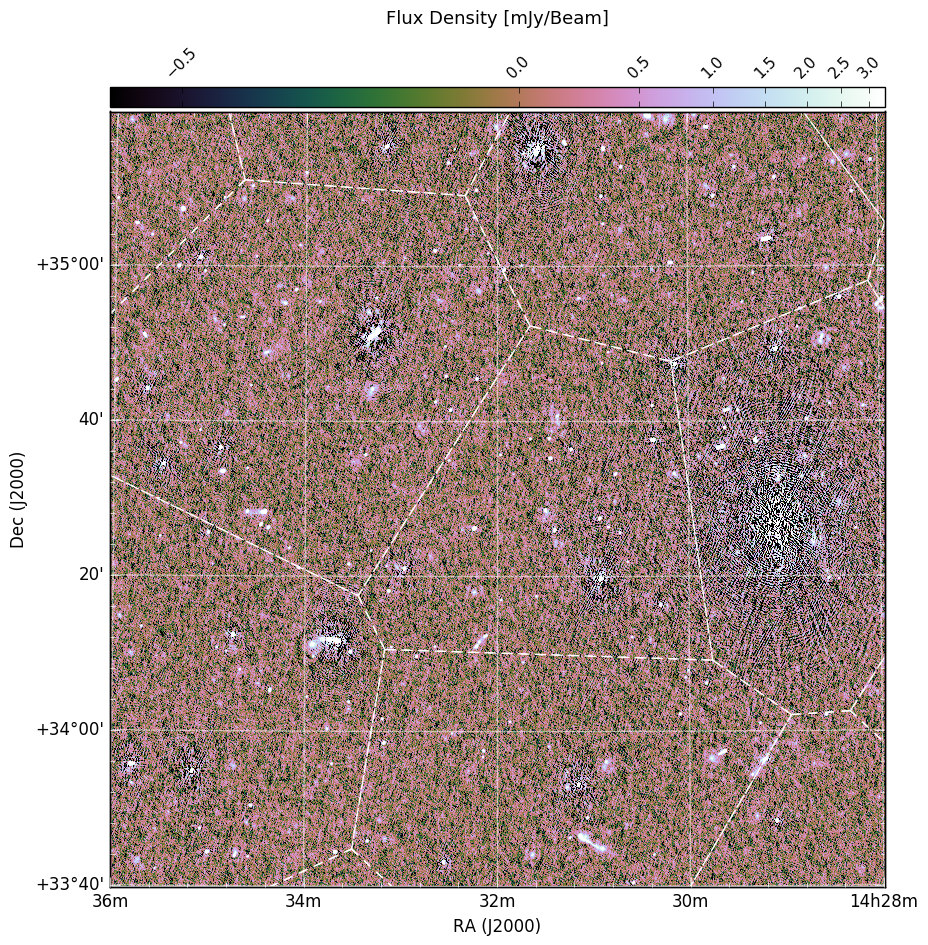}
\includegraphics[width=\columnwidth]{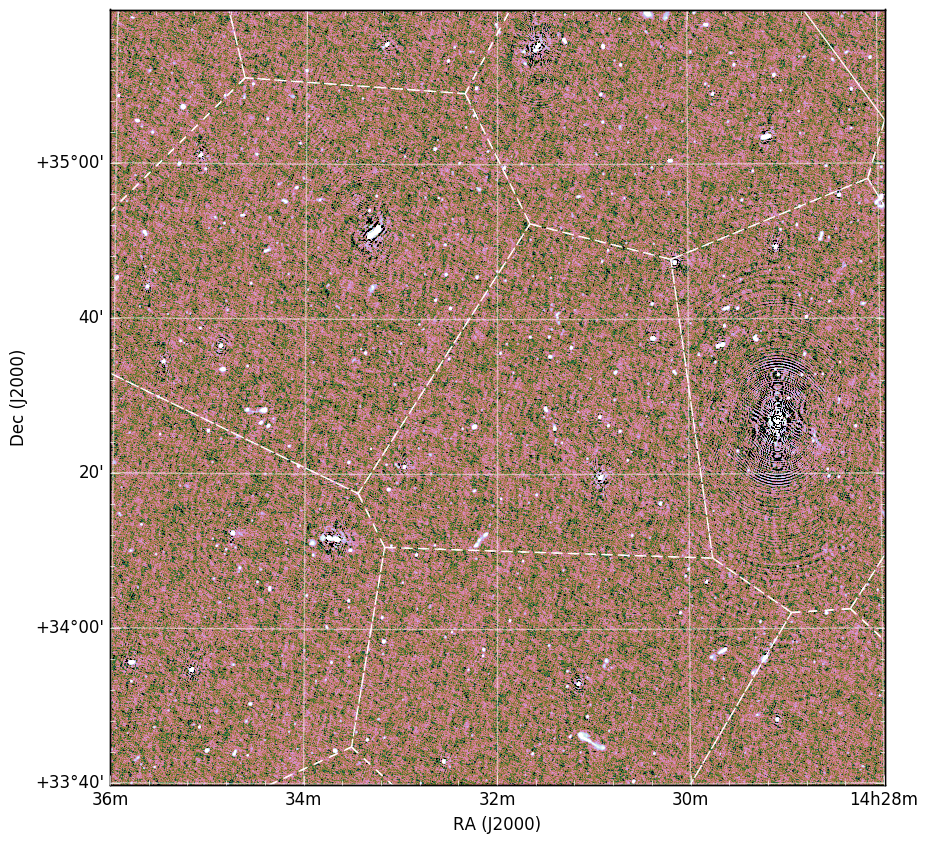}
\includegraphics[width=\columnwidth]{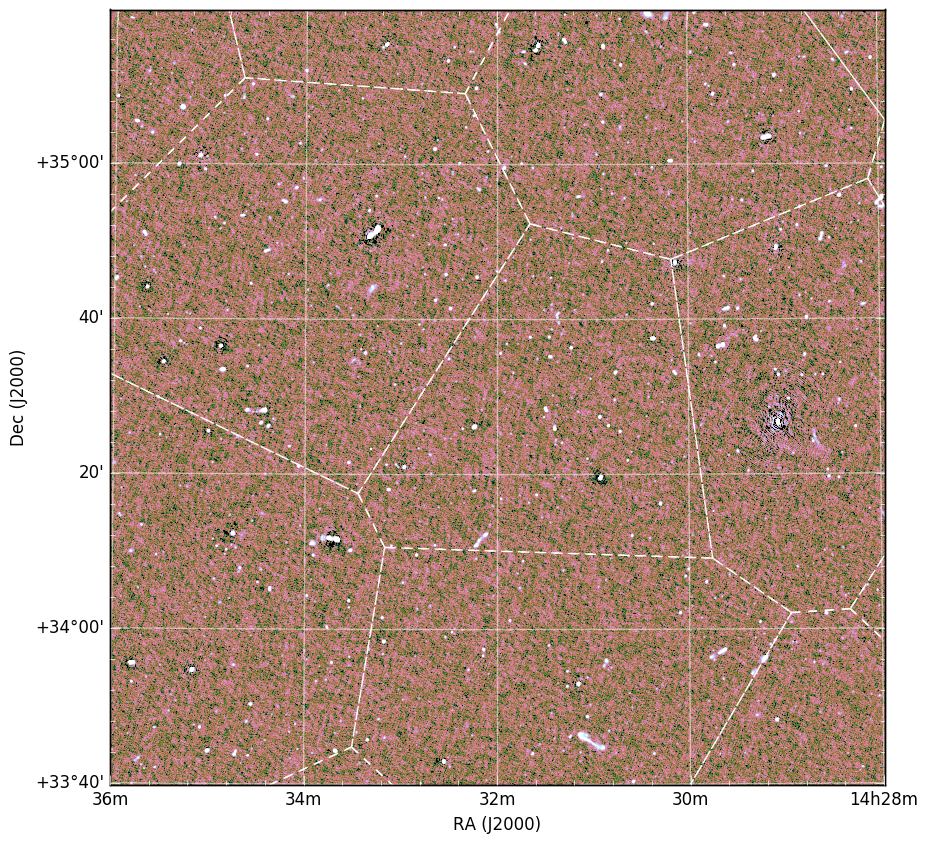}
\includegraphics[width=\columnwidth]{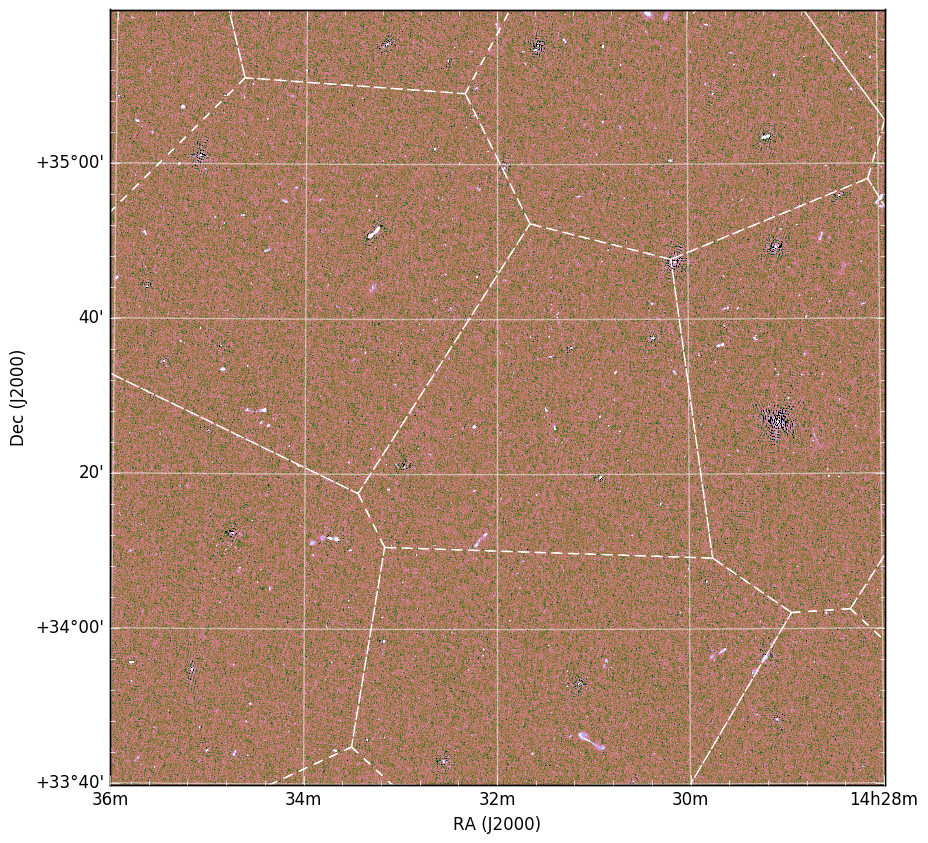}
\caption{\label{fig:SelfCal_DR2} Some of the images produced at different
  steps in the \DD/-self-calibration loop implemented as
  Alg. \ref{alg:DR2}. The maps correspond (from left to right, top to
  bottom) to Steps \ref{step:VII_I_DI0}, \ref{step:VII_I_DD0},
  \ref{step:VII_I_DD1b} and \ref{step:VII_I_DD4} respectively. The
  white lines show the facets' locations. The colorscale is the same on
  all panels, and diplayed using an inverse hyperbolic sine function
  to render both the low level artifacts and some bright sources.}
\end{center}
\end{figure*}

\ALGODRTWOB/


\ANSW{
  In this section we describe in detail a \DD/ calibration and imaging
algorithm that aims to make the overall \RIME/ imaging and calibration
solver more robust against artifacts around the brightest sources
(Sec. \ref{sec:DIDR}) and unmodeled flux absorption
(Sec. \ref{sec:regularisation} and Sec. \ref{sec:SlowSolve}). An
overview of this approach is shown in Alg. \ref{alg:DR2} (the
implementation of which is referred as \PipeVII/), and the
corresponding \DD/ self-calibration loop is displayed in
Fig. \ref{fig:SelfCal_DR2}. The \PipeVII/ products are described in
Sec. \ref{sec:AdditionalProds}. In Sec. \ref{sec:ComparisonVIVII} we
show that \PipeVII/ produces improved images as compared to those
previously made with \PipeVI/ 
\citep[see Sec.  \ref{sec:PipeVI} and][]{Shimwell18}. In Sec. \ref{sec:Profiling} we discuss the
\PipeVII/ computing efficiency.
}

\subsection{Dynamic range issue}
\label{sec:DIDR}

With \LOFAR/'s very large field of view, it is quite common to
observe bright sources within the station's primary beam. \ANSW{It turns out from tests we conducted
on fields containing bright sources (such as 3C\,295 which has a flux density of $\sim100$ Jy
at $150$ MHz) that the related residual errors create powerful
artifacts that largely dominate the thermal noise, thereby driving a
dynamic range limit.} As
explained in Sec. \ref{sec:DICalibration} the initial phase calibration is
done against \TGSS/ at $150$ MHz \citep[][]{Intema17}. However, since \LOTSS/ resolution is
much higher than \TGSS/'s ($6$\arcsec$\times6$\arcsec against
$25$\arcsec$\times25$\arcsec respectively), small spatial
uncertainties on how the individual bright sources are modeled lead to
large Jones matrix errors.\ANSW{ Specifically, this effect can be severe
when the true point sources are erroneously found to be resolved by \TGSS/, as this leads to large
calibration errors for the long baselines. In these situations, our experience
shows that the initial \DI/ calibration is not good
enough to start doing a \DD/ calibration (that due to ill conditioning
issues has to be done on larger time-frequency solutions intervals).
In the following we study the \DI/ calibration solutions and assess
whether they need to be recomputed using a high angular resolution sky model.
When using the \PREFACTOR/ \DI/-calibrated \LOTSS/ data and associated
imaging products, this amounts to doing a round of \DI/ self-calibration at the beginning of \PipeVII/.
}

\begin{figure}[ht!]
\begin{center}
\includegraphics[width=\columnwidth]{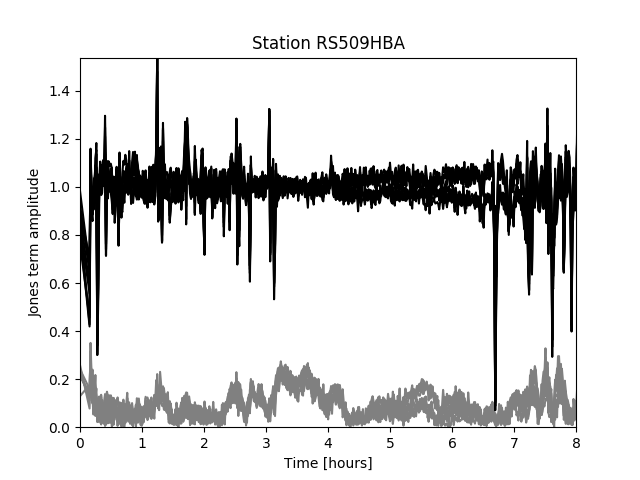}
\end{center}
\caption{\label{fig:DI_Solutions} \ANSW{This plot shows the amplitude of the diagonal (black)
  and off-diagonal (gray) terms of the estimated Jones matrices for a
  remote station, using the \LOFAR/'s observation synthesized $6\arcsec$ image as
  the sky model (self-calibration). If the initial \DI/ calibration and correction by
  \PREFACTOR/ on the lower resolution \TGSS/ sky model would be
  good enough, the calibration solutions found here would be the unity
  matrix at all times and frequency. Therefore, in \PipeVII/ we carry out a few
  full-Jones \DI/-only self-calibration steps. This approach seems
  very efficient in increasing the dynamic range around the brightest
  radio sources.} }
\end{figure}

Solution time and frequency variability is however hard to
interpret. Indeed, because the \RIME/ formalism is subject to unitary ambiguity
\citep[see][for a detailed discussion]{Hamaker2000}, the off-diagonal or
absolute phase terms found by a solver are not
meaningful. Instead, these are given with respect to a reference
antenna. When Jones matrices are scalar, this amounts to zeroing the
phases $\varphi_0$ of the reference antenna, by subtracting
$\varphi_0$ from all phases of all
antennas. To do this in the general case of non-diagonal Jones matrices, we use a polar decomposition on the Jones
matrix $\JonesMat_0$ of the reference antenna such that
$\JonesMat_0=\bm{\mathrm{U}}\bm{\mathrm{P}}_0$ where $\bm{\mathrm{U}}$
is a unitary matrix\footnote{The unitary matrix $\bm{\mathrm{U}}$ is found by doing a
  singular value decomposition
  $\JonesMat_0=\bm{\mathrm{W}}\bm{\Sigma}\bm{\mathrm{V}}$ and is then
  built as $\bm{\mathrm{U}}=\bm{\mathrm{W}}\bm{\mathrm{V}}^H$}. We
then apply $\bm{\mathrm{U}}$ to all Jones
matrices as
$\JonesMat_p\leftarrow\bm{\mathrm{U}}^H\JonesMat_p$. Intuitively, when the Jones matrices are all scalar, the unitary matrix
$\bm{\mathrm{U}}$ is simply $\exp{\left(i\varphi_0\right)}\Unity$,
and that step makes the phases of all $\JonesMat_p$ relative to the reference antenna (and specifically zeros the
phases of $\JonesMat_0$). In the case of non-trivial $2\times2$ Jones
matrices, finding and applying $\bm{\mathrm{U}}$ has the effect of
removing a common rotation from all Jones matrices, and orthogonalises
them.

\ANSW{We apply this in Fig. \ref{fig:DI_Solutions} where we show the typical \DI/
Jones matrices we can estimate at stage \ref{step:VII_I_DI1} for a
given remote station and frequency, with respect to a reference
station in the \LOFAR/ core. They are estimated using \kMS/ and the
skymodel synthesized by \DDFacet/ from the visibilities corrected by
\PREFACTOR/.}
Since the polar transform has been
applied, the variations of the amplitude of the off-diagonal Jones matrices are genuine.
These are interpretable in terms of differential Faraday rotation: the
rotation of the electric field polarisation changes across the \LOFAR/
array. This demonstrates the need to conduct a full-Jones
calibration on the \PREFACTOR/-calibrated \LOTSS/ data. 

Therefore in Step \ref{step:VII_K_DI0} the visibilities are calibrated
against modeled visibilities generated by \DDFacet/ in Step
\ref{step:VII_I_DI1}. The sky being mostly unpolarised, in this
full-Jones \DI/
calibration step, we assume $Q=U=V=0$ Jy (see
Sec. \ref{sec:AdditionalProds} for a discussion of polarisation
related data products). The solution intervals $\delta t_0$ and
$\delta\nu_0$ along time and frequency are determined such that
$n_b\propto\left(T/\left<|\vec{x}_\nu|\right>\right)^2\textrm{Var}\{\textbf{n}\}$
where $n_b$ is the number of points in the
$\delta t_0\times\delta\nu_0$ time-frequency domain, $T$ is the target
solution SNR, and $\textrm{Var}\{\textbf{n}\}$ is the
variance of the visibilities' noise (see Mbou Sob et al. in
preparation for a justification).

Note that after the initial \DD/ calibration solutions have been
obtained in Steps \ref{step:VII_K_DD0} and \ref{step:VII_K_DD1}, a
more accurate \DI/ calibration can be performed. Specifically, in
the \DI/ calibration Steps \ref{step:VII_K_DI1} and
\ref{step:VII_K_DI2}, on any baseline $\bl$ the model visibilities
$\widehat{\textbf{v}^\Sigma_{\bl}}$ (Eq. \ref{eq:ME_Facet1}) 
are predicted based on the previously 
estimated \DD/-Jones matrices  $\widehat{\JonesMat}$
\citep[as is done by][]{Smirnov2011_3}. 


\subsection{Regularisation}
\label{sec:regularisation}

\begin{figure*}[]
\begin{center}
\includegraphics[width=5.5cm]{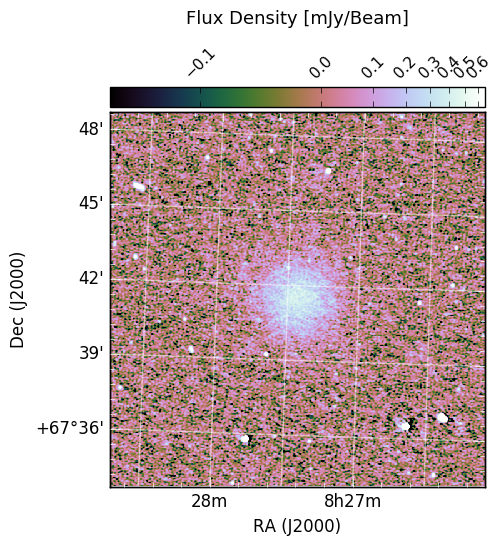}
\includegraphics[width=5.5cm]{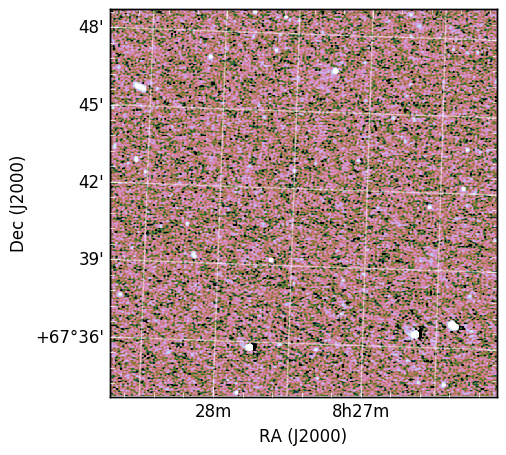}
\includegraphics[width=5.5cm]{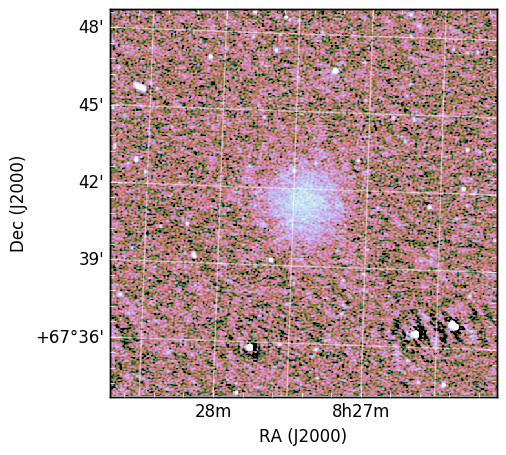}
\end{center}
\caption{\label{fig:SimulCluster} In order to test the robustness of
  the algorithm described in Sec. \ref{sec:PipeVII} and implemented in
  \PipeVII/, we have simulated an unmodeled extended emission (left
  panel). The emission is absorbed by the \DD/-calibration step
  (middle), while it can be partially recovered (right panel) by decreasing the
  effective size of the unknown solutions space (Sec. \ref{sec:regularisation} and \ref{sec:SlowSolve}).
  }
\end{figure*}

The absorption of unmodeled flux by calibration is a well known issue
connected to the calibration of \DDE/. Intuitively speaking, when real
flux is missing from the modeled sky $\widehat{\SkyX_i}$ of $\SkyX$ at
step $i$, and since the \RIME/ inversion is often ill-posed, the
estimates $\widehat{\JonesMat_i}$ of $\JonesMat$ can be biased in a
systematic way. Experience and simulations show that building a new
estimate $\widehat{\SkyX_{i+1}}$ from $\widehat{\JonesMat_i}$ can be
biased in that the unmodeled emission is not and will never
be recovered (Fig. \ref{fig:SimulCluster}). This effect is especially severe when the extended
emission is poorly modeled or unmodeled since this is detected only by the
shortest baselines. Effectively, during the inversion of the \RIME/ system of
equations, the \DD/-self-calibration algorithm has fallen into the wrong (local)
minimum.

\begin{figure*}[]
\begin{center}
\includegraphics[width=8.5cm]{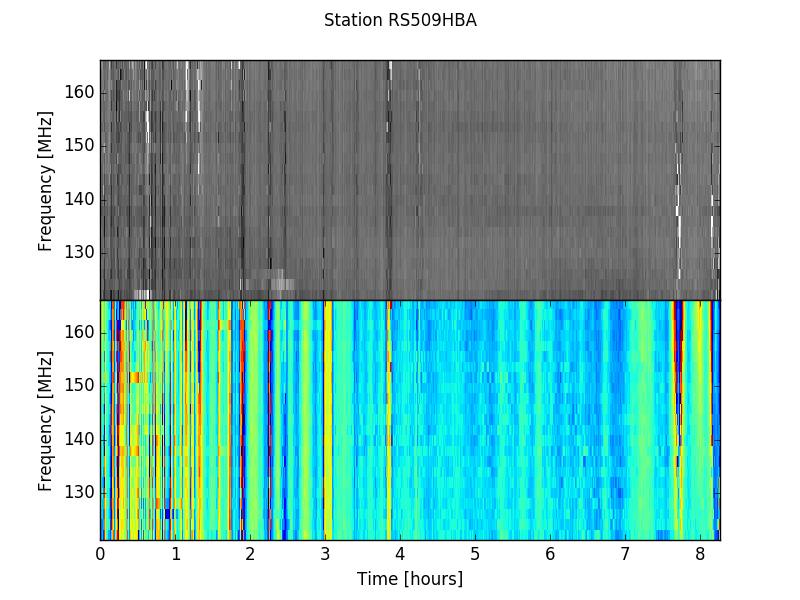}
\includegraphics[width=8.5cm]{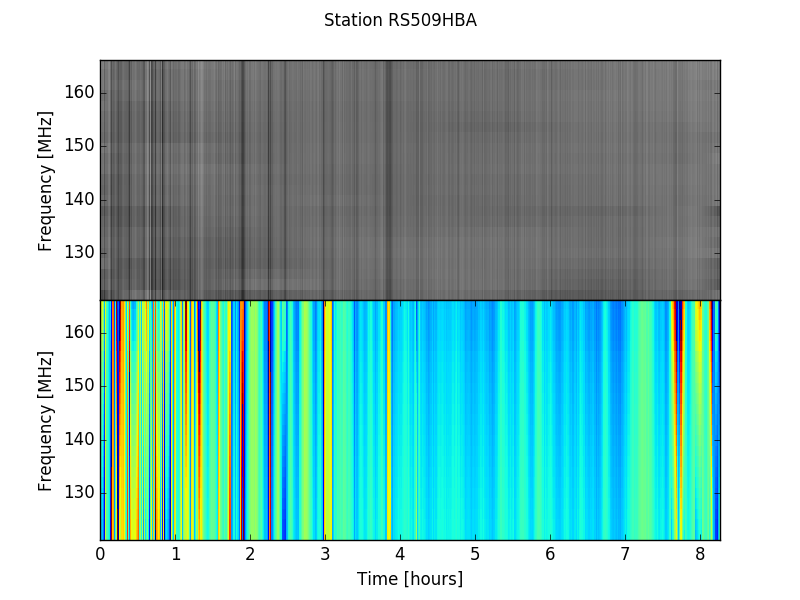}
\end{center}
\caption{\label{fig:FitSols} This figure shows the amplitude and phase
  (top/bottom respectively)
  of a scalar Jones matrix for a given station in a given direction in
  the example observation. The left panel shows the solution as
  estimated by the \kMS/ solver. The right panel shows the regularised
  solution, as updated by the $\OpFit$ function. The amplitude color
  scale ranges from 0 to 1.5.}
\end{figure*}

In order to address this problem, one idea is to reduce the effective number of
free parameters used to describe the Jones matrices in the
$\{pdt\nu\}$-space \citep[see for
  example][]{Tasse14b,Yatawatta15,Weeren16,Repetti17,Birdi20}. Forcing the estimated Jones
matrices' shape to look like that of the real underlying ones improves
the conditioning of the inverse problem.
In Alg. \ref{alg:DR2} (implemented in \PipeVII/) we have replaced that normalisation method by a smoothing of the
\kMS/-estimated Jones matrices. This function $\OpFit$ updates the
Jones matrices $\JonesMat\leftarrow\OpFit\left(\JonesMat\right)$ \ANSW{in
each direction independently} by
imposing on them a certain behavior in the time-frequency space \ANSW{(see below)},
effectively reducing the size of the unknown stochastic process. This
can be thought of as a
regularisation. This is done independently on the phases and
amplitudes on the scalar Jones matrices generated at Steps \ref{step:VII_K_DD1}, \ref{step:VII_K_DD2},
\ref{step:VII_K_DD3}. The updated Jones matrices take the analytical
form 

\def\DTEC{\Delta^T_{pd,t}}
\def\TEC/{{\sc tec}}


\begin{alignat}{2}
  \widehat{\JonesMat_{pd,t\nu}}=&\widehat{a_{pd,t}}\mathcal{P}\left(t,\widehat{\bm{\theta}_{pd,\nu}}\right)\exp{\left(iK\nu^{-1}\widehat{\DTEC}\right)}\Unity
\end{alignat}

\noindent where $\widehat{\DTEC}$ is the differential \TEC/ (see also
Sec. \ref{sec:DICalibration} and Eq. \ref{eq:PhaseTEC}), $\mathcal{P}$ is a polynomial
parametrised by the coefficients in $\bm{\theta}_{pd,\nu}$ (of size
$10$), and $\widehat{a_{pd,t}}$ is a scalar meant to describe the loss of
correlation due to ionospheric scintillation as seen in the left panel
of Fig. \ref{fig:FitSols}. Typically, for the $\sim8$
hours' integration of \LOTSS/ pointings and solving every $30$ sec. and $2$ MHz, this parametrisation of the
Jones matrices reduces the number of free parameters by a factor $\gtrsim20$.

To assess the recovery of unmodelled flux in \PipeVII/ a series of
simulations were conducted in which faint simulated sources of various
fluxes and extents were injected into real LOFAR data that had been
fully processed with the \PipeVII/ strategy. The properties of the
injected sources were chosen to be typical for large extragalactic
objects such as radio halos of galaxy clusters. After the injection of
the artificial extended sources the steps \ref{step:VII_K_DD3} and
\ref{step:VII_K_DD4} were repeated using the sky model derived at step
\ref{step:VII_I_DD4} prior to the injection of the sources. These
simulations will be discussed further by Shimwell et al. (in preparation) but
in each simulation the recovered flux of the completely unmodelled
emission exceeded $60$\%. Examples of the injected and recovered
emission are shown in Fig \ref{fig:SimulCluster}.


As suggested by the results of simulations, the effect on real data is
in general very satisfactory and allows us to recover \ANSW{a good fraction
of} the unmodeled
extended emission even when it is quite faint and  extended. This is shown in Fig. \ref{fig:Holes} for a typical
\LOTSS/ observation. Here the extended emission is about $10\arcmin$
across, with a mean flux density at the peak of only $0.7$ of the local standard deviation.

\subsection{Conditioning and solution interval}
\label{sec:SlowSolve}

\def\IMCAL/{{\sc Im+Cal}}

\begin{figure*}[]
\begin{center}
\includegraphics[width=5.5cm]{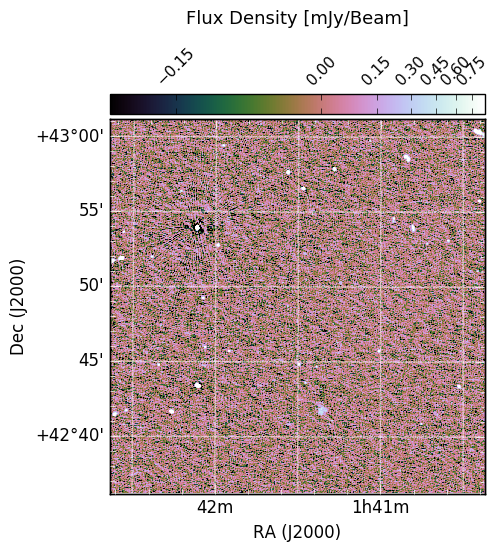}
\includegraphics[width=5.5cm]{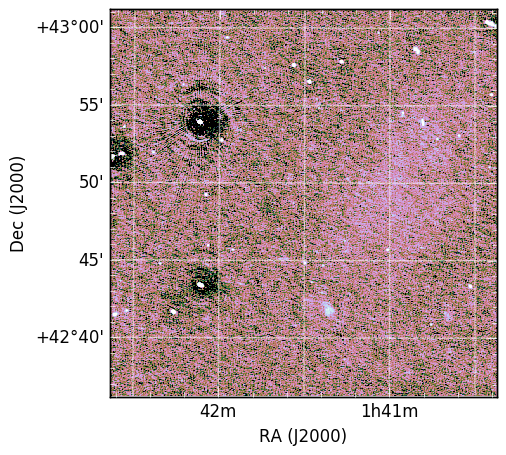}
\includegraphics[width=5.5cm]{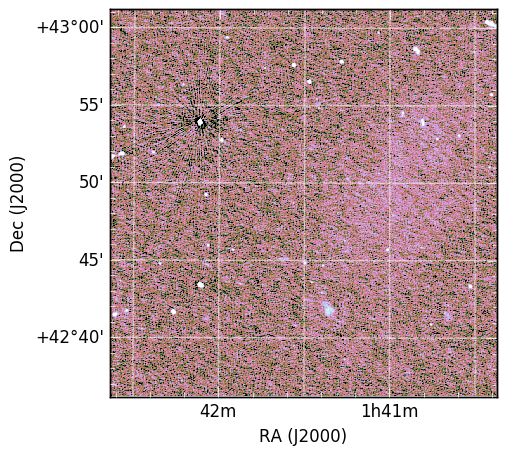}
\end{center}
\caption{\label{fig:Holes} Conserving the unmodeled extended emission
  while keeping high dynamic range is extremely challenging in the
  context of \DD/ calibration and imaging. The left panel shows that a
  faint and unmodeled extended
  emission (on the level of $\sim0.7\sigma$ here) can be totally
  absorbed. While regularising the
  \DD/ calibration solutions can help in recovering the unmodeled
  emission (typically after Step \ref{step:VII_K_DD3}), it can also produce negative imaging artifacts and 'holes'
  around bright sources (middle panel). The right panel shows that solving the
  residuals on longer time intervals (Step \ref{step:VII_K_DD4})
  corrects for this issue.}
\end{figure*}

The additional issue of arcmin-scale negative haloes appearing around
bright compact sources (at a level of $\lesssim 1\%$ or the peak)
could be seen however in $\sim10-20\%$ of the \LOTSS/ pointings
processed with \PipeVI/. As
shown in Fig. \ref{fig:Holes}, we believe this to be connected to the
solution regularisation itself. This issue is hard to understand in
detail because of the non-linearity in the \CRIME/ inversion, but is
likely to be due to the pointings showing these issues being more severely affected by
the incompleteness of the sky model. Specifically, conducting several
experiments, we were able to observe that the situation was improved by
deconvolving deeper or taking into account sources outside the
synthesized image field of view.

An additional way to
improve the conditioning of the problem is to increase the amount of data used to
contrain the estimated Jones matrices. For the \DD/ calibration steps
presented in Alg. \ref{alg:DR2} we use solution
intervals of $0.5-1$ minute. \ANSW{Following
\citet{Weeren16}, we add an extra calibration step
\ref{step:VII_K_DD4}, where the visibilities are modeled using the
latest available skymodel together with the smoothed Jones-matrices estimated in Step \ref{step:VII_K_DD3} and
that are defined over the finer time and frequency mesh. Intuitively, since the negative haloes are
produced by some systematic effects, the idea is to calibrate for a
slowly varying differential effect. The time interval is set to $\sim43$
minutes in \PipeVII/ giving $11$ solution intervals in the $8$ hours
\LoTSS/ pointings. This interval has to be long enough to reach a good
conditioning for the \CRIME/ inversion, and short enough to sample
the Jones matrices remaining physical variations. As shown in the right panel of
Fig. \ref{fig:Holes} this method is very efficient in reducing the
negative haloes.}

\def\MainFigWidth{0.55\textwidth}
\def\SmallFigWidth{0.27\textwidth}

\begin{figure*}[]
    \centering
    \begin{tabular}[t]{cc}
\begin{subfigure}{\MainFigWidth}
    \centering
    \includegraphics[width=1\linewidth,height=1\textwidth]{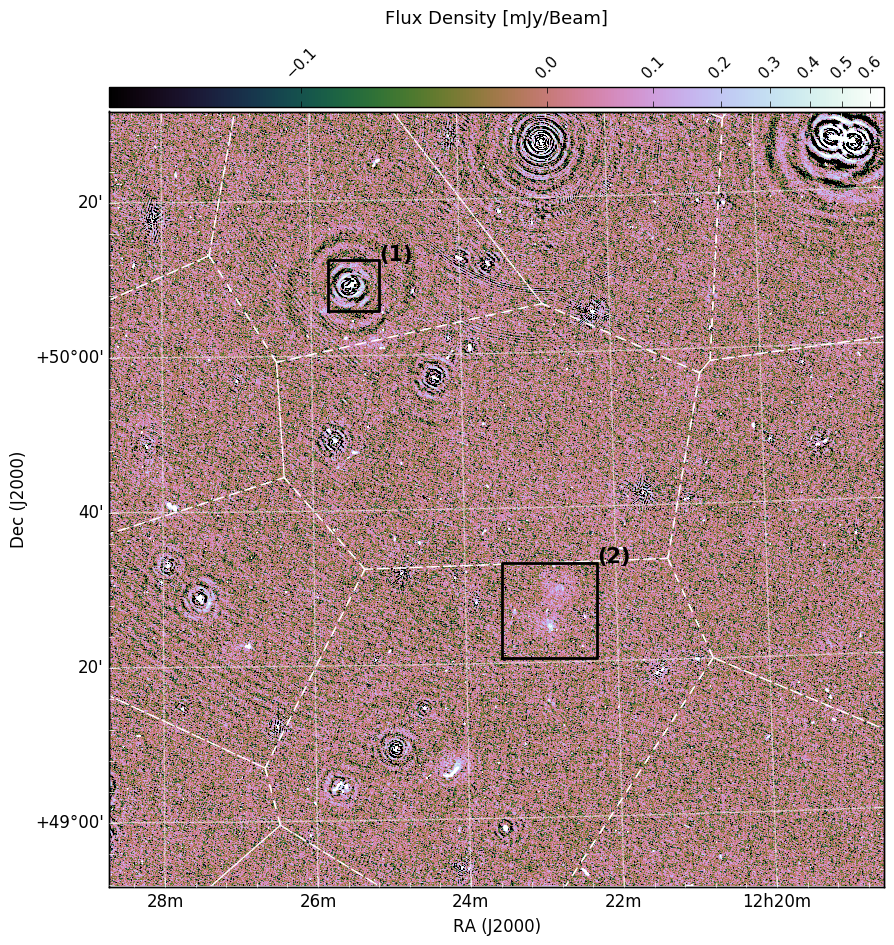}
    \caption{\label{fig:CompareDR1DR2_DR1main} The central part of the {\sc P26Hetdex03} $8$ hours
      \LOFAR/-\HBA/ scan as imaged by Alg. \ref{alg:DR1}.} 
\end{subfigure}
    &
        \begin{tabular}{c}
        \smallskip
            \begin{subfigure}[t]{\SmallFigWidth}
                \centering
                \includegraphics[width=1\textwidth]{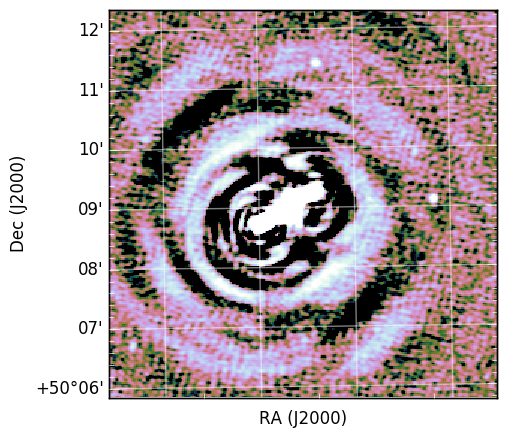}
                \caption{\label{fig:CompareDR1DR2_DR1sub1} Region (1) as imaged by Alg. \ref{alg:DR1}}
            \end{subfigure}\\
            \begin{subfigure}[t]{\SmallFigWidth}
                \centering
                \includegraphics[width=1\textwidth]{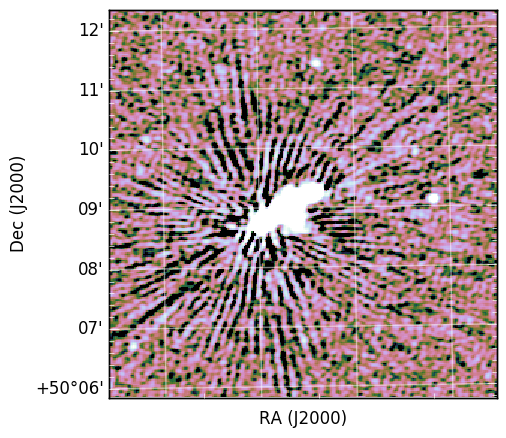}
                \caption{\label{fig:CompareDR1DR2_DR2sub1} Region (1) as imaged by Alg. \ref{alg:DR2}}
            \end{subfigure}
        \end{tabular}\\

\begin{subfigure}{\MainFigWidth}
    \centering
    \includegraphics[width=1\linewidth,height=1\textwidth]{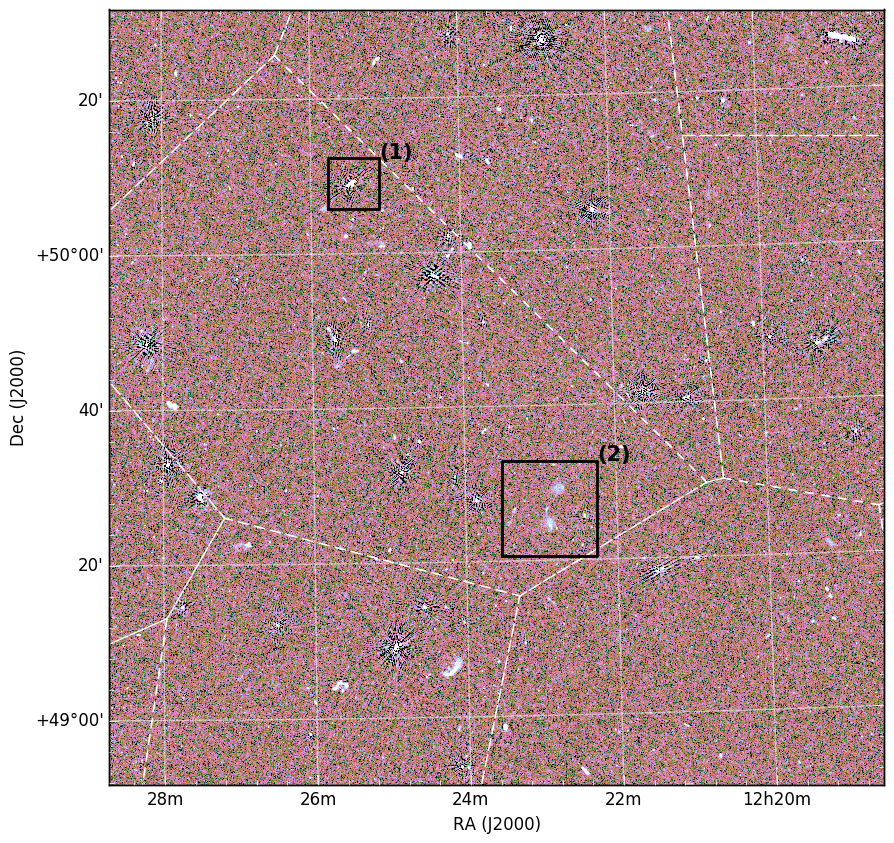}
    \caption{The central part of the {\sc P26Hetdex03} $8$ hours
      \LOFAR/-\HBA/ scan as imaged by Alg. \ref{alg:DR2}.} 
\end{subfigure}
    &
        \begin{tabular}{c}
        \smallskip
            \begin{subfigure}[t]{\SmallFigWidth}
                \centering
                \includegraphics[width=1\textwidth]{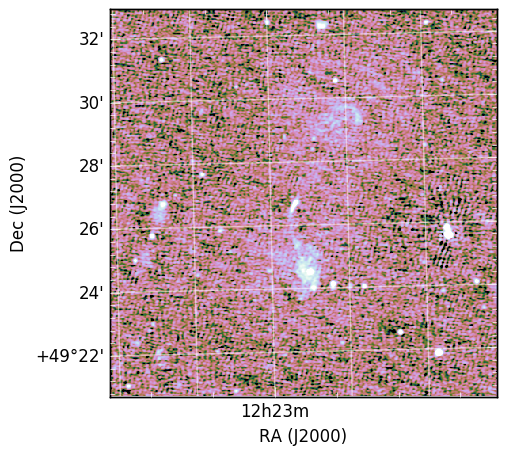}
                \caption{\label{fig:CompareDR1DR2_DR1sub2} Region (2) as imaged by Alg. \ref{alg:DR1}}
            \end{subfigure}\\
            \begin{subfigure}[t]{\SmallFigWidth}
                \centering
                \includegraphics[width=1\textwidth]{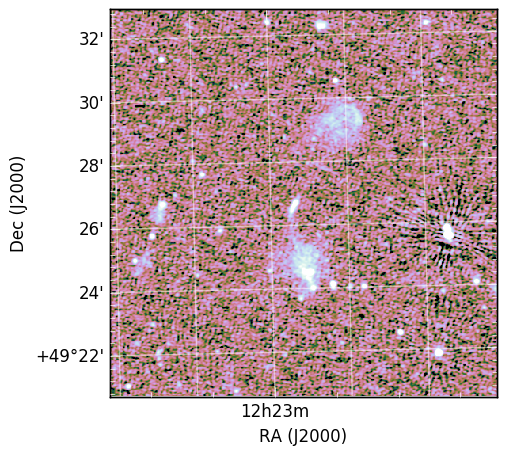}
                \caption{\label{fig:CompareDR1DR2_DR2sub2} Region (2) as imaged by Alg. \ref{alg:DR2}}
            \end{subfigure}
        \end{tabular}\\
    \end{tabular}
    \caption{\label{fig:CompareDR1DR2} This figure shows the
      differences between the maps produced by Alg. \ref{alg:DR1} and
      Alg. \ref{alg:DR2} from a typical $8$ hour scans \citep[here the
        {\sc P26Hetdex03} pointing in the HETDEX field, see][]{Shimwell17}. The colorscale is the same on
  all panels, and diplayed using an inverse hyperbolic sine function
  to render both the low level artifacts and some bright sources.}
\end{figure*}

\subsection{Data products}
\label{sec:AdditionalProds}

\subsubsection{Unpolarised flux}

Once the estimated \DD/-Jones matrices and skymodel $\widehat{\SkyXnu}$ have been obtained at
the highest available spatial resolution following the \DI//\DD/-self-calibration steps presented in
Alg. \ref{alg:DR2}, additional data products are
formed.

Users can adapt the weighting scheme depending on the scientific
exploitation they want to make of the interferometric data. This is
very much tied to how the calibration and deconvolution algorithms are
working, and concurrent effects take place along the self calibration
loop. Extended emission is hard to properly model since the
deconvolution problem is more ill-posed in these cases (more pixels
are non-zero). To tackle this issue the \PSF/ can be modified to make
the convolution matrix more diagonal and the deconvolution problem
correspondingly better conditioned. \ANSWII{This is done at the cost of a
lower sensitivity, that can drive, in return, systematic errors in the calibration
solutions estimates, because extended emission is poorly modeled on the shorter baselines.}

For all these concurring reasons the faint and extended flux in the highest
resolution maps produced by Alg. \ref{alg:DR2} is either poorly
modeled or not deconvolved at all. Since the pixel values of extended sources
are not interpretable in the residual maps, the flux density of the
radio sources cannot be measured if they are not deconvolved. We therefore intentionally degrade the resolution of some of the
imaging to allow survey users to choose a resolution based on the broad
scientific topic that they need to
address. Also, we store the sub-space deconvolution masks \citep[\SSD/
  hereafter, see][]{Tasse18} as residual
images so the end user can know if any given source has been
deconvolved. With this in mind, the following Stokes I products are generated: 

\begin{enumerate}
\item High resolution ($6$\arcsec, $20.000\times20.000$ pixels) wide-bandwidth Stokes I image (Step
  \ref{step:VII_I_DD4})
\item Low resolution ($20$\arcsec) wide-bandwidth Stokes I image (Step 2b.1)
\item High resolution ($6$\arcsec) Stokes I image in $3$ frequency
  chunks spread over
  the whole \HBA/ bandwidth (Step 2b.2)
\end{enumerate}

The \DI/-calibrated visibilities as well as the final skymodels and
\DD/ calibration solutions are stored. This allows for additional
postprocessing to be made such as better calibration towards a
particular point on the sky (van Weeren et
al in prep.), and also reimaging at different resolutions if required.

\subsubsection{QUV images}

The \DDFacet/ \DD/-imager only deals with
I-Stokes deconvolution. As discussed by \citet{Tasse18}, estimating
the QUV Stokes parameters is complex in the context of \DD/-imaging due
to the leakage terms. Indeed, for the problem to be properly addressed,
16 \PSF/ have to be computed (as there are 16 terms in the quadartic
mean of the Mueller matrices).
As most of the sources are unpolarised, the leakage terms are
properly taken into account in the \DD/-predict \ANSW{(i.e. the forward
mapping from sky and Jones matrices estimates to modeled visibilities)}. Instead of deconvolving the polarised flux, we grid the IQUV residual
data. The polarised flux is directly interpretable when the sources
are unresolved. Hence we also generate the following additional products:

\begin{enumerate}
  \setcounter{enumi}{3}
\item Low resolution ($20$\arcsec) spectral Stokes QU cubes (480
  planes - Step 2b.3)
\item Very low resolution Stokes QU cubes (480 planes -
  Step 2b.4), by cutting the baselines $>1.6$ km, giving an effective
  resolution of $\sim3\arcmin$
\item Low resolution ($20$\arcsec) wide-bandwidth Stokes V image (Step
  2b.5)
\end{enumerate}

The output QU cubes are processed using Faraday rotation
measure (RM) synthesis
\citep{bdb2005} to find polarised sources and their RM with the sensitivity of the full bandwidth. The wide
bandwidth (120 to 168 MHz) combined with the narrow channel width
(97.6 kHz) provides a resolution in RM space of $\sim$1.1 rad/m$^2$
and an ability to measure RMs of up to $\sim$450 rad/m$^2$ \citep[e.g.][]{osullivan2020}.

\begin{figure}
\begin{center}
\includegraphics[width=8cm]{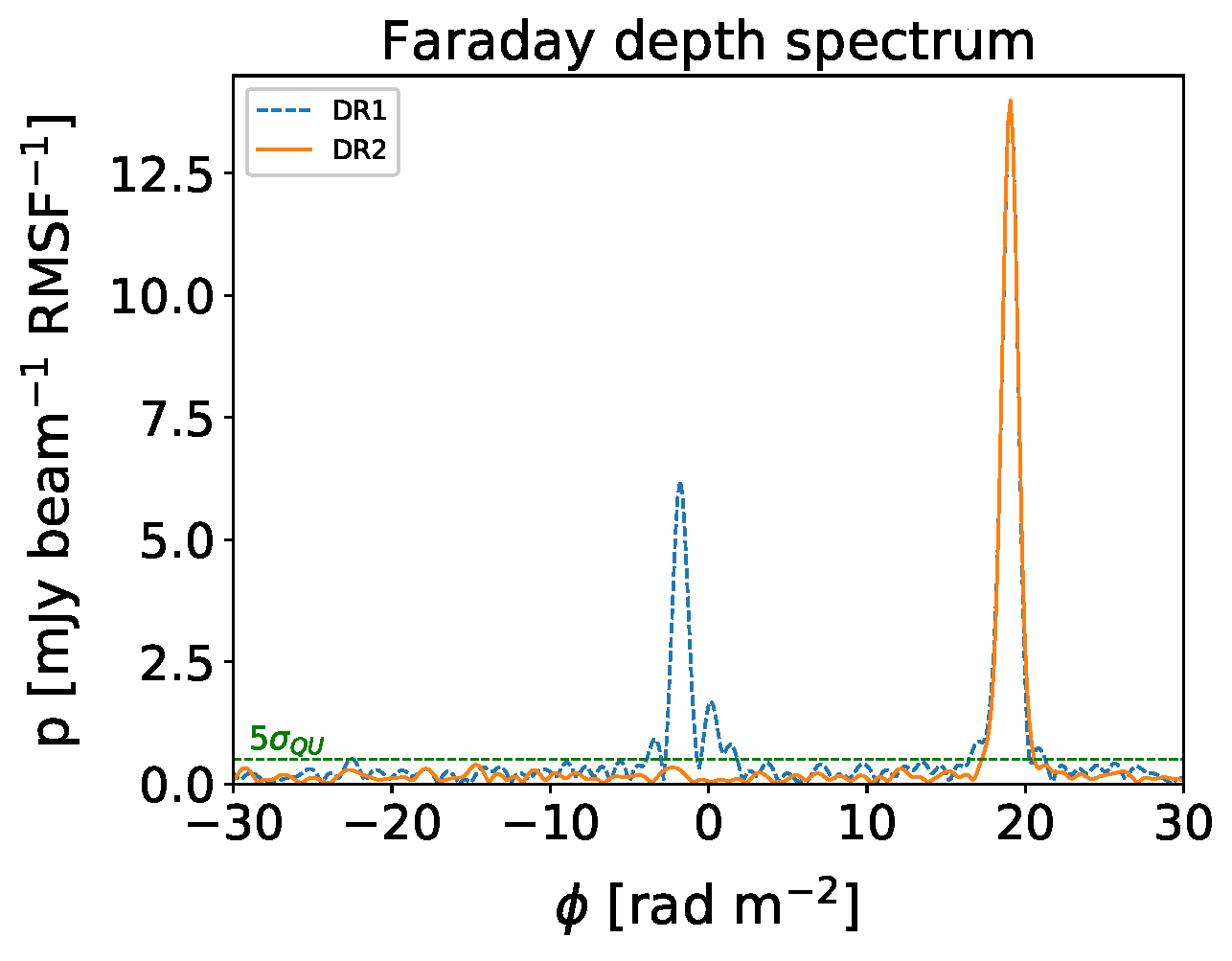}
\end{center}
\caption{\label{fig:leakage} A plot of the Faraday depth spectrum, or
  Faraday dispersion function (FDF), for a radio galaxy in both DR1
  and DR2 datasets, showing the improvement in the suppression of the
  instrumental polarisation signal. The blue dashed line shows the FDF
  from the DR1 data with a strong instrumental polarisation feature
  near Faraday depths of $\phi\sim 0$ rad/m$^2$, while the orange
  solid line shows the FDF from the DR2 data in which the instrumental
  feature is suppressed below the noise level. In both cases, the
  Faraday depth of the real astrophysical signal is the same.}
\end{figure}

The $3\arcmin$ QU cubes are sensitive to the large-scale polarised emission
from the Milky Way, while the $20\arcsec$ QU cubes are excellent for finding
compact polarised sources. However, detailed studies of the
polarisation and RM structure of resolved extragalactic sources will
require deconvolution of the Q and U data. The \PipeVII/ output
provides significantly better performance in correcting for the
effect of the instrumental polarisation (Fig. \ref{fig:leakage}),
which is typically at the level of 1\% or less for bright total
intensity sources (O'Sullivan et al., in prep).

There is no absolute polarisation angle calibration for each LoTSS
observation, meaning that while the RM values of sources in
overlapping fields are consistent, the polarisation angles are
not. Therefore, to avoid unnecessary depolarisation for both mosaicing
and the deep fields, the polarisation angles between the observations
need to be aligned. The simplest way to do this is by choosing a reference angle of a
polarised source in a single observation and applying a polarisation
angle correction to all other observations to align with this
reference angle, as presented in \citet{Herrera20}. An alternative
approach is to use the diffuse polarised emission that is present in
the $\sim3\arcmin$ QU cubes.


Bright polarised sources are rare in the LoTSS data, with only three
sources having a polarised intensity greater than 50 mJy beam$^{-1}$ in the
DR1 HETDEX sky area \citep{vaneck2018,osullivan2018}. However, in the
fields containing these bright polarised sources the \PipeVII/
output becomes unreliable for polarised sources. This limitation
likely arises from assuming $Q=U=V=0$ Jy for a field in the \DI/
calibration step. While only a few percent of fields are strongly
affected, the exact extent of this issue is being investigated further
through simulations, where bright polarised sources are inserted into
existing LoTSS uv-datasets. Possible solutions will be tested in
future pipeline developments.



\begin{figure}[ht!]
\begin{center}
\includegraphics[width=\columnwidth]{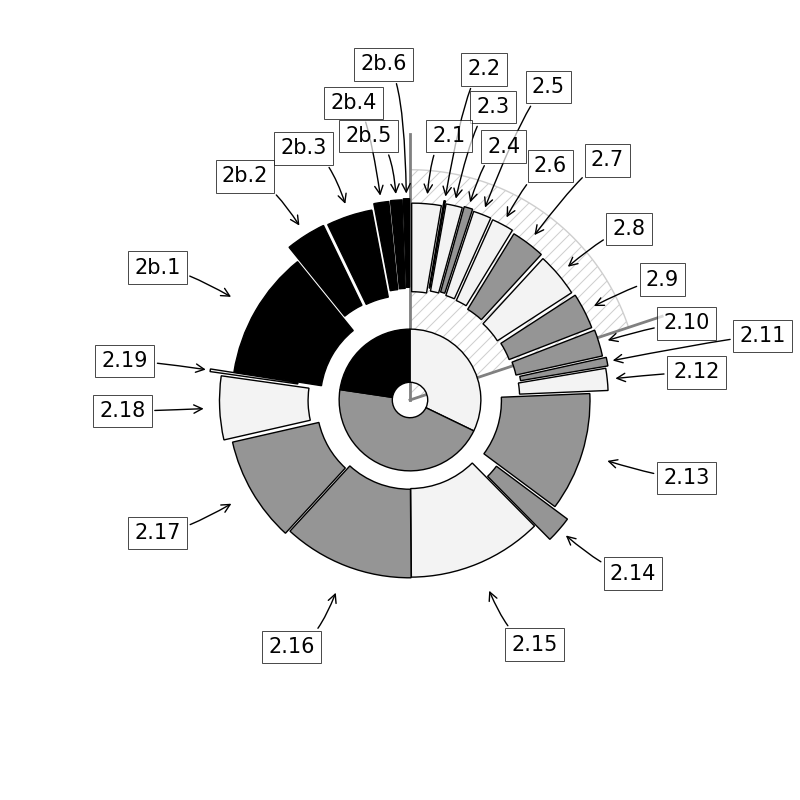}
\end{center}
\caption{\label{fig:Cheese} This pie graph shows the nature and ordering of the different steps of Alg. \ref{alg:DR2} and how the computing time
  is distributed across them. The
  lighter and darker grey areas represent the imaging and calibration
  steps respectively. The black area are the miscellaneous tasks
  (additional data products, see Sec. \ref{sec:AdditionalProds}) that
  are done once the \DI/ and \DD/ self-calibration loops have
  completed. It has been created from a \PipeVII/ run on a
  node equipped with 192 GBytes RAM and 2 Intel Xeon Gold 6130
  CPU@2.10GHz, giving 32 physical compute cores. The dashed area is a
  quadrant representing a day, while the inner pie shows the total
  contributions of the imaging, calibration and miscellaneous tasks.}
\end{figure}

\subsection{Comparison between \PipeVI/ and  \PipeVII/}
\label{sec:ComparisonVIVII}

\ANSW{
  Fig. \ref{fig:CompareDR1DR2} shows the comparison between the final
  high resolution images produced by \PipeVI/ and \PipeVII/ for an 8
  hours integration \LoTSS/ pointing ({\sc P26Hetdex03}).
  Many
  processes are involved in the sky reconstruction from radio
  interferometric data. Imaging and calibration affect the final
  synthesized maps and introduce complex and
  systematic residual errors. It is therefore difficult to find a good and absolute
  metric to compare the final imaging products. 

  As discussed in Sec. \ref{sec:DIDR}, the quality of the initial \DI/
  calibration proved to be quite crucial for the feasibility of the following
  \DD/ calibration and imaging steps. \PipeVI/ was indeed failing at imaging
  certain fields with very bright sources, while artifacts were
  present around most moderately bright ones, thereby driving the dynamic
  range limit in large areas. In Fig. \ref{fig:CompareDR1DR2_DR1sub1}
  and \ref{fig:CompareDR1DR2_DR2sub1} we show a radio source imaged by
  \PipeVI/ and \PipeVII/.
  
  Another important issue with the approach we presented in
  \citet[][]{Shimwell18} was the presence of a low spatial frequency
  pattern corresponding to a positive or negative halo around radio
  sources. Although the effect is complex to analyze, we concluded
  from various experiments that these systematics were due to
  the combination of (i) skymodel incompleteness, (ii) a uv-distance cut used during the calibration
  and (iii) the $\OpNorm$ normalisation function (see
  Sec. \ref{sec:PipeDR_I} for details), that we had introduced for
  \PipeVI/ to be robust against the absorption of extended extended
  emission. As shown in Fig. \ref{fig:CompareDR1DR2_DR1sub2} and
  \ref{fig:CompareDR1DR2_DR2sub2}, the approach developed in
  Sec. \ref{sec:regularisation} and \ref{sec:SlowSolve} to conserve
  unmodeled extended emission and implemented in \PipeVII/ does not produce any
  significant low spatial frequencies systematics.
}

\subsection{\PipeVII/ robustness and performance}
\label{sec:Profiling}

As explained above \PipeVII/ is a high level script interfacing \kMS/
and \DDFacet/. Both of the underlying software packages are
efficiently parallelised using a custom version of the Python {\tt
  multiprocessing} package for process-level parallelism, and using
the
SharedArray\footnote{\url{https://pypi.python.org/pypi/SharedArray}}
module. As explained by \citet{Tasse18}, this pythonic approach
minimizes the process interconnections for both the \kMS/ and \DDFacet/
software.


This paper considers the application of \PipeVII/ to the \LoTSSd/. The pipeline is also being used to process data from the
wider and shallower \LoTSS/ survey. The \LoTSS/ project is presently observing at a rate of up to
1,500\,hrs every 6 month cycle which corresponds to approximately two
8\,hr pointings (observed simultaneously) each day. The \PipeVII/ compute time is roughly split equally between
calibration and imaging tasks (see Fig. \ref{fig:Cheese}). The
total run time for an 8 hour pointing is $\sim5$ days (on a
  node equipped with 192 GBytes RAM and 2 Intel Xeon Gold 6130
  CPU@2.10GHz, giving 32 physical compute cores), and
takes an extra $\sim30\%$ of computing time to completion as compared
to \PipeVI/. Hence 10 compute nodes are
sufficient to keep up with the observing rate. However, in practice
more compute nodes are used because \LoTSS/ has been observing since
2014 and as of June 1st 2019 over 1,000 pointings exist in the
archive. Over $\sim1000$ pointings and $\sim12$ PB of averaged
and compressed \LOFAR/ data ($\sim40$ PB uncompressed) have now been
processed with \PipeVII/.


\section{LoTSS deep fields data and processing}
\label{sec:DeepObs}


\newcommand{\MyWidth}{.8\columnwidth}
\begin{table*}[t]\footnotesize
\centering
  \caption{\label{tab:Observations} Overview of the deep fields
    pointings used to synthetise the images on the \Bootes/ and \LH/
    extragalactic fields. Columns $f_{flag}$ and nMS stand for the
    fraction of flagged data and number of measurement sets present in
    the archives.}
\begin{tabular}{ccccccccc}
\bottomrule
Project ID & LOFAR Obs. ID & Obs. Date & Start time & Integration & $\nu_{min}$ & $\nu_{max}$ & $f_{flag}$ & nMS\\
 & & & UTC & time [h] & [MHz] & [MHz] &  &\\
\midrule
\multicolumn{9}{c}{{\bf \Bootes/}}\\
LC2\_038 &   L236786 &  21/07/2014 &  10:44:00 &  8.0 &  120.0 &  168.7 &  37.8 &  25 \\
LC2\_038 &   L243561 &  15/09/2014 &  10:22:42 &  8.0 &  120.0 &  168.7 &  19.2 &  25 \\
LC4\_034 &   L346004 &  11/06/2015 &  16:04:35 &  8.0 &  120.2 &  167.0 &  10.9 &  24 \\
LC4\_034 &   L373377 &  17/09/2015 &  10:21:18 &  8.0 &  120.2 &  168.9 &  20.1 &  25 \\
LC4\_034 &   L374583 &  24/09/2015 &  10:09:57 &  8.0 &  120.2 &  168.9 &  10.6 &  25 \\
LC4\_034 &   L387597 &  29/09/2015 &  09:13:00 &  8.0 &  120.2 &  168.9 &  27.4 &  25 \\
LC4\_034 &   L387569 &  01/10/2015 &  09:00:00 &  8.0 &  120.2 &  168.9 &  32.3 &  25 \\
LC4\_034 &   L400135 &  10/10/2015 &  08:46:22 &  8.0 &  120.2 &  168.9 &  13.6 &  25 \\
LC4\_034 &   L401825 &  21/10/2015 &  08:00:30 &  8.0 &  120.2 &  168.9 &   9.0 &  25 \\
LC4\_034 &   L401839 &  22/10/2015 &  07:55:23 &  8.0 &  120.2 &  168.9 &   8.0 &  25 \\

\midrule
\multicolumn{9}{c}{{\bf \LH/}}\\
  LC3\_008 &  L274099 & 08/03/2015 &  20:11:00 &  8.7 &  120.2 &  168.9 &  12.4 &  25 \\
  LC3\_008 &  L281008 & 14/03/2015 &  18:26:39 &  8.7 &  120.4 &  169.1 &   8.1 &  25 \\
  LC3\_008 &  L294287 & 21/03/2015 &  19:11:00 &  8.7 &  120.2 &  168.9 &  18.7 &  25 \\
  LC3\_008 &  L299961 & 24/03/2015 &  17:47:20 &  8.7 &  120.2 &  168.9 &  12.1 &  25 \\
  LC3\_008 &  L340794 & 25/04/2015 &  17:08:00 &  8.7 &  120.2 &  168.9 &  14.5 &  25 \\
  LC3\_008 &  L342938 & 08/05/2015 &  14:50:24 &  8.7 &  120.2 &  168.9 &  16.6 &  25 \\
 LT10\_012 &  L659554 & 10/07/2018 &  11:11:00 &  8.0 &  120.2 &  168.9 &   9.9 &  25 \\
 LT10\_012 &  L659948 & 12/07/2018 &  11:08:10 &  8.0 &  120.2 &  168.9 &  11.9 &  25 \\
 LT10\_012 &  L664320 & 15/08/2018 &  08:49:00 &  8.0 &  120.2 &  168.9 &  11.0 &  25 \\
 LT10\_012 &  L664480 & 19/08/2018 &  08:38:46 &  8.0 &  120.2 &  168.9 &  11.3 &  25 \\
 LT10\_012 &  L667204 & 12/09/2018 &  07:06:09 &  8.0 &  120.2 &  168.9 &  11.2 &  25 \\
 LT10\_012 &  L667218 & 13/09/2018 &  07:05:12 &  8.0 &  120.2 &  168.9 &  10.3 &  25 \\
\bottomrule
\end{tabular}
\end{table*}

\subsection{Observations}

\LoTSSd/ observations are being carried out over the four 
northern fields with high-Galactic latitude and the highest-quality multi-degree-scale ancillary
data across the electromagnetic spectrum: the \Bootes/ field, the
Lockman Hole, \ELAIS/ and the North Ecliptic Pole fields. The
ultimate aim of the LoTSS Deep Fields project is to reach noise levels
of 10-15 \uJypb/ in each of these fields (requiring $\sim500$ hours of
integration). The first \LoTSSd/ data release consists of
initial observations in three of these fields: \Bootes/ ($\sim80$ hrs) and
Lockman Hole ($\sim112$ hrs) presented in the current paper, and \ELAIS/
\citep[presented by][for an integration time of $\sim170$ hrs in paper 2]{LoTSSDeepII}. This first data
release also includes an extensive effort of
optical/IR cross-matching, which has obtained host galaxy
identifications for over 97\% of the $\sim$80,000 radio sources detected
within the  $\sim25$ \sqdeg/ overlap with the high-quality
multi-wavelength data \citep[][Paper 3]{LoTSSDeepIII}. This is
supplemented by high quality photometric redshifts, and
characterisation of host galaxy properties \citep[][Paper 4]{LoTSSDeepIV},
and source classification \citep[e.g. star-forming vs AGN:][Paper 5]{LoTSSDeepV}.

In order to put the \LoTSS/-deep observations in a wider context, in this section we briefly describe the multi-wavelength data available on the \Bootes/
and \LH/ fields, focusing on the radio coverage \citep[for a more detailed description see][Paper 3]{LoTSSDeepIII}.

\begin{figure}[ht!]
\begin{center}
\includegraphics[width=\columnwidth]{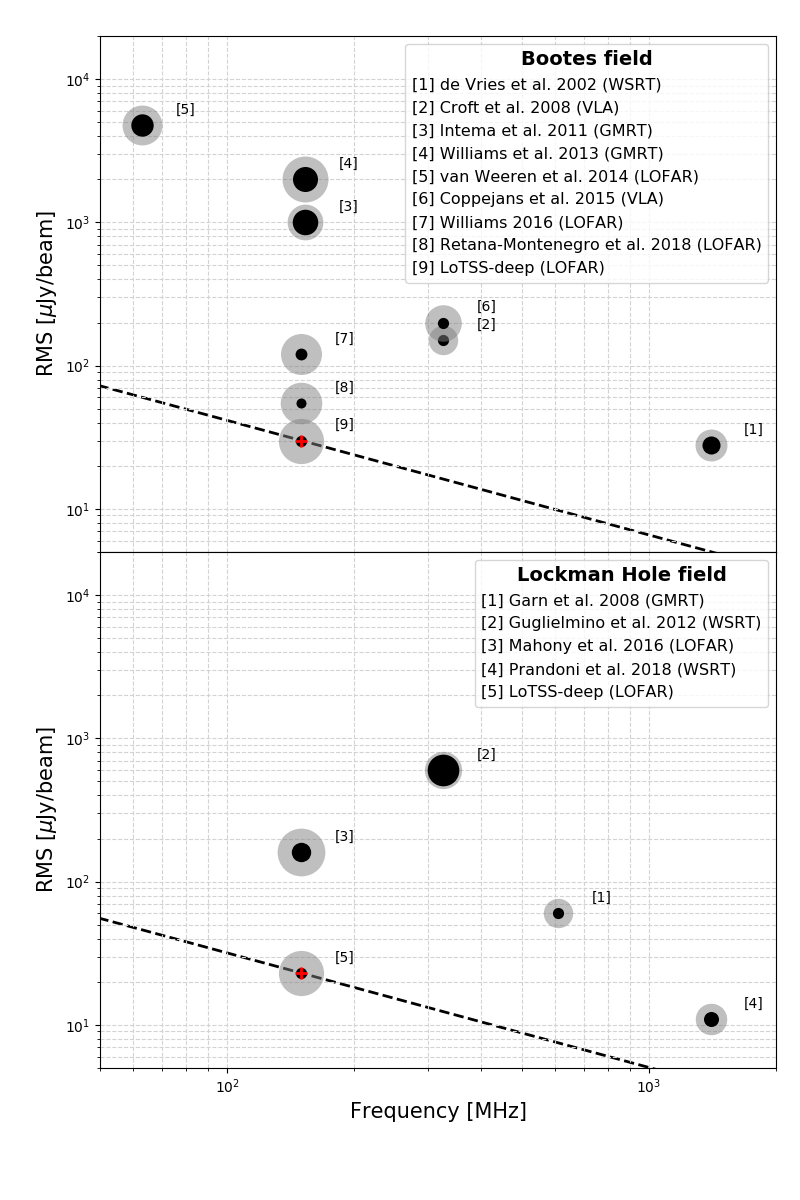}
\end{center}
\caption{\label{fig:Sensitivity} The sensitivity of the various deep dedicated
  surveys covering the \Bootes/ (top) and \LH/ fields (bottom) as a
  function of observing frequency. The resolution of the various
  surveys corresponds to the radius of the black dot, while the diameter
  of the corresponding surveyed area is encoded in the size of the
  gray circle. The \LoTSS/-deep pointings are marked with a red
  cross, the dashed line corresponding to a source having a spectral
  index of $-0.7$.}
\end{figure}

\subsubsection{\Bootes/ field}
\label{sec:ObsBootes}

The \Bootes/ field is one of the NOAO Deep Wide Field Survey
\citep[NDWFS][]{Jannuzi99} fields covering $\sim9.2$ deg$^2$. It contains
multi-wavelength data including infrared \citep[{\it Spitzer} space
  telescope, see][]{Ashby09,Jannuzi10}, X-rays \citep[{\it Chandra} space
  telescope, see][]{Murray05,Kenter05}, optical data
\citep[][]{Jannuzi99,Cool07,Brown07,Brown08}
. At radio frequencies it has been mapped with
the Westerbork Radio Telescope \citep[WSRT, see][]{deVries02}, the
Very Large Array \citep[VLA, see][]{Croft08,Coppejans15}, the Giant
Meterwave Radio Telescope \citep[GMRT, see][]{Intema11,Williams13} and
LOFAR \citep[][]{vanWeeren14,Williams16,Retana18} at various depths,
frequencies, resolutions and covered areas (see Fig.
\ref{fig:Sensitivity} for an overview of the available radio data).

The \Bootes/ pointings data that are presented in this paper are centered on
$(\alpha,\delta)=$(\RA{14}{32}{00},\DEC{+34}{30}{00})
and were observed with
the \LOFAR/-\HBA/ in \HBADUALINNER/ mode during Cycle 2 and Cycle 4,
with a bandwidth of $48$ MHz (see Tab. \ref{tab:Observations}). The total integration time
of $\sim80$ hours is spread over $10$ scans of 8 hours.

\subsubsection{Lockman hole}
\label{sec:LH}

The Lockman Hole field is also covered by a large variety of
multiwavelength data. Specifically, it has been observed by the {\it Spitzer} Wide-area
Infrared Extragalactic survey \citep[SWIRE][]{Lonsdale03} over
$\sim11$ \sqdeg/, and over $16$ \sqdeg/ by the {\it Herschel} Multi-tiered
Extragalactic Survey \citep{Oliver12}. It has also been observed in UV
\citep{Martin05}, optical \citep{Gonzalez11}, near IR 
\citep[UK Infrared Deep Sky Survey Deep Extragalactic Survey UKIDSS-DXS, see][]{Lawrence07}, and with the Submillimetre
Common-User Bolometer Array \citep[][]{Coppin06,Geach17}. At higher energy, it
has been observed with {\it XMM-Newton} \citep{Brunner08}, and {\it Chandra}
\citep{Polletta06}. In the radio domain, the \LH/ has been observed
over the two deep aforementioned X-ray fields over small sub-\sqdeg/
areas \citep{deRuiter97,Ciliegi03,Biggs06,Ibar09}. Wide surveys of the
\LH/ have been done with GMRT \citep{Garn10}, VLA \citep{Owen09}, WSRT
\citep{Guglielmino12,Prandoni18} and LOFAR at 150 MHz
\citep{Mahony16}. Fig.
\ref{fig:Sensitivity} presents an overview of the available radio data on
the \LH/.

Our \LH/ observation that we are presenting in this paper consists of $12$ pointings of $\sim8$ hours centered on
$(\alpha,\delta)=$(\RA{10}{47}{00},\DEC{+58}{05}{00}) and
observed from March 2015 (Cycle 3) to November 2018 (Cycle 4). As for
the \Bootes/ field observation, we observe in \HBADUALINNER/ with
$\sim48$ MHz bandwidth, while the integration time depends on the
\LOFAR/ cycle ($8.7$ hours in cycle 3, $8$ hours in cycle 10, see Tab. \ref{tab:Observations}). The
total integration time is $\sim100$ hours.

\subsection{Image synthesis}
\label{sec:DeepImageSynthesis}

\begin{figure}[]
\begin{center}
\includegraphics[width=\columnwidth]{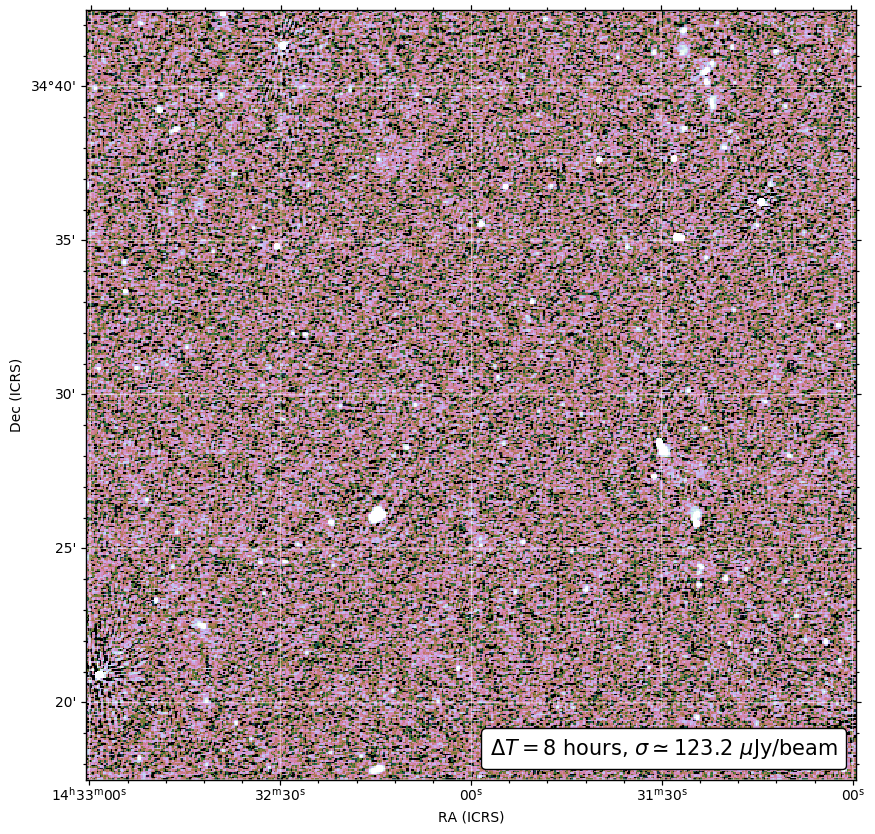}
\includegraphics[width=\columnwidth]{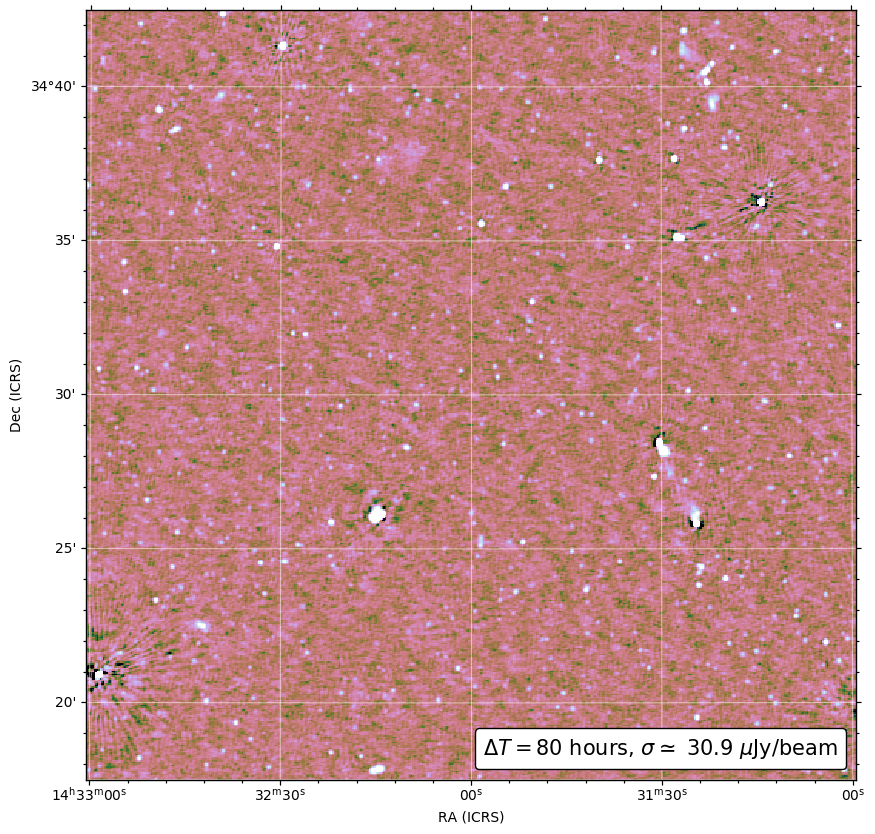}
\caption{\label{fig:Comp880} This figure shows the restored high
  resolution image towards the center of the \Bootes/ field for the
  $8$ hours image produced with Alg. \ref{alg:DR2} (top panel) and
  the $80$ hours image produced with
  Alg. \ref{alg:DR2_DEEP} (bottom panel). Both images are thermal noise
limited, with the same colorscale being used on both. }
\end{center}
\end{figure}

\def\MainFigWidth{0.55\textwidth}
\def\SmallFigWidth{0.25\textwidth}

\begin{figure*}[!htb]
    \centering
    \begin{tabular}[t]{cc}
\begin{subfigure}{\MainFigWidth}
    \centering
    \includegraphics[width=1\linewidth,height=1\textwidth]{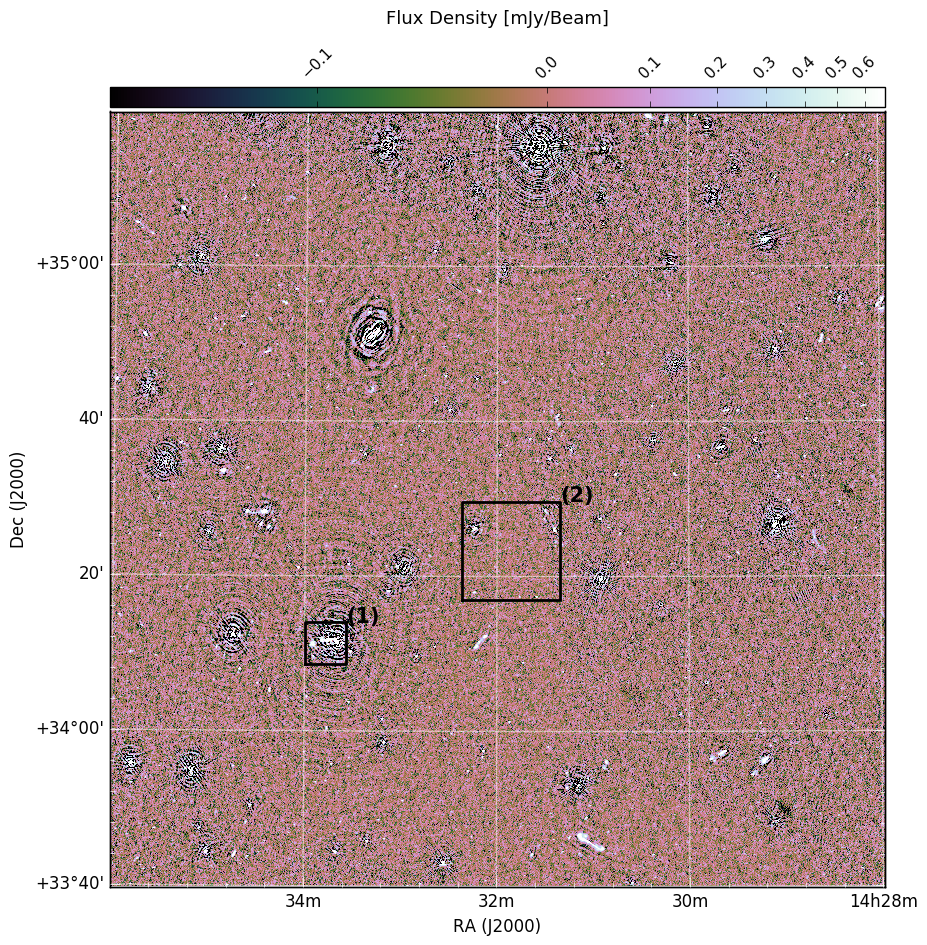}
    \caption{\label{fig:FactorCentral} The central $\gtrsim2$ \sqdeg/ part of the Bootes field as imaged by the
      direction dependent \FACTOR/ algorithm \citep{Retana18}.}
\end{subfigure}
    &
        \begin{tabular}{c}
        \smallskip
            \begin{subfigure}[t]{\SmallFigWidth}
                \centering
                \includegraphics[width=1\textwidth]{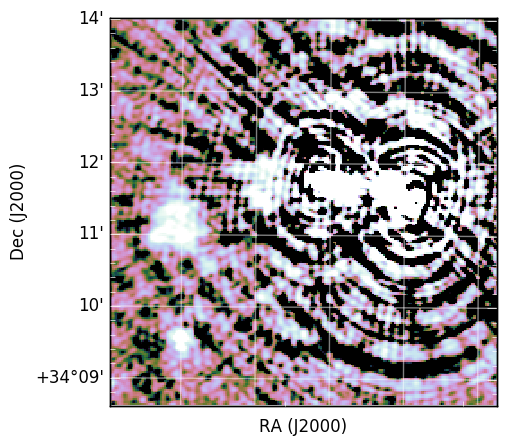}
                \caption{Zoom in on region (1) of the map
                  synthesised by \citet{Retana18}.}
            \end{subfigure}\\
            \begin{subfigure}[t]{\SmallFigWidth}
                \centering
                \includegraphics[width=1\textwidth]{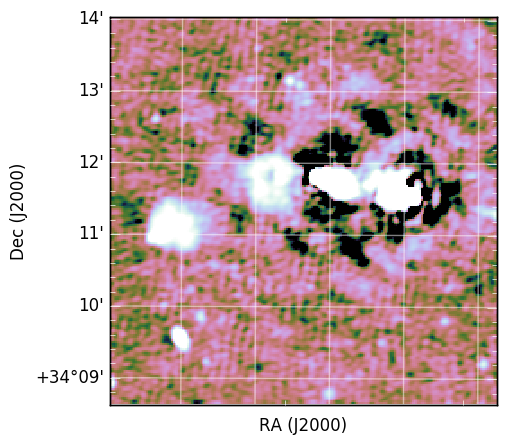}
                \caption{Zoom in on region (1) of the map
                  synthesised by \kMS/-\DDFacet/ (this work).}
            \end{subfigure}
        \end{tabular}\\

\begin{subfigure}{\MainFigWidth}
    \centering
    \includegraphics[width=1\linewidth,height=1\textwidth]{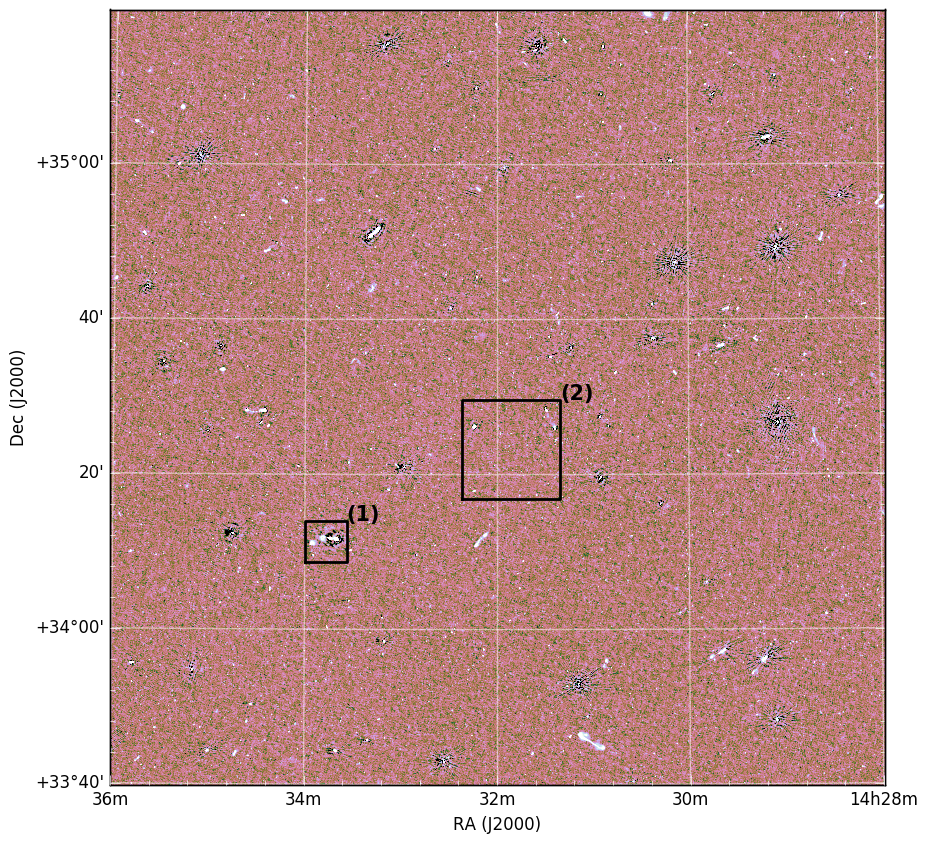}
    \caption{The same as in \ref{fig:FactorCentral}, but imaged with
      Alg. \ref{alg:DR2_DEEP}.} 
\end{subfigure}
    &
        \begin{tabular}{c}
        \smallskip
            \begin{subfigure}[t]{\SmallFigWidth}
                \centering
                \includegraphics[width=1\textwidth]{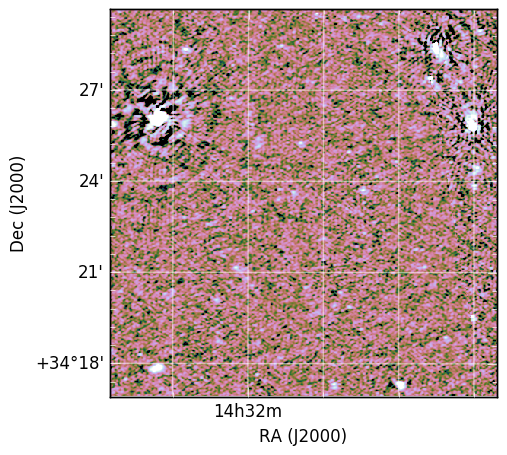}
                \caption{Zoom in on region (2) of the map
                  synthesised by \citet{Retana18}.}
            \end{subfigure}\\
            \begin{subfigure}[t]{\SmallFigWidth}
                \centering
                \includegraphics[width=1\textwidth]{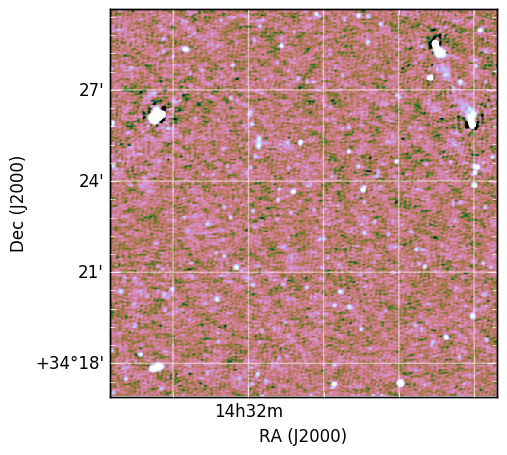}
                \caption{Zoom in on region (2) of the map
                  synthesised by \kMS/-\DDFacet/ (this work).}
            \end{subfigure}
        \end{tabular}\\
    \end{tabular}
    \caption{\label{fig:CompBootesFactor} Comparison between the
      \LOFAR/-\HBA/ maps generated at $150$ MHz by \citet{Retana18} and
      in the current work. The colorscale is the same on
  all panels, and diplayed using an inverse hyperbolic sine function
  to render both the low level artifacts and some bright sources.}
\end{figure*}

\def\MainFigWidth{0.55\textwidth}
\def\SmallFigWidth{0.25\textwidth}

\begin{figure*}[!htb]
    \centering
    \begin{tabular}[t]{cc}
\begin{subfigure}{\MainFigWidth}
    \centering
    \includegraphics[width=1\linewidth,height=1\textwidth]{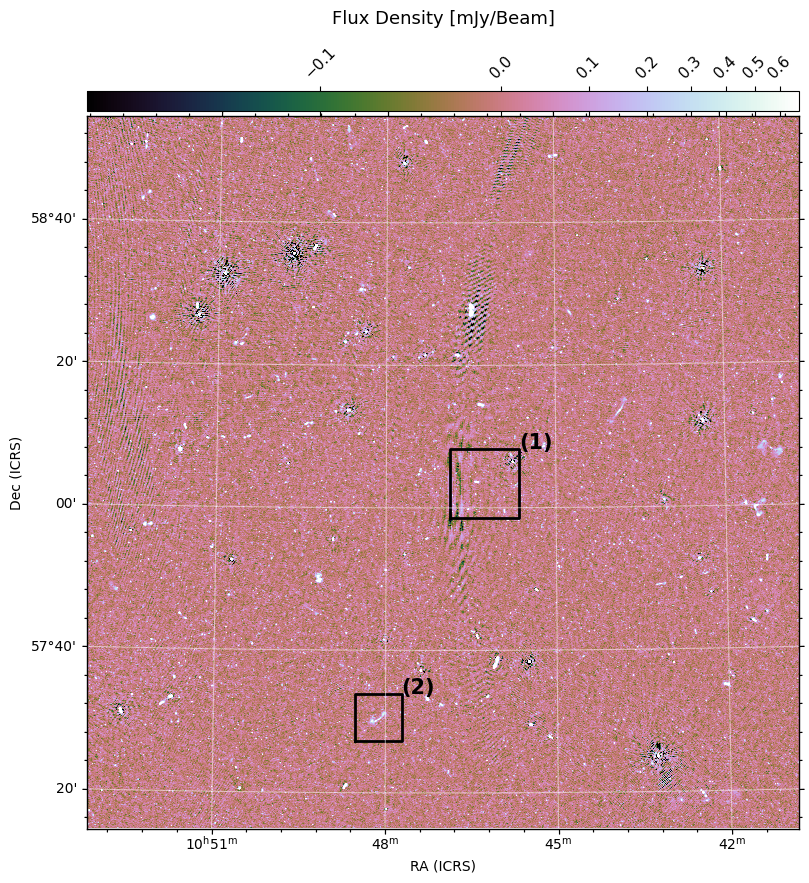}
    \caption{\label{fig:LHCentral} The central $\gtrsim2$ \sqdeg/ part of the \LH/ field as imaged by Alg. \ref{alg:DR2_DEEP} (Sec. \ref{sec:DeepImageSynthesis}).}
\end{subfigure}
    &
        \begin{tabular}{c}
        \smallskip
            \begin{subfigure}[t]{\SmallFigWidth}
                \centering
                \includegraphics[width=1\textwidth]{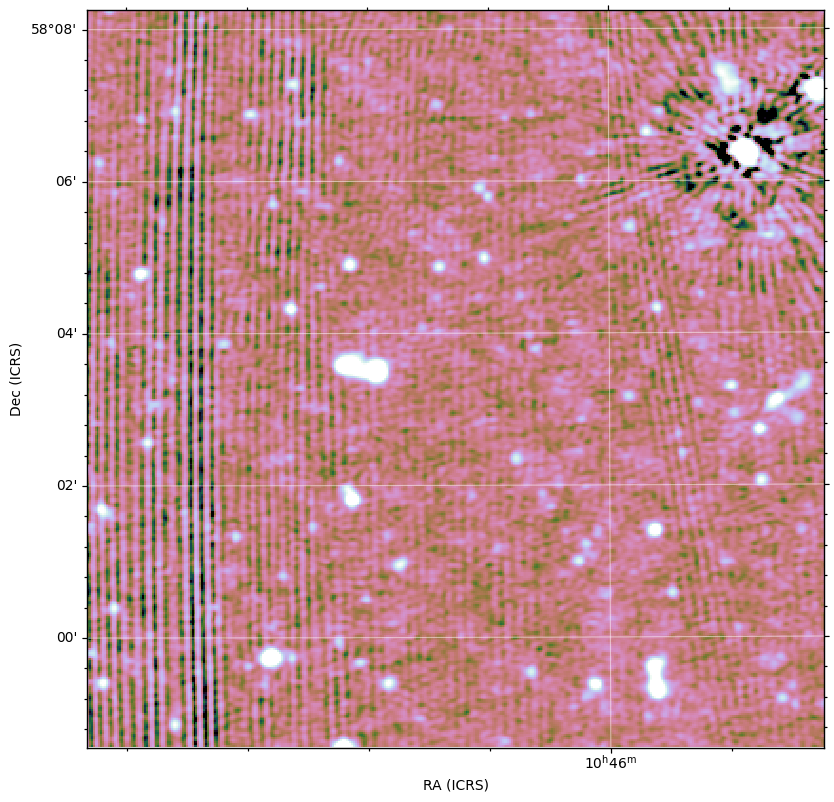}
                \caption{\label{fig:StripesLHCentral} Zoom in on region (1) of the map shown in Fig. \ref{fig:LHCentral}.}
            \end{subfigure}\\
            \begin{subfigure}[t]{\SmallFigWidth}
                \centering
                \includegraphics[width=1\textwidth]{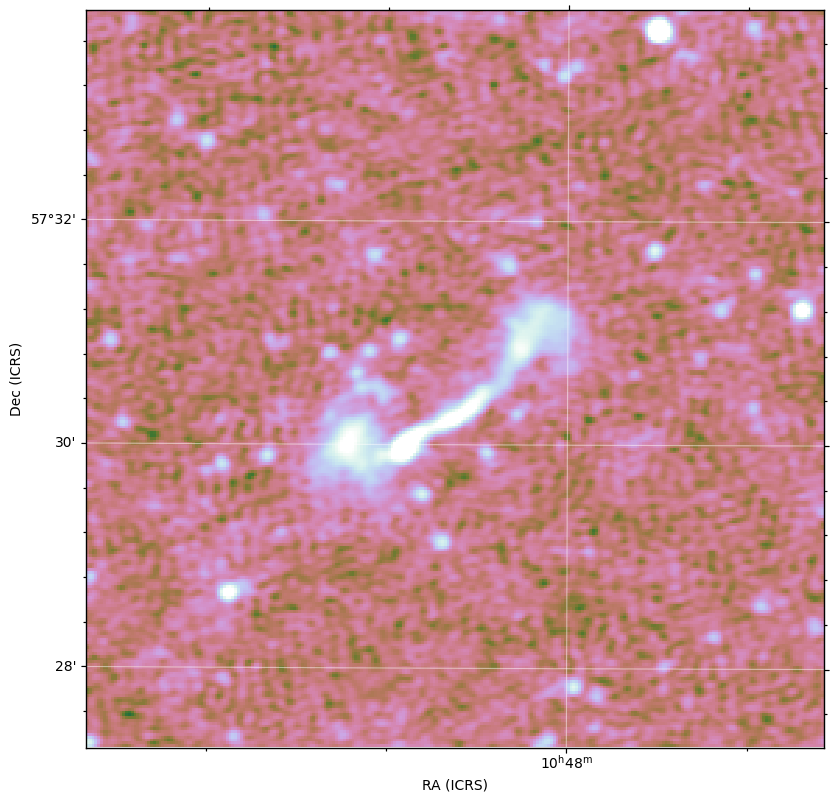}
                \caption{Zoom in on region (2) of the map shown in Fig. \ref{fig:LHCentral}.}
            \end{subfigure}
        \end{tabular}
        \end{tabular}
    \caption{\label{fig:LHMap} This figure shows the central region of the deep 
      \LOFAR/-\HBA/ maps of the \LH/ field generated at $150$ MHz. The colorscale is the same on
  all panels, and diplayed using an inverse hyperbolic sine function
  to render both the low level artifacts and some bright sources. \ANSWII{The
        stripy artifact seen in the zoomout Fig. \ref{fig:StripesLHCentral} seems to be
        produced by the residual deconvolution and calibration errors
        of a few $\gtrsim10$ \mJypb/ bright sources that are a few degrees away from the center of the field.}}
        
\end{figure*}

\begin{figure}[ht!]
\begin{center}
\includegraphics[width=\columnwidth]{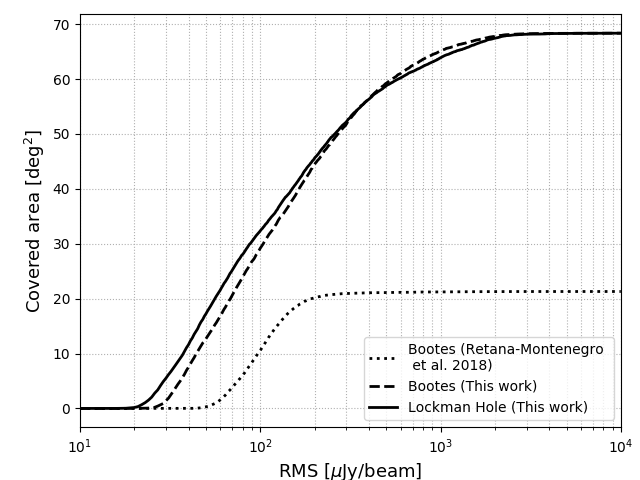}
\end{center}
\caption{\label{fig:SensitivityArea} The cumulative distribution of
  the local noise estimates in the various maps discussed
  here. As shown here, we have imaged a larger fraction of
  \LOFAR/'s \HBA/ primary beam than the image presented in \citet{Retana18}.
}
\end{figure}

\ALGODRTWOBDEEP/

The \LH/ and \Bootes/ fields data have been both reduced using Alg. \ref{alg:DR2_DEEP}. In this approach we first build a
wide-band \DI/+\DD/ self-calibrated sky model $\SkyXnu$ from a
single wide band $\sim8$ hours observation using
Alg. \ref{alg:DR2}. This model is then used to \DI/+\DD/ calibrate all the $n_p$ pointings
(with $n_p=10$ and $n_p=12$ for the \Bootes/ and \LH/ datasets
respectively) following Alg. \ref{alg:DR2_DEEP}. This amounts to
repeating Steps \ref{step:VII_K_DD2} to \ref{step:VII_I_DD4} of
Alg. \ref{alg:DR2} on a larger dataset. A comparison between the
images synthetised from $8$
and $80$ hours datasets is presented in Fig. \ref{fig:Comp880}. On a single node equipped with $\sim500$ GB of
$2.4$ GHz RAM 
and $2$ Intel Xeon CPU E5-2660 v4@2.00GHz with $14$ physical
cores each, Alg. \ref{alg:DR2_DEEP} took $\sim21$ days to process the
$80$ hours of \Bootes/ data. Fig. \ref{fig:CompBootesFactor} (further
discussed in Sec. \ref{sec:ComparisionFactor}) and
\ref{fig:LHMap} show the central parts of the of these deep \LOFAR/ \Bootes/ and
\LH/ observations.


Estimating the noise in radio maps is not straightforward since noise
is correlated and non-Gaussian. Also, while the covariance
matrix should be entirely described by the \PSF/, the real covariance
matrix is hard to estimate due to the calibration artifacts \citep[see][for a
  detailed discussion]{Tasse18,Bonnassieux18}. Here, in order to estimate the
local noise we use the statistics of the $\min{\{.\}}$
estimator (that returns the minimum value of a given sample). Intuitively, while the I-Stokes image $\max{\{.\}}$ statistics has
contributions from both artifacts and real sources, the $\min{\{.\}}$
only accounts for the artifacts. A $\min{\{.\}}$ filter with a given
box size is therefore run through a restored image, and depending on
the box size\footnote{The cumulative distribution $\mathcal{F}$ of 
  $Y=\min{\{X\}}$ with $X\sim\mathcal{N}\{\mu=0,\sigma=1\}$ is
  $\mathcal{F}\{y\}=1-\left[\frac{1}{2}\left(1-\mathrm{erf}\left\{\frac{y}{\sqrt{2}}\right\}\right)\right]^n$, where $n$ in the number of pixels in a given box. Finding
  $y_{\sigma}$ such
that $\mathcal{F}\{y_{\sigma}\}=1/2$ given the box size gives us a
conversion factor from the minimum estimate to the standard deviation.},
the effective standard deviation is derived.

Fig. \ref{fig:SensitivityArea} shows the cumulative distribution of
the local noise in the \LH/ and \Bootes/ fields maps, reaching
$\lesssim23$ and $\lesssim30$ \uJypb/ respectively.
Taking into account the number of pointings with their respective
amount of flagged data, we get total integration times of $\sim65$ and
$\sim88$ hours on the \Bootes/ and \LH/ fields respectively, giving a
theoretical thermal noise difference of a factor $\sim1.16$ compatible
with the observed value of $\sim1.3$. Other factors to be taken
into account to compare noise properties include the
bootstrapping errors, the individual fields' average elevation, and the
Galactic noise differences.


\subsection{Comparison with deep \FACTOR/ image synthesis}
\label{sec:ComparisionFactor}


The image of the \Bootes/ field based on $55$ hours of \LOFAR/
\HBA/ data and presented \citet{Retana18} reaches an unprecedented noise
level image of
$\sim55$ \uJypb/ at $150$ MHz.
To achieve such high
sensitivity, \citet{Retana18} have applied third generation calibration and imaging to
correct for the \DDE/ using the \FACTOR/ package
\citep[developped by][see Sec. \ref{sec:3GC} for more detail]{Weeren16}.
Because the set of \LOFAR/ datasets used by
\citet{Retana18} is different\footnote{Out of the sets of $7$ and $10$
  observations used in \citet{Retana18} and in this work respectively,
  $4$ are common, namely L243561, L374583, L400135, L401825.} the
comparison can only be approximate. In
Fig. \ref{fig:CompBootesFactor} we compare the images produced by
\citet{Retana18} and by Alg. \ref{alg:DR2_DEEP}. While the noise
difference should be on the order of $20\%$, as shown in
Fig. \ref{fig:SensitivityArea} the measured one is on the level of
$\sim60\%$. Consistently artifacts around bright sources are also much less severe in the maps generated
by Alg. \ref{alg:DR2_DEEP} and implemented in \PipeVII/. 

\subsection{Cataloguing}


In order to extract astrophysical information we build a catalogue of
radio sources from the images produced by Alg. \ref{alg:DR2_DEEP} and
the data described in Sec. \ref{sec:DeepObs}. Even in the apparent
flux maps, because of the imperfect calibration and imaging, the
\LoTSS/-deep images have spatially variable noise, and to deal with
this issue we use
PyBDSF\footnote{\url{https://www.astron.nl/citt/pybdsf}} \citep[Python
  Blob Detector and Source Finder, see][]{Mohan15} since it measures
noise locally rather than globally. The sources were detected with a
$3$ and $5\sigma$ for the island and peak detection threshold
respectively. The position-dependent noise was estimated using a
sliding box algorithm with a size of $40\times40$ synthesised beams,
except around bright sources where the box size was decreased to
$15\times15$ beams to more accurately capture the increased noise in
these regions. The columns kept in the final catalogue are the source
position, peak and integrated flux density, source size and
orientation, the associated uncertainties, the estimated local rms at
the source position, as well as a code describing the type of
structure fitted by PyBDSF. As described in \citet{LoTSSDeepII}, the
peak and integrated flux densities of the final catalogs and images
are corrected from overall scaling factors of $0.920$ and $0.859$ for
the for \LH/ and \Bootes/ fields respectively. \ANSW{These 
numbers were estimated from the comparison between the \LoTSS/-deep flux
densities and a variety of radio data available at other frequencies.}
The full catalogues
cover out to 0.3 of the power primary beam and contain 36,767 entries
over 26.5 square degrees and 50,112 over 25.0 square degrees for
\Bootes/ and \LH/ respectively. These raw PyBDSF catalogues are
available online on the \LOFAR/ survey webpage
\url{https://www.lofar-surveys.org/} and a thorough analysis of the
source catalogues will be presented by Mandal et al. in preparation.


\section{Conclusion and future plans}

Imaging low-frequency \LOFAR/ data at high resolution and over wide fields of
view is extremely challenging. This is mainly due to the
\RIME/ system being complex in this regime: the background
wide-band sky is unknown, as are the time-frequency-antenna
\DD/-Jones matrices. Due to the high number of free parameters in that
system, and to the finite amount of data points in the non-linear
\RIME/ system, the inversion can be subject to ill-conditioning and
the \DD/-\CRIME/ solver can absorb unmodeled extended flux.

In order to address this robustness issue we have developed a strategy that aims at conserving the unmodeled emission
without affecting the final dynamic range. The method we have
developed has similarities with those presented by
\citet{Yatawatta15,Weeren16,Repetti17,Birdi20}, and relies on
reducing the effective size of the unknown stochastic process. We show that this allows us to recover most of
the faint unmodeled extended emission.

We have applied this third generation calibration and imaging \DD/
algorithm both to the wide-field imaging of the \LoTSS/ survey and to
the synthesis of deep $150$ MHz resolution images on the \Bootes/ and
\LH/ fields. The synthesized images are the deepest ever obtained at
these frequencies. \ANSW{Detailed analysis of the \LoTSS/-deep
  catalogues (including the source counts of the
  \LH/, \Bootes/ and \ELAIS/ fields) are
  presented in \citet{Mandal20}}. In the future we plan to continue increasing the
depth of these fields: data are already in hand, or scheduled, to
double the integration time on each field, with a further aim to
increase this to $500$ hours in each field.


\section{Acknowledgements}

This paper is based (in part) on data obtained with the
International LOFAR Telescope (ILT). LOFAR \citep{vanHaarlem13} is the Low
Frequency Array designed and constructed by ASTRON.
It has observing, data processing, and data storage facilities in several countries, which are owned by various parties
(each with their own funding sources), and which are collectively operated by the ILT foundation under a joint scientific policy. The ILT resources have benefitted from the following recent major funding sources: CNRS-INSU, Observatoire de Paris and Universit\'e d'Orl\'eans, France; BMBF,
MIWF-NRW, MPG, Germany; Science Foundation Ireland
(SFI), Department of Business, Enterprise and Innovation
(DBEI), Ireland; NWO, The Netherlands; The Science and
Technology Facilities Council, UK; Ministry of Science and
Higher Education, Poland.

This work makes use of kern astronomical software package
\citep[available at \url{https://kernsuite.info} and presented
  in][]{molenaar2018kern}.

MB acknowledges support from INAF under PRIN SKA/CTA FORECaST. MB
acknowledges the support from the Ministero degli Affari Esteri della
Cooperazione Internazionale - Direzione Generale per la Promozione del
Sistema Paese Progetto di Grande Rilevanza ZA18GR02.

MJJ acknowledges support from the UK Science and Technology Facilities
Council [ST/N000919/1] and the Oxford Hintze Centre for Astrophysical
Surveys which is funded through generous support from the Hintze
Family Charitable Foundation.

PNB and JS are grateful for support from the UK STFC via grant ST/R000972/1.

MJH acknowledges support from STFC via grant ST/R000905/1.

WLW acknowledges support from the ERC Advanced Investigator programme NewClusters 321271. WLW also acknowledges support from the CAS-NWO programme for radio astronomy with project number 629.001.024, which is financed by the Netherlands Organisation for Scientific Research (NWO).

AB acknowledges support from the VIDI research programme with project number 639.042.729, which is financed by the Netherlands Organisation for Scientific Research (NWO).

IP acknowledges support from INAF under the SKA/CTA PRIN “FORECaST” and the PRIN MAIN STREAM “SAuROS” projects

MB acknowledges support from INAF under PRIN SKA/CTA FORECaST and from
the Ministero degli Affari Esteri della Cooperazione Internazionale -
Direzione Generale per la Promozione del Sistema Paese Progetto di
Grande Rilevanza ZA18GR02.

RK acknowledges support from the Science and Technology Facilities
Council (STFC) through an STFC studentship.

\begin{appendices}
\section{\LOTSS/ first data release: overview of \PipeVI/}
\label{sec:PipeVI}
\label{sec:PipeDR_I}

The data processing strategy of the \LOTSS/ first data release (\DRI/) has been
extensively described by \citet[][]{Shimwell18}. Since addressing the
issues described in Sec. \ref{sec:3GC} involves making improvements
relative to this approach,
we give here a brief description of the data reduction strategy in
\PipeVI/ (the various steps are outlined in
Alg. \ref{alg:DR1}).

\ALGODRONEB/
As discussed in Sec. \ref{sec:3GC}, the calibration and imaging problem is
non-convex and ill-posed. Beyond the computational issues, the great
difficulty of the calibration of the \DDE/ is sky incompleteness,
because the \DD/-\CRIME/ non-linear system can be subject to
ill-conditioning. This is due to the fact that the
extended emission (i) is hard to model in the deconvolution step, and
(ii) is seen by only the shortest baselines, and therefore sky
incompleteness biases the Jones matrices in the calibration
step. Experience shows that this leads to some of the unmodeled extended emission being
absorbed when running a \DD/ deconvolution with \DDFacet/.

To try to compensate for this effect, in \PipeVI/ (Alg. \ref{alg:DR1})
we introduced an inner $uv$-distance cut during calibration, as well as
a normalization of the Jones matrix. With this the \PipeVI/ was able
to recover some of the unmodeled extended emission. The underlying
idea was to assume the sky incompleteness was generating some
baseline-dependent systematic errors. \ANSW{So for every given direction and
solution interval in \citet{Shimwell18} we were trying to find a gain vector $\vec{g}$ such that
$\vec{g}\vec{g}^H\sim\left<\vec{g}_{t\nu}\vec{g}_{t\nu}^H\right>$ (where $\vec{\mathrm{A}}^H$ is the hermitian transpose of matrix $\vec{\mathrm{A}}$).
This amounts to constraining the baseline-dependent error to be solely antenna-dependent. 
This normalization (described by the function symbol $\OpNorm$ in Alg. \ref{alg:DR1}) was able to recover some extended emission otherwise
absorbed in the calibration solution.} However, as shown in
Fig. \ref{fig:CompareDR1DR2_DR1main} and explained by
\citet{Shimwell18} it also produced large scale fake haloes centered
on extended sources together with artifacts around bright sources. On
fields having a bright $\gtrsim1$ Jy source within the primary beam
(such as 3C sources), \PipeVI/ was not able to converge.


\end{appendices}

\bibliographystyle{aa}
\bibliography{references,referencesBoote,referencesLockman,referenceOtherSurveys}

\begin{thebibliography}{91}
\expandafter\ifx\csname natexlab\endcsname\relax\def\natexlab#1{#1}\fi

\bibitem[{{Ashby} {et~al.}(2009){Ashby}, {Stern}, {Brodwin}, {Griffith},
  {Eisenhardt}, {Koz{\l}owski}, {Kochanek}, {Bock}, {Borys}, {Brand}, {Brown},
  {Cool}, {Cooray}, {Croft}, {Dey}, {Eisenstein}, {Gonzalez}, {Gorjian},
  {Grogin}, {Ivison}, {Jacob}, {Jannuzi}, {Mainzer}, {Moustakas},
  {R{\"o}ttgering}, {Seymour}, {Smith}, {Stanford}, {Stauffer}, {Sullivan},
  {van Breugel}, {Willner}, \& {Wright}}]{Ashby09}
{Ashby}, M.~L.~N., {Stern}, D., {Brodwin}, M., {et~al.} 2009, \apj, 701, 428

\bibitem[{{Becker} {et~al.}(1995){Becker}, {White}, \& {Helfand}}]{FIRST}
{Becker}, R.~H., {White}, R.~L., \& {Helfand}, D.~J. 1995, \apj, 450, 559

\bibitem[{{Best et al.}(2020)}]{LoTSSDeepV}
{Best et al.} 2020, \mnras, 463, 2997

\bibitem[{{Bhatnagar} \& {Cornwell}(2017)}]{Bhatnagar17}
{Bhatnagar}, S. \& {Cornwell}, T.~J. 2017, \aj, 154, 197

\bibitem[{{Bhatnagar} {et~al.}(2008){Bhatnagar}, {Cornwell}, {Golap}, \&
  {Uson}}]{Bhatnagar08}
{Bhatnagar}, S., {Cornwell}, T.~J., {Golap}, K., \& {Uson}, J.~M. 2008, \aap,
  487, 419

\bibitem[{{Biggs} \& {Ivison}(2006)}]{Biggs06}
{Biggs}, A.~D. \& {Ivison}, R.~J. 2006, \mnras, 371, 963

\bibitem[{{Birdi} {et~al.}(2020){Birdi}, {Repetti}, \& {Wiaux}}]{Birdi20}
{Birdi}, J., {Repetti}, A., \& {Wiaux}, Y. 2020, \mnras, 492, 3509

\bibitem[{{Bonnassieux} {et~al.}(2018){Bonnassieux}, {Tasse}, {Smirnov}, \&
  {Zarka}}]{Bonnassieux18}
{Bonnassieux}, E., {Tasse}, C., {Smirnov}, O., \& {Zarka}, P. 2018, \aap, 615,
  A66

\bibitem[{{Brentjens} \& {de Bruyn}(2005)}]{bdb2005}
{Brentjens}, M.~A. \& {de Bruyn}, A.~G. 2005, \aap, 441, 1217

\bibitem[{{Brown} {et~al.}(2007){Brown}, {Dey}, {Jannuzi}, {Brand }, {Benson},
  {Brodwin}, {Croton}, \& {Eisenhardt}}]{Brown07}
{Brown}, M. J.~I., {Dey}, A., {Jannuzi}, B.~T., {et~al.} 2007, \apj, 654, 858

\bibitem[{{Brown} {et~al.}(2008){Brown}, {Zheng}, {White}, {Dey}, {Jannuzi},
  {Benson}, {Brand }, {Brodwin}, \& {Croton}}]{Brown08}
{Brown}, M. J.~I., {Zheng}, Z., {White}, M., {et~al.} 2008, \apj, 682, 937

\bibitem[{{Brunner} {et~al.}(2008){Brunner}, {Cappelluti}, {Hasinger},
  {Barcons}, {Fabian}, {Mainieri}, \& {Szokoly}}]{Brunner08}
{Brunner}, H., {Cappelluti}, N., {Hasinger}, G., {et~al.} 2008, \aap, 479, 283

\bibitem[{{Butler} {et~al.}(2018){Butler}, {Huynh}, {Delhaize},
  {Smol{\v{c}}i{\'c}}, {Kapi{\'n}ska}, {Milakovi{\'c}}, {Novak}, {Baran},
  {O'Brien}, {Chiappetti}, {Desai}, {Fotopoulou}, {Horellou}, {Lidman}, \&
  {Pierre}}]{XXL}
{Butler}, A., {Huynh}, M., {Delhaize}, J., {et~al.} 2018, \aap, 620, A3

\bibitem[{{Ciliegi} {et~al.}(2003){Ciliegi}, {Zamorani}, {Hasinger}, {Lehmann},
  {Szokoly}, \& {Wilson}}]{Ciliegi03}
{Ciliegi}, P., {Zamorani}, G., {Hasinger}, G., {et~al.} 2003, \aap, 398, 901

\bibitem[{{Condon} {et~al.}(1998){Condon}, {Cotton}, {Greisen}, {Yin},
  {Perley}, {Taylor}, \& {Broderick}}]{NVSS}
{Condon}, J.~J., {Cotton}, W.~D., {Greisen}, E.~W., {et~al.} 1998, \aj, 115,
  1693

\bibitem[{{Cool}(2007)}]{Cool07}
{Cool}, R.~J. 2007, \apjs, 169, 21

\bibitem[{{Coppejans} {et~al.}(2015){Coppejans}, {Cseh}, {Williams}, {van
  Velzen}, \& {Falcke}}]{Coppejans15}
{Coppejans}, R., {Cseh}, D., {Williams}, W.~L., {van Velzen}, S., \& {Falcke},
  H. 2015, \mnras, 450, 1477

\bibitem[{{Coppin} {et~al.}(2006){Coppin}, {Chapin}, {Mortier}, {Scott},
  {Borys}, {Dunlop}, {Halpern}, {Hughes}, {Pope}, {Scott}, {Serjeant}, {Wagg},
  {Alexander}, {Almaini}, {Aretxaga}, {Babbedge}, {Best}, {Blain}, {Chapman},
  {Clements}, {Crawford}, {Dunne}, {Eales}, {Edge}, {Farrah}, {Gazta{\~n}aga},
  {Gear}, {Granato}, {Greve}, {Fox}, {Ivison}, {Jarvis}, {Jenness}, {Lacey},
  {Lepage}, {Mann}, {Marsden}, {Martinez-Sansigre}, {Oliver}, {Page},
  {Peacock}, {Pearson}, {Percival}, {Priddey}, {Rawlings}, {Rowan-Robinson},
  {Savage}, {Seigar}, {Sekiguchi}, {Silva}, {Simpson}, {Smail}, {Stevens},
  {Takagi}, {Vaccari}, {van Kampen}, \& {Willott}}]{Coppin06}
{Coppin}, K., {Chapin}, E.~L., {Mortier}, A.~M.~J., {et~al.} 2006, \mnras, 372,
  1621

\bibitem[{{Croft} {et~al.}(2008){Croft}, {van Breugel}, {Brown}, {de Vries},
  {Dey}, {Eisenhardt}, {Jannuzi}, {R{\"o}ttgering}, {Stanford}, {Stern}, \&
  {Willner}}]{Croft08}
{Croft}, S., {van Breugel}, W., {Brown}, M.~J.~I., {et~al.} 2008, \aj, 135,
  1793

\bibitem[{{de Gasperin} {et~al.}(2019){de Gasperin}, {Dijkema}, {Drabent},
  {Mevius}, {Rafferty}, {van Weeren}, {Br{\"u}ggen}, {Callingham}, {Emig},
  {Heald}, {Intema}, {Morabito}, {Offringa}, {Oonk}, {Orr{\`u}},
  {R{\"o}ttgering}, {Sabater}, {Shimwell}, {Shulevski}, \&
  {Williams}}]{deGasperin19}
{de Gasperin}, F., {Dijkema}, T.~J., {Drabent}, A., {et~al.} 2019, \aap, 622,
  A5

\bibitem[{{de Ruiter} {et~al.}(1997){de Ruiter}, {Zamorani}, {Parma},
  {Hasinger}, {Hartner}, {Truemper}, {Burg}, {Giacconi}, \&
  {Schmidt}}]{deRuiter97}
{de Ruiter}, H.~R., {Zamorani}, G., {Parma}, P., {et~al.} 1997, \aap, 319, 7

\bibitem[{{de Vries} {et~al.}(2002){de Vries}, {Morganti}, {R{\"o}ttgering},
  {Vermeulen}, {van Breugel}, {Rengelink}, \& {Jarvis}}]{deVries02}
{de Vries}, W.~H., {Morganti}, R., {R{\"o}ttgering}, H.~J.~A., {et~al.} 2002,
  \aj, 123, 1784

\bibitem[{{Dewdney} {et~al.}(2009){Dewdney}, {Hall}, {Schilizzi}, \&
  {Lazio}}]{Dewdney_2009}
{Dewdney}, P.~E., {Hall}, P.~J., {Schilizzi}, R.~T., \& {Lazio}, T.~J.~L.~W.
  2009, IEEE Proceedings, 97, 1482

\bibitem[{{Duncan et al.}(2020)}]{LoTSSDeepIV}
{Duncan et al.} 2020, \mnras, 463, 2997

\bibitem[{{Fomalont} {et~al.}(2006){Fomalont}, {Kellermann}, {Cowie}, {Capak},
  {Barger}, {Partridge}, {Windhorst}, \& {Richards}}]{SSA13}
{Fomalont}, E.~B., {Kellermann}, K.~I., {Cowie}, L.~L., {et~al.} 2006, \apjs,
  167, 103

\bibitem[{{Garn} {et~al.}(2010){Garn}, {Green}, {Riley}, \&
  {Alexander}}]{Garn10}
{Garn}, T.~S., {Green}, D.~A., {Riley}, J.~M., \& {Alexander}, P. 2010,
  Bulletin of the Astronomical Society of India, 38, 103

\bibitem[{{Geach} {et~al.}(2017){Geach}, {Dunlop}, {Halpern}, {Smail}, {van der
  Werf}, {Alexander}, {Almaini}, {Aretxaga}, {Arumugam}, {Asboth}, {Banerji},
  {Beanlands}, {Best}, {Blain}, {Birkinshaw}, {Chapin}, {Chapman}, {Chen},
  {Chrysostomou}, {Clarke}, {Clements}, {Conselice}, {Coppin}, {Cowley},
  {Danielson}, {Eales}, {Edge}, {Farrah}, {Gibb}, {Harrison}, {Hine}, {Hughes},
  {Ivison}, {Jarvis}, {Jenness}, {Jones}, {Karim}, {Koprowski}, {Knudsen},
  {Lacey}, {Mackenzie}, {Marsden}, {McAlpine}, {McMahon}, {Meijerink},
  {Micha{\l}owski}, {Oliver}, {Page}, {Peacock}, {Rigopoulou}, {Robson},
  {Roseboom}, {Rotermund}, {Scott}, {Serjeant}, {Simpson}, {Simpson}, {Smith},
  {Spaans}, {Stanley}, {Stevens}, {Swinbank}, {Targett}, {Thomson}, {Valiante},
  {Wake}, {Webb}, {Willott}, {Zavala}, \& {Zemcov}}]{Geach17}
{Geach}, J.~E., {Dunlop}, J.~S., {Halpern}, M., {et~al.} 2017, \mnras, 465,
  1789

\bibitem[{{Gonz{\'a}lez-Solares} {et~al.}(2011){Gonz{\'a}lez-Solares}, {Irwin},
  {McMahon}, {Hodgkin}, {Lewis}, {Walton}, {Jarvis}, {Marchetti}, {Oliver},
  {P{\'e}rez-Fournon}, {Siana}, {Surace}, \& {Vaccari}}]{Gonzalez11}
{Gonz{\'a}lez-Solares}, E.~A., {Irwin}, M., {McMahon}, R.~G., {et~al.} 2011,
  \mnras, 416, 927

\bibitem[{{Guglielmino} {et~al.}(2012){Guglielmino}, {Prandoni}, {Morganti}, \&
  {Heald}}]{Guglielmino12}
{Guglielmino}, G., {Prandoni}, I., {Morganti}, R., \& {Heald}, G. 2012, in
  Resolving The Sky - Radio Interferometry: Past, Present and Future, 22

\bibitem[{{Hamaker}(2000)}]{Hamaker2000}
{Hamaker}, J.~P. 2000, \aaps, 143, 515

\bibitem[{{Hamaker} {et~al.}(1996){Hamaker}, {Bregman}, \& {Sault}}]{Hamaker96}
{Hamaker}, J.~P., {Bregman}, J.~D., \& {Sault}, R.~J. 1996, \aaps, 117, 137

\bibitem[{{Herrera Ruiz et al.}(2020)}]{Herrera20}
{Herrera Ruiz et al.} 2020, \mnras, 463, 2997

\bibitem[{{Heywood} {et~al.}(2016){Heywood}, {Jarvis}, {Baker}, {Bannister},
  {Carvalho}, {Hardcastle}, {Hilton}, {Moodley}, {Smirnov}, {Smith}, {White},
  \& {Wollack}}]{Stripe82}
{Heywood}, I., {Jarvis}, M.~J., {Baker}, A.~J., {et~al.} 2016, \mnras, 460,
  4433

\bibitem[{{Ibar} {et~al.}(2009){Ibar}, {Ivison}, {Biggs}, {Lal}, {Best}, \&
  {Green}}]{Ibar09}
{Ibar}, E., {Ivison}, R.~J., {Biggs}, A.~D., {et~al.} 2009, \mnras, 397, 281

\bibitem[{{Intema} {et~al.}(2017){Intema}, {Jagannathan}, {Mooley}, \&
  {Frail}}]{Intema17}
{Intema}, H.~T., {Jagannathan}, P., {Mooley}, K.~P., \& {Frail}, D.~A. 2017,
  \aap, 598, A78

\bibitem[{{Intema} {et~al.}(2011){Intema}, {van Weeren}, {R{\"o}ttgering}, \&
  {Lal}}]{Intema11}
{Intema}, H.~T., {van Weeren}, R.~J., {R{\"o}ttgering}, H.~J.~A., \& {Lal},
  D.~V. 2011, \aap, 535, A38

\bibitem[{{Jannuzi} {et~al.}(2010){Jannuzi}, {Weiner}, {Block}, {Borys},
  {Eisenstein}, {Kochanek}, {Rieke}, {Rieke}, {Armus}, {Brodwin}, {Brown},
  {Cool}, {Desai}, {Dey}, {Dickinson}, {Dole}, {Herrera}, {Le Floc'h},
  {Morrison}, {Papovich}, {P{\'e}rez-Gonz{\'a}lez}, {Stern}, {Rujopakarn}, \&
  {Zehavi}}]{Jannuzi10}
{Jannuzi}, B., {Weiner}, B., {Block}, M., {et~al.} 2010, in Bulletin of the
  American Astronomical Society, Vol.~42, American Astronomical Society Meeting
  Abstracts \#215, 513

\bibitem[{{Jannuzi} \& {Dey}(1999)}]{Jannuzi99}
{Jannuzi}, B.~T. \& {Dey}, A. 1999, in Astronomical Society of the Pacific
  Conference Series, Vol. 193, The Hy-Redshift Universe: Galaxy Formation and
  Evolution at High Redshift, ed. A.~J. {Bunker} \& W.~J.~M. {van Breugel}, 258

\bibitem[{{Kazemi} {et~al.}(2011){Kazemi}, {Yatawatta}, {Zaroubi},
  {Lampropoulos}, {de Bruyn}, {Koopmans}, \& {Noordam}}]{Kazemi11}
{Kazemi}, S., {Yatawatta}, S., {Zaroubi}, S., {et~al.} 2011, \mnras, 414, 1656

\bibitem[{{Kenter} {et~al.}(2005){Kenter}, {Murray}, {Forman}, {Jones},
  {Green}, {Kochanek}, {Vikhlinin}, {Fabricant}, {Fazio}, {Brand}, {Brown},
  {Dey}, {Jannuzi}, {Najita}, {McNamara}, {Shields}, \& {Rieke}}]{Kenter05}
{Kenter}, A., {Murray}, S.~S., {Forman}, W.~R., {et~al.} 2005, \apjs, 161, 9

\bibitem[{{Kondapally et al.}(2020)}]{LoTSSDeepIII}
{Kondapally et al.} 2020, \mnras, 463, 2997

\bibitem[{{Lacy} {et~al.}(2020){Lacy}, {Baum}, {Chandler}, {Chatterjee},
  {Clarke}, {Deustua}, {English}, {Farnes}, {Gaensler}, {Gugliucci},
  {Hallinan}, {Kent}, {Kimball}, {Law}, {Lazio}, {Marvil}, {Mao}, {Medlin},
  {Mooley}, {Murphy}, {Myers}, {Osten}, {Richards}, {Rosolowsky}, {Rudnick},
  {Schinzel}, {Sivakoff}, {Sjouwerman}, {Taylor}, {White}, {Wrobel},
  {Andernach}, {Beasley}, {Berger}, {Bhatnager}, {Birkinshaw}, {Bower},
  {Brandt}, {Brown}, {Burke-Spolaor}, {Butler}, {Comerford}, {Demorest}, {Fu},
  {Giacintucci}, {Golap}, {G{\"u}th}, {Hales}, {Hiriart}, {Hodge}, {Horesh},
  {Ivezi{\'c}}, {Jarvis}, {Kamble}, {Kassim}, {Liu}, {Loinard}, {Lyons},
  {Masters}, {Mezcua}, {Moellenbrock}, {Mroczkowski}, {Nyland},
  {O{\textquoteright}Dea}, {O{\textquoteright}Sullivan}, {Peters}, {Radford},
  {Rao}, {Robnett}, {Salcido}, {Shen}, {Sobotka}, {Witz}, {Vaccari}, {van
  Weeren}, {Vargas}, {Williams}, \& {Yoon}}]{VLASS}
{Lacy}, M., {Baum}, S.~A., {Chandler}, C.~J., {et~al.} 2020, \pasp, 132, 035001

\bibitem[{{Lawrence} {et~al.}(2007){Lawrence}, {Warren}, {Almaini}, {Edge},
  {Hambly}, {Jameson}, {Lucas}, {Casali}, {Adamson}, {Dye}, {Emerson},
  {Foucaud}, {Hewett}, {Hirst}, {Hodgkin}, {Irwin}, {Lodieu}, {McMahon},
  {Simpson}, {Smail}, {Mortlock}, \& {Folger}}]{Lawrence07}
{Lawrence}, A., {Warren}, S.~J., {Almaini}, O., {et~al.} 2007, \mnras, 379,
  1599

\bibitem[{{Lonsdale} {et~al.}(2003){Lonsdale}, {Smith}, {Rowan-Robinson},
  {Surace}, {Shupe}, {Xu}, {Oliver}, {Padgett}, {Fang}, {Conrow},
  {Franceschini}, {Gautier}, {Griffin}, {Hacking}, {Masci}, {Morrison},
  {O'Linger}, {Owen}, {P{\'e}rez-Fournon}, {Pierre}, {Puetter}, {Stacey},
  {Castro}, {Polletta}, {Farrah}, {Jarrett}, {Frayer}, {Siana}, {Babbedge},
  {Dye}, {Fox}, {Gonzalez-Solares}, {Salaman}, {Berta}, {Condon}, {Dole}, \&
  {Serjeant}}]{Lonsdale03}
{Lonsdale}, C.~J., {Smith}, H.~E., {Rowan-Robinson}, M., {et~al.} 2003, \pasp,
  115, 897

\bibitem[{{Mahony} {et~al.}(2016){Mahony}, {Morganti}, {Prandoni}, {van
  Bemmel}, {Shimwell}, {Brienza}, {Best}, {Br{\"u}ggen}, {Calistro Rivera}, {de
  Gasperin}, {Hardcastle}, {Harwood}, {Heald}, {Jarvis}, {Mandal}, {Miley},
  {Retana-Montenegro}, {R{\"o}ttgering}, {Sabater}, {Tasse}, {van Velzen}, {van
  Weeren}, {Williams}, \& {White}}]{Mahony16}
{Mahony}, E.~K., {Morganti}, R., {Prandoni}, I., {et~al.} 2016, \mnras, 463,
  2997

\bibitem[{{Mandal et al.}(2020)}]{Mandal20}
{Mandal et al.} 2020, \mnras, 463, 2997

\bibitem[{{Martin} \& {GALEX Team}(2005)}]{Martin05}
{Martin}, C. \& {GALEX Team}. 2005, in IAU Symposium, Vol. 216, Maps of the
  Cosmos, ed. M.~{Colless}, L.~{Staveley-Smith}, \& R.~A. {Stathakis}, 221

\bibitem[{{Mauch} {et~al.}(2020){Mauch}, {Cotton}, {Condon}, {Matthews},
  {Abbott}, {Adam}, {Aldera}, {Asad}, {Bauermeister}, {Bennett}, {Bester},
  {Botha}, {Brederode}, {Brits}, {Buchner}, {Burger}, {Camilo}, {Chalmers},
  {Cheetham}, {de Villiers}, {de Villiers}, {Dikgale-Mahlakoana}, {du Toit},
  {Esterhuyse}, {Fadana}, {Fanaroff}, {Fataar}, {February}, {Frank},
  {Gamatham}, {Geyer}, {Goedhart}, {Gounden}, {Gumede}, {Heywood}, {Hlakola},
  {Horrell}, {Hugo}, {Isaacson}, {J{\'o}zsa}, {Jonas}, {Julie}, {Kapp},
  {Kasper}, {Kenyon}, {Kotz{\'e}}, {Kriek}, {Kriel}, {Kusel}, {Lehmensiek},
  {Loots}, {Lord}, {Lunsky}, {Madisa}, {Magnus}, {Main}, {Malan}, {Manley},
  {Marais}, {Martens}, {Merry}, {Millenaar}, {Mnyandu}, {Moeng}, {Mokone},
  {Monama}, {Mphego}, {New}, {Ngcebetsha}, {Ngoasheng}, {Ockards}, {Oozeer},
  {Otto}, {Patel}, {Peens-Hough}, {Perkins}, {Ramaila}, {Ramudzuli}, {Renil},
  {Richter}, {Robyntjies}, {Salie}, {Schollar}, {Schwardt}, {Serylak},
  {Siebrits}, {Sirothia}, {Smirnov}, {Sofeya}, {Stone}, {Taljaard}, {Tasse},
  {Theron}, {Tiplady}, {Toruvanda}, {Twum}, {van Balla}, {van der Byl}, {van
  der Merwe}, {Van Tonder}, {Wallace}, {Welz}, {Williams}, \& {Xaia}}]{DEEP2}
{Mauch}, T., {Cotton}, W.~D., {Condon}, J.~J., {et~al.} 2020, \apj, 888, 61

\bibitem[{{Mohan} \& {Rafferty}(2015)}]{Mohan15}
{Mohan}, N. \& {Rafferty}, D. 2015, {PyBDSF: Python Blob Detection and Source
  Finder}

\bibitem[{Molenaar \& Smirnov(2018)}]{molenaar2018kern}
Molenaar, G. \& Smirnov, O. 2018, Astronomy and Computing

\bibitem[{{Murray} {et~al.}(2005){Murray}, {Kenter}, {Forman}, {Jones},
  {Green}, {Kochanek}, {Vikhlinin}, {Fabricant}, {Fazio}, {Brand}, {Brown},
  {Dey}, {Jannuzi}, {Najita}, {McNamara}, {Shields}, \& {Rieke}}]{Murray05}
{Murray}, S.~S., {Kenter}, A., {Forman}, W.~R., {et~al.} 2005, \apjs, 161, 1

\bibitem[{{Norris}(2010)}]{EMU}
{Norris}, R. 2010, in American Astronomical Society Meeting Abstracts, Vol.
  215, American Astronomical Society Meeting Abstracts \#215, 604.05

\bibitem[{{Offringa} {et~al.}(2012){Offringa}, {van de Gronde}, \&
  {Roerdink}}]{Offringa12}
{Offringa}, A.~R., {van de Gronde}, J.~J., \& {Roerdink}, J.~B.~T.~M. 2012,
  \aap, 539, A95

\bibitem[{{Oliver} {et~al.}(2012){Oliver}, {Bock}, {Altieri}, {Amblard},
  {Arumugam}, {Aussel}, {Babbedge}, {Beelen}, {B{\'e}thermin}, {Blain},
  {Boselli}, {Bridge}, {Brisbin}, {Buat}, {Burgarella},
  {Castro-Rodr{\'{\i}}guez}, {Cava}, {Chanial}, {Cirasuolo}, {Clements},
  {Conley}, {Conversi}, {Cooray}, {Dowell}, {Dubois}, {Dwek}, {Dye}, {Eales},
  {Elbaz}, {Farrah}, {Feltre}, {Ferrero}, {Fiolet}, {Fox}, {Franceschini},
  {Gear}, {Giovannoli}, {Glenn}, {Gong}, {Gonz{\'a}lez Solares}, {Griffin},
  {Halpern}, {Harwit}, {Hatziminaoglou}, {Heinis}, {Hurley}, {Hwang}, {Hyde},
  {Ibar}, {Ilbert}, {Isaak}, {Ivison}, {Lagache}, {Le Floc'h}, {Levenson},
  {Faro}, {Lu}, {Madden}, {Maffei}, {Magdis}, {Mainetti}, {Marchetti},
  {Marsden}, {Marshall}, {Mortier}, {Nguyen}, {O'Halloran}, {Omont}, {Page},
  {Panuzzo}, {Papageorgiou}, {Patel}, {Pearson}, {P{\'e}rez-Fournon}, {Pohlen},
  {Rawlings}, {Raymond}, {Rigopoulou}, {Riguccini}, {Rizzo}, {Rodighiero},
  {Roseboom}, {Rowan-Robinson}, {S{\'a}nchez Portal}, {Schulz}, {Scott},
  {Seymour}, {Shupe}, {Smith}, {Stevens}, {Symeonidis}, {Trichas}, {Tugwell},
  {Vaccari}, {Valtchanov}, {Vieira}, {Viero}, {Vigroux}, {Wang}, {Ward},
  {Wardlow}, {Wright}, {Xu}, \& {Zemcov}}]{Oliver12}
{Oliver}, S.~J., {Bock}, J., {Altieri}, B., {et~al.} 2012, \mnras, 424, 1614

\bibitem[{{O'Sullivan} {et~al.}(2018){O'Sullivan}, {Br{\"u}ggen}, {Van Eck},
  {Hardcastle}, {Haverkorn}, {Shimwell}, {Tasse}, {Vacca}, {Horellou}, \&
  {Heald}}]{osullivan2018}
{O'Sullivan}, S., {Br{\"u}ggen}, M., {Van Eck}, C., {et~al.} 2018, Galaxies, 6,
  126

\bibitem[{{O'Sullivan} {et~al.}(2020){O'Sullivan}, {Brueggen}, \&
  {Vazza}}]{osullivan2020}
{O'Sullivan}, S.~P., {Brueggen}, M., \& {Vazza}, F. 2020, \mnras

\bibitem[{{Owen} \& {Morrison}(2008)}]{SWIRE}
{Owen}, F.~N. \& {Morrison}, G.~E. 2008, \aj, 136, 1889

\bibitem[{{Owen} {et~al.}(2009){Owen}, {Morrison}, {Klimek}, \&
  {Greisen}}]{Owen09}
{Owen}, F.~N., {Morrison}, G.~E., {Klimek}, M.~D., \& {Greisen}, E.~W. 2009,
  \aj, 137, 4846

\bibitem[{{Padovani}(2016)}]{Padovani16}
{Padovani}, P. 2016, \aapr, 24, 13

\bibitem[{{Pearson} \& {Readhead}(1984)}]{Pearson+Readhead84}
{Pearson}, T.~J. \& {Readhead}, A.~C.~S. 1984, \araa, 22, 97

\bibitem[{{Polletta} {et~al.}(2006){Polletta}, {Wilkes}, {Siana}, {Lonsdale},
  {Kilgard}, {Smith}, {Kim}, {Owen}, {Efstathiou}, {Jarrett}, {Stacey},
  {Franceschini}, {Rowan-Robinson}, {Babbedge}, {Berta}, {Fang}, {Farrah},
  {Gonz{\'a}lez-Solares}, {Morrison}, {Surace}, \& {Shupe}}]{Polletta06}
{Polletta}, M.~d.~C., {Wilkes}, B.~J., {Siana}, B., {et~al.} 2006, \apj, 642,
  673

\bibitem[{{Prandoni} {et~al.}(2018){Prandoni}, {Guglielmino}, {Morganti},
  {Vaccari}, {Maini}, {R{\"o}ttgering}, {Jarvis}, \& {Garrett}}]{Prandoni18}
{Prandoni}, I., {Guglielmino}, G., {Morganti}, R., {et~al.} 2018, \mnras, 481,
  4548

\bibitem[{{Prandoni} \& {Seymour}(2015{\natexlab{a}})}]{Prandoni15}
{Prandoni}, I. \& {Seymour}, N. 2015{\natexlab{a}}, in Advancing Astrophysics
  with the Square Kilometre Array (AASKA14), 67

\bibitem[{{Prandoni} \& {Seymour}(2015{\natexlab{b}})}]{Prandoni_2015}
{Prandoni}, I. \& {Seymour}, N. 2015{\natexlab{b}}, in Advancing Astrophysics
  with the Square Kilometre Array (AASKA14), 67

\bibitem[{{Rengelink} {et~al.}(1997){Rengelink}, {Tang}, {de Bruyn}, {Miley},
  {Bremer}, {Roettgering}, \& {Bremer}}]{WENSS}
{Rengelink}, R.~B., {Tang}, Y., {de Bruyn}, A.~G., {et~al.} 1997, \aaps, 124,
  259

\bibitem[{{Repetti} {et~al.}(2017){Repetti}, {Birdi}, {Dabbech}, \&
  {Wiaux}}]{Repetti17}
{Repetti}, A., {Birdi}, J., {Dabbech}, A., \& {Wiaux}, Y. 2017, \mnras, 470,
  3981

\bibitem[{{Retana-Montenegro} {et~al.}(2018){Retana-Montenegro},
  {R{\"o}ttgering}, {Shimwell}, {van Weeren}, {Prandoni}, {Brunetti}, {Best},
  \& {Br{\"u}ggen}}]{Retana18}
{Retana-Montenegro}, E., {R{\"o}ttgering}, H.~J.~A., {Shimwell}, T.~W.,
  {et~al.} 2018, \aap, 620, A74

\bibitem[{{Richards}(2000)}]{Richards00}
{Richards}, E.~A. 2000, \apj, 533, 611

\bibitem[{{Sabater et al.}(2020)}]{LoTSSDeepII}
{Sabater et al.} 2020, \mnras, 463, 2997

\bibitem[{{Schinnerer} {et~al.}(2004){Schinnerer}, {Carilli}, {Scoville},
  {Bondi}, {Ciliegi}, {Vettolani}, {Le F{\`e}vre}, {Koekemoer}, {Bertoldi}, \&
  {Impey}}]{VLACOSMOS}
{Schinnerer}, E., {Carilli}, C.~L., {Scoville}, N.~Z., {et~al.} 2004, \aj, 128,
  1974

\bibitem[{{Shimwell} {et~al.}(2017{\natexlab{a}}){Shimwell}, {R{\"o}ttgering},
  {Best}, {Williams}, {Dijkema}, {de Gasperin}, {Hardcastle}, {Heald}, {Hoang},
  {Horneffer}, {Intema}, {Mahony}, {Mandal}, {Mechev}, {Morabito}, {Oonk},
  {Rafferty}, {Retana-Montenegro}, {Sabater}, {Tasse}, {van Weeren},
  {Br{\"u}ggen}, {Brunetti}, {Chy{\.z}y}, {Conway}, {Haverkorn}, {Jackson},
  {Jarvis}, {McKean}, {Miley}, {Morganti}, {White}, {Wise}, {van Bemmel},
  {Beck}, {Brienza}, {Bonafede}, {Calistro Rivera}, {Cassano}, {Clarke},
  {Cseh}, {Deller}, {Drabent}, {van Driel}, {Engels}, {Falcke}, {Ferrari},
  {Fr{\"o}hlich}, {Garrett}, {Harwood}, {Heesen}, {Hoeft}, {Horellou},
  {Israel}, {Kapi{\'n}ska}, {Kunert-Bajraszewska}, {McKay}, {Mohan},
  {Orr{\'u}}, {Pizzo}, {Prandoni}, {Schwarz}, {Shulevski}, {Sipior}, {Smith},
  {Sridhar}, {Steinmetz}, {Stroe}, {Varenius}, {van der Werf}, {Zensus}, \&
  {Zwart}}]{Shimwell17}
{Shimwell}, T.~W., {R{\"o}ttgering}, H.~J.~A., {Best}, P.~N., {et~al.}
  2017{\natexlab{a}}, \aap, 598, A104

\bibitem[{{Shimwell} {et~al.}(2017{\natexlab{b}}){Shimwell}, {R{\"o}ttgering},
  {Best}, {Williams}, {Dijkema}, {de Gasperin}, {Hardcastle}, {Heald}, {Hoang},
  {Horneffer}, {Intema}, {Mahony}, {Mandal}, {Mechev}, {Morabito}, {Oonk},
  {Rafferty}, {Retana-Montenegro}, {Sabater}, {Tasse}, {van Weeren},
  {Br{\"u}ggen}, {Brunetti}, {Chy{\.z}y}, {Conway}, {Haverkorn}, {Jackson},
  {Jarvis}, {McKean}, {Miley}, {Morganti}, {White}, {Wise}, {van Bemmel},
  {Beck}, {Brienza}, {Bonafede}, {Calistro Rivera}, {Cassano}, {Clarke},
  {Cseh}, {Deller}, {Drabent}, {van Driel}, {Engels}, {Falcke}, {Ferrari},
  {Fr{\"o}hlich}, {Garrett}, {Harwood}, {Heesen}, {Hoeft}, {Horellou},
  {Israel}, {Kapi{\'n}ska}, {Kunert-Bajraszewska}, {McKay}, {Mohan},
  {Orr{\'u}}, {Pizzo}, {Prandoni}, {Schwarz}, {Shulevski}, {Sipior}, {Smith},
  {Sridhar}, {Steinmetz}, {Stroe}, {Varenius}, {van der Werf}, {Zensus}, \&
  {Zwart}}]{LOTSSDR1}
{Shimwell}, T.~W., {R{\"o}ttgering}, H.~J.~A., {Best}, P.~N., {et~al.}
  2017{\natexlab{b}}, \aap, 598, A104

\bibitem[{{Shimwell} {et~al.}(2019){Shimwell}, {Tasse}, {Hardcastle}, {Mechev},
  {Williams}, {Best}, {R{\"o}ttgering}, {Callingham}, {Dijkema}, {de Gasperin},
  {Hoang}, {Hugo}, {Mirmont}, {Oonk}, {Prandoni}, {Rafferty}, {Sabater},
  {Smirnov}, {van Weeren}, {White}, {Atemkeng}, {Bester}, {Bonnassieux},
  {Br{\"u}ggen}, {Brunetti}, {Chy{\.z}y}, {Cochrane}, {Conway}, {Croston},
  {Danezi}, {Duncan}, {Haverkorn}, {Heald}, {Iacobelli}, {Intema}, {Jackson},
  {Jamrozy}, {Jarvis}, {Lakhoo}, {Mevius}, {Miley}, {Morabito}, {Morganti},
  {Nisbet}, {Orr{\'u}}, {Perkins}, {Pizzo}, {Schrijvers}, {Smith}, {Vermeulen},
  {Wise}, {Alegre}, {Bacon}, {van Bemmel}, {Beswick}, {Bonafede}, {Botteon},
  {Bourke}, {Brienza}, {Calistro Rivera}, {Cassano}, {Clarke}, {Conselice},
  {Dettmar}, {Drabent}, {Dumba}, {Emig}, {En{\ss}lin}, {Ferrari}, {Garrett},
  {G{\'e}nova-Santos}, {Goyal}, {G{\"u}rkan}, {Hale}, {Harwood}, {Heesen},
  {Hoeft}, {Horellou}, {Jackson}, {Kokotanekov}, {Kondapally},
  {Kunert-Bajraszewska}, {Mahatma}, {Mahony}, {Mandal}, {McKean}, {Merloni},
  {Mingo}, {Miskolczi}, {Mooney}, {Nikiel-Wroczy{\'n}ski}, {O'Sullivan},
  {Quinn}, {Reich}, {Roskowi{\'n}ski}, {Rowlinson}, {Savini}, {Saxena},
  {Schwarz}, {Shulevski}, {Sridhar}, {Stacey}, {Urquhart}, {van der Wiel},
  {Varenius}, {Webster}, \& {Wilber}}]{Shimwell18}
{Shimwell}, T.~W., {Tasse}, C., {Hardcastle}, M.~J., {et~al.} 2019, \aap, 622,
  A1

\bibitem[{{Smirnov}(2011)}]{Smirnov2011_3}
{Smirnov}, O.~M. 2011, \aap, 527, A108

\bibitem[{{Smirnov} \& {Tasse}(2015)}]{Smirnov15}
{Smirnov}, O.~M. \& {Tasse}, C. 2015, \mnras, 449, 2668

\bibitem[{{Smol{\v{c}}i{\'c}} {et~al.}(2017{\natexlab{a}}){Smol{\v{c}}i{\'c}},
  {Novak}, {Bondi}, {Ciliegi}, {Mooley}, {Schinnerer}, {Zamorani}, {Navarrete},
  {Bourke}, {Karim}, {Vardoulaki}, {Leslie}, {Delhaize}, {Carilli}, {Myers},
  {Baran}, {Delvecchio}, {Miettinen}, {Banfield}, {Balokovi{\'c}}, {Bertoldi},
  {Capak}, {Frail}, {Hallinan}, {Hao}, {Herrera Ruiz}, {Horesh}, {Ilbert},
  {Intema}, {Jeli{\'c}}, {Kl{\"o}ckner}, {Krpan}, {Kulkarni}, {McCracken},
  {Laigle}, {Middleberg}, {Murphy}, {Sargent}, {Scoville}, \&
  {Sheth}}]{Smolcic17}
{Smol{\v{c}}i{\'c}}, V., {Novak}, M., {Bondi}, M., {et~al.} 2017{\natexlab{a}},
  \aap, 602, A1

\bibitem[{{Smol{\v{c}}i{\'c}} {et~al.}(2017{\natexlab{b}}){Smol{\v{c}}i{\'c}},
  {Novak}, {Delvecchio}, {Ceraj}, {Bondi}, {Delhaize}, {Marchesi}, {Murphy},
  {Schinnerer}, {Vardoulaki}, \& {Zamorani}}]{VLACOSMOS3}
{Smol{\v{c}}i{\'c}}, V., {Novak}, M., {Delvecchio}, I., {et~al.}
  2017{\natexlab{b}}, \aap, 602, A6

\bibitem[{{Tasse}(2014{\natexlab{a}})}]{Tasse14b}
{Tasse}, C. 2014{\natexlab{a}}, ArXiv e-prints

\bibitem[{{Tasse}(2014{\natexlab{b}})}]{Tasse14a}
{Tasse}, C. 2014{\natexlab{b}}, \aap, 566, A127

\bibitem[{{Tasse} {et~al.}(2018){Tasse}, {Hugo}, {Mirmont}, {Smirnov},
  {Atemkeng}, {Bester}, {Hardcastle}, {Lakhoo}, {Perkins}, \&
  {Shimwell}}]{Tasse18}
{Tasse}, C., {Hugo}, B., {Mirmont}, M., {et~al.} 2018, \aap, 611, A87

\bibitem[{{Tasse} {et~al.}(2010){Tasse}, {R{\"o}ttgering}, \& {Best}}]{Tasse11}
{Tasse}, C., {R{\"o}ttgering}, H., \& {Best}, P.~N. 2010, \aap, 525, A127+

\bibitem[{{Van Eck} {et~al.}(2018){Van Eck}, {Haverkorn}, {Alves}, {Beck},
  {Best}, {Carretti}, {Chy{\.z}y}, {Farnes}, {Ferri{\`e}re}, {Hardcastle},
  {Heald}, {Horellou}, {Iacobelli}, {Jeli{\'c}}, {Mulcahy}, {O'Sullivan},
  {Polderman}, {Reich}, {Riseley}, {R{\"o}ttgering}, {Schnitzeler}, {Shimwell},
  {Vacca}, {Vink}, \& {White}}]{vaneck2018}
{Van Eck}, C.~L., {Haverkorn}, M., {Alves}, M.~I.~R., {et~al.} 2018, \aap, 613,
  A58

\bibitem[{{van Haarlem} {et~al.}(2013){van Haarlem}, {Wise}, {Gunst}, {Heald},
  {McKean}, {Hessels}, {de Bruyn}, {Nijboer}, {Swinbank}, {Fallows},
  {Brentjens}, {Nelles}, {Beck}, {Falcke}, {Fender}, {H{\"o}randel},
  {Koopmans}, {Mann}, {Miley}, {R{\"o}ttgering}, {Stappers}, {Wijers},
  {Zaroubi}, {van den Akker}, {Alexov}, {Anderson}, {Anderson}, {van Ardenne},
  {Arts}, {Asgekar}, {Avruch}, {Batejat}, {B{\"a}hren}, {Bell}, {Bell}, {van
  Bemmel}, {Bennema}, {Bentum}, {Bernardi}, {Best}, {B{\^\i}rzan}, {Bonafede},
  {Boonstra}, {Braun}, {Bregman}, {Breitling}, {van de Brink}, {Broderick},
  {Broekema}, {Brouw}, {Br{\"u}ggen}, {Butcher}, {van Cappellen}, {Ciardi},
  {Coenen}, {Conway}, {Coolen}, {Corstanje}, {Damstra}, {Davies}, {Deller},
  {Dettmar}, {van Diepen}, {Dijkstra}, {Donker}, {Doorduin}, {Dromer}, {Drost},
  {van Duin}, {Eisl{\"o}ffel}, {van Enst}, {Ferrari}, {Frieswijk}, {Gankema},
  {Garrett}, {de Gasperin}, {Gerbers}, {de Geus}, {Grie{\ss}meier}, {Grit},
  {Gruppen}, {Hamaker}, {Hassall}, {Hoeft}, {Holties}, {Horneffer}, {van der
  Horst}, {van Houwelingen}, {Huijgen}, {Iacobelli}, {Intema}, {Jackson},
  {Jelic}, {de Jong}, {Juette}, {Kant}, {Karastergiou}, {Koers}, {Kollen},
  {Kondratiev}, {Kooistra}, {Koopman}, {Koster}, {Kuniyoshi}, {Kramer},
  {Kuper}, {Lambropoulos}, {Law}, {van Leeuwen}, {Lemaitre}, {Loose}, {Maat},
  {Macario}, {Markoff}, {Masters}, {McFadden}, {McKay-Bukowski}, {Meijering},
  {Meulman}, {Mevius}, {Middelberg}, {Millenaar}, {Miller-Jones}, {Mohan},
  {Mol}, {Morawietz}, {Morganti}, {Mulcahy}, {Mulder}, {Munk}, {Nieuwenhuis},
  {van Nieuwpoort}, {Noordam}, {Norden}, {Noutsos}, {Offringa}, {Olofsson},
  {Omar}, {Orr{\'u}}, {Overeem}, {Paas}, {Pand ey-Pommier}, {Pandey}, {Pizzo},
  {Polatidis}, {Rafferty}, {Rawlings}, {Reich}, {de Reijer}, {Reitsma},
  {Renting}, {Riemers}, {Rol}, {Romein}, {Roosjen}, {Ruiter}, {Scaife}, {van
  der Schaaf}, {Scheers}, {Schellart}, {Schoenmakers}, {Schoonderbeek},
  {Serylak}, {Shulevski}, {Sluman}, {Smirnov}, {Sobey}, {Spreeuw}, {Steinmetz},
  {Sterks}, {Stiepel}, {Stuurwold}, {Tagger}, {Tang}, {Tasse}, {Thomas},
  {Thoudam}, {Toribio}, {van der Tol}, {Usov}, {van Veelen}, {van der Veen},
  {ter Veen}, {Verbiest}, {Vermeulen}, {Vermaas}, {Vocks}, {Vogt}, {de Vos},
  {van der Wal}, {van Weeren}, {Weggemans}, {Weltevrede}, {White}, {Wijnholds},
  {Wilhelmsson}, {Wucknitz}, {Yatawatta}, {Zarka}, {Zensus}, \& {van
  Zwieten}}]{vanHaarlem13}
{van Haarlem}, M.~P., {Wise}, M.~W., {Gunst}, A.~W., {et~al.} 2013, \aap, 556,
  A2

\bibitem[{{van Weeren} {et~al.}(2016){van Weeren}, {Williams}, {Hardcastle},
  {Shimwell}, {Rafferty}, {Sabater}, {Heald}, {Sridhar}, {Dijkema}, {Brunetti},
  {Br{\"u}ggen}, {Andrade-Santos}, {Ogrean}, {R{\"o}ttgering}, {Dawson},
  {Forman}, {de Gasperin}, {Jones}, {Miley}, {Rudnick}, {Sarazin}, {Bonafede},
  {Best}, {B{\^i}rzan}, {Cassano}, {Chy{\.z}y}, {Croston}, {Ensslin},
  {Ferrari}, {Hoeft}, {Horellou}, {Jarvis}, {Kraft}, {Mevius}, {Intema},
  {Murray}, {Orr{\'u}}, {Pizzo}, {Simionescu}, {Stroe}, {van der Tol}, \&
  {White}}]{Weeren16}
{van Weeren}, R.~J., {Williams}, W.~L., {Hardcastle}, M.~J., {et~al.} 2016,
  \apjs, 223, 2

\bibitem[{{van Weeren} {et~al.}(2014){van Weeren}, {Williams}, {Tasse},
  {R{\"o}ttgering}, {Rafferty}, {van der Tol}, {Heald}, {White}, {Shulevski},
  {Best}, {Intema}, {Bhatnagar}, {Reich}, {Steinmetz}, {van Velzen},
  {En{\ss}lin}, {Prandoni}, {de Gasperin}, {Jamrozy}, {Brunetti}, {Jarvis},
  {McKean}, {Wise}, {Ferrari}, {Harwood}, {Oonk}, {Hoeft},
  {Kunert-Bajraszewska}, {Horellou}, {Wucknitz}, {Bonafede}, {Mohan}, {Scaife},
  {Kl{\"o}ckner}, {van Bemmel}, {Merloni}, {Chyzy}, {Engels}, {Falcke},
  {Pandey-Pommier}, {Alexov}, {Anderson}, {Avruch}, {Beck}, {Bell}, {Bentum},
  {Bernardi}, {Breitling}, {Broderick}, {Brouw}, {Br{\"u}ggen}, {Butcher},
  {Ciardi}, {de Geus}, {de Vos}, {Deller}, {Duscha}, {Eisl{\"o}ffel},
  {Fallows}, {Frieswijk}, {Garrett}, {Grie{\ss}meier}, {Gunst}, {Hamaker},
  {Hassall}, {H{\"o}randel}, {van der Horst}, {Iacobelli}, {Jackson}, {Juette},
  {Kondratiev}, {Kuniyoshi}, {Maat}, {Mann}, {McKay-Bukowski}, {Mevius},
  {Morganti}, {Munk}, {Offringa}, {Orr{\`u}}, {Paas}, {Pandey}, {Pietka},
  {Pizzo}, {Polatidis}, {Renting}, {Rowlinson}, {Schwarz}, {Serylak}, {Sluman},
  {Smirnov}, {Stappers}, {Stewart}, {Swinbank}, {Tagger}, {Tang}, {Thoudam},
  {Toribio}, {Vermeulen}, {Vocks}, \& {Zarka}}]{vanWeeren14}
{van Weeren}, R.~J., {Williams}, W.~L., {Tasse}, C., {et~al.} 2014, \apj, 793,
  82

\bibitem[{{Wayth} {et~al.}(2015){Wayth}, {Lenc}, {Bell}, {Callingham},
  {Dwarakanath}, {Franzen}, {For}, {Gaensler}, {Hancock}, {Hindson},
  {Hurley-Walker}, {Jackson}, {Johnston-Hollitt}, {Kapi{\'n}ska}, {McKinley},
  {Morgan}, {Offringa}, {Procopio}, {Staveley-Smith}, {Wu}, {Zheng}, {Trott},
  {Bernardi}, {Bowman}, {Briggs}, {Cappallo}, {Corey}, {Deshpande}, {Emrich},
  {Goeke}, {Greenhill}, {Hazelton}, {Kaplan}, {Kasper}, {Kratzenberg},
  {Lonsdale}, {Lynch}, {McWhirter}, {Mitchell}, {Morales}, {Morgan}, {Oberoi},
  {Ord}, {Prabu}, {Rogers}, {Roshi}, {Shankar}, {Srivani}, {Subrahmanyan},
  {Tingay}, {Waterson}, {Webster}, {Whitney}, {Williams}, \&
  {Williams}}]{GLEAM}
{Wayth}, R.~B., {Lenc}, E., {Bell}, M.~E., {et~al.} 2015, \pasa, 32, e025

\bibitem[{{Williams} {et~al.}(2013){Williams}, {Intema}, \&
  {R{\"o}ttgering}}]{Williams13}
{Williams}, W.~L., {Intema}, H.~T., \& {R{\"o}ttgering}, H.~J.~A. 2013, \aap,
  549, A55

\bibitem[{{Williams} {et~al.}(2016){Williams}, {van Weeren}, {R{\"o}ttgering},
  {Best}, {Dijkema}, {de Gasperin}, {Hardcastle}, {Heald}, {Prandoni},
  {Sabater}, {Shimwell}, {Tasse}, {van Bemmel}, {Br{\"u}ggen}, {Brunetti},
  {Conway}, {En{\ss}lin}, {Engels}, {Falcke}, {Ferrari}, {Haverkorn},
  {Jackson}, {Jarvis}, {Kapi{\'n}ska}, {Mahony}, {Miley}, {Morabito},
  {Morganti}, {Orr{\'u}}, {Retana-Montenegro}, {Sridhar}, {Toribio}, {White},
  {Wise}, \& {Zwart}}]{Williams16}
{Williams}, W.~L., {van Weeren}, R.~J., {R{\"o}ttgering}, H.~J.~A., {et~al.}
  2016, \mnras

\bibitem[{{Yatawatta}(2015)}]{Yatawatta15}
{Yatawatta}, S. 2015, \mnras, 449, 4506

\bibitem[{{Yatawatta} {et~al.}(2008){Yatawatta}, {Zaroubi}, {de Bruyn},
  {Koopmans}, \& {Noordam}}]{Yatawatta08}
{Yatawatta}, S., {Zaroubi}, S., {de Bruyn}, G., {Koopmans}, L., \& {Noordam},
  J. 2008, ArXiv e-prints

\bibitem[{{Zwart} {et~al.}(2015){Zwart}, {Wall}, {Karim}, {Jackson}, {Norris},
  {Condon}, {Afonso}, {Heywood}, {Jarvis}, \& {Navarrete}}]{Zwart_2015}
{Zwart}, J., {Wall}, J., {Karim}, A., {et~al.} 2015, in Advancing Astrophysics
  with the Square Kilometre Array (AASKA14), 172

\end{thebibliography}


\end{document}